%% file: bec_journal.tex
\documentclass[11pt,draftcls,journal,letterpaper,oneside,onecolumn]{IEEEtran}
\usepackage[cmex10]{amsmath}
\usepackage{algorithm}
\usepackage{algpseudocode}
\usepackage{amssymb}
\usepackage{graphicx}
\usepackage[noadjust]{cite}
\usepackage{color}
\usepackage{url}
\usepackage{verbatim}
\usepackage{psfrag}

\renewcommand{\vec}[1]{\mbox{\boldmath$#1$}}

\newcommand{\beq}{\begin{equation}}
\newcommand{\eeq}{\end{equation}}
\newcommand{\eq}[1]{(\ref{#1})}

\newcommand{\bup}{\begin{upshape}}
\newcommand{\eup}{\end{upshape}}
\newcommand{\mc}{\mathcal}
\newcommand{\abs}[1]{\lvert#1\rvert}
\newcommand{\fig}[1]{Fig.~\ref{#1}}

\newcommand{\beqsub}{\begin{subequations}}
\newcommand{\eeqsub}{\end{subequations}}

\newcommand{\epsi}{\epsilon}

\newcommand{\NEW}{\textcolor{blue}}

\def\argmax{\mathop{\rm arg\,max}}
\def\argmin{\mathop{\rm arg\,min}}

\newtheorem{lemma}{Lemma}
\newtheorem{corollary}{Corollary}

\newtheorem{theorem}{Theorem}
\newtheorem{definition}{Definition}

\newtheorem{proposition}{Proposition}
\newtheorem{criterion}{Criterion}

\begin{document}

\title{Multiuser Broadcast Erasure Channel with Feedback --- Capacity and Algorithms}

\author{Marios~Gatzianas,~\IEEEmembership{Member,~IEEE,} Leonidas~Georgiadis,~\IEEEmembership{Senior~Member,~IEEE,} and Leandros~Tassiulas,~\IEEEmembership{Fellow,~IEEE}%
\thanks{Part of this work was presented in the 4th Workshop on Network Control and Optimization 
(NetCoop), Ghent, Belgium, Nov.~29--Dec.~1, 2010 and the IEEE International Symposium on Information Theory, Saint Petersburg, Russia, Jul.~31--Aug.~5, 2011.}%
\thanks{M.~Gatzianas is with the Center for Research and Technology Hellas, Informatics \& Telematics 
Institute (CERTH-ITI), 6th km Charilaou-Thermi road, Thessaloniki, 57 001, Greece (e-mail:mgatzia@ee.auth.gr).}%
\thanks{L.~Georgiadis is with the Department of Electrical and Computer Engineering, Division of 
Telecommunications, Aristotle University of Thessaloniki, Thessaloniki, 54 124, Greece and with 
CERTH-ITI (e-mail: leonid@auth.gr).}%
\thanks{L.~Tassiulas is with the Computer Engineering and Telecommunications Department, 
University of Thessaly, Volos, 38 221, Greece and with CERTH-ITI (e-mail: leandros@uth.gr).}}

\maketitle

\begin{abstract} 
  We consider the $N$-user broadcast erasure channel with $N$ unicast
  sessions (one for each user) where receiver feedback is regularly
  sent to the transmitter in the form of ACK/NACK messages. We first
  provide a generic outer bound to the capacity of this system; we
  then propose a virtual-queue-based inter-session mixing coding
  algorithm, determine its rate region and show that it achieves
  capacity under certain conditions on channel statistics, assuming
  that instantaneous feedback is known to all users.  Removing this
  assumption results in a rate region that asymptotically differs from
  the outer bound by 1 bit as $L\to \infty$, where $L$ is the number
  of bits per packet (packet length). For the case of arbitrary
  channel statistics, we present a modification of the previous
  algorithm whose rate region is identical to the outer bound for
  $N=3$, when instant feedback is known to all users, and differs from
  the bound by 1 bit as $L\to \infty$, when the 3 users know only
  their own ACK.  The proposed algorithms do not require any prior
  knowledge of channel statistics.
\end{abstract}


\section{Introduction} \label{intro}

Broadcast channels have been extensively studied by the information
theory community since their introduction in \cite{Co72}. Although
their capacity remains unknown in the general case, special cases have
been solved, including the important category of ``degraded'' channels
\cite{Bergmans_first}. Another class of channels that has received
significant attention is erasure channels, where either the receiver
receives the input symbol unaltered or the input symbol is erased
(i.e.~not received at all) at the receiver. The class of erasure
channels is usually employed as a model for lossy packet networks.

Combining the above classes, a broadcast packet erasure channel
(BPEC) is a suitable abstraction for wireless communications
modeling since it captures the essentially broadcast nature of the
medium as well as the potential for packet loss (due to fading, packet
collision etc). Since this channel is not necessarily degraded, the
computation of its feedback capacity region is an open
problem. Numerous variations of this channel, under different
assumptions, have been studied, a brief summary of which follows.

For multicast traffic, an outer bound to the capacity region of
erasure channels is derived in \cite{DaGo06}, in the form of a
suitably defined minimum cut, and it is proved that the bound can be
achieved by linear coding at intermediate nodes. The broadcast nature
is captured by requiring each node to transmit the same signal on all
its outgoing links, while it is assumed that the destinations have
complete knowledge of any erasures that occurred on all
source-destination paths. In a sense, \cite{DaGo06} is the
``wireless'' counterpart to the classical network coding paradigm of
\cite{Ahl00}, since it carries all results of \cite{Ahl00} (which were
based on the assumption of error-free channels) into the wireless
regime.

The concept of combining packets for efficient transmission based on
receiver feedback is also used in \cite{KeDr08}, where broadcast
traffic is assumed and a rate-optimal, zero-delay, offline algorithm
is presented for 3 users. Online heuristics that attempt to minimize the
decoding delay are also presented. Reference~\cite{Koe09} expands on
this work by presenting an online algorithm that solves at each slot
a (NP-hard) set packing problem in order to decide which packets to
combine. This algorithm also aims in minimizing delay.

Multiple unicast flows, which are traditionally difficult to handle
within the network coding paradigm, are studied in \cite{Wa10} for a
network where each source is connected to a relay as well as to all
destinations, other than its own, and all connections are modeled as
BPECs. A capacity outer bound is presented for an arbitrary number of 
users $N$ and is shown to be achievable for $N=3$ and almost achievable 
for $N=4,5$. The capacity-achieving algorithm operates in two stages
with the relay having knowledge of the destination message side
information at the end of the first stage but not afterward (i.e.~once
the second stage starts, the relay does not receive feedback from the
destinations).

A similar setting is studied in \cite{LaJo06}, where ACK-based packet
combining is proposed and emphasis is placed on the overhead and
complexity requirements of the proposed scheme. An actual
implementation of packet XORing in an intermediate layer
between the IP and 802.11 MAC layers is presented and evaluated in
\cite{KaRa08}, while \cite{Roz07} proposes a replacement for the
802.11 retransmission scheme based on exploiting knowledge of
previously received packets.

This paper expands upon earlier work in \cite{Leo_2user},
\cite{SagEph06}, (which studied the case $N=2$) and differs from the
aforementioned works in that, although it also uses the idea of packet
mixing (in the network coding sense), it introduces additional
concepts and tools that generalize the results concerning achievable
rates to more than 2 users and provides explicit performance
guarantees. Specifically, an outer bound to the feedback capacity
region for multiple unicast flows (one for each user) is computed and,
assuming public feedback is available, two online algorithms (named
$\mbox{\texttt{CODE1}}_{pub}$ and $\mbox{\texttt{CODE2}}_{pub}$) are
presented that achieve this bound under certain conditions on rates
and channel statistics. If public feedback is not available, we
propose modifications to these algorithms that achieve rates within 1
bit/transmission of the outer bound asymptotically in the size of
packet length.

The algorithms do not require any knowledge of channel parameters
(such as erasure probabilities) or future events so that they can be
applied to any BPEC. They use receiver feedback to combine packets
intended for different users into a single packet which is then
transmitted. The combining scheme (i.e.~choosing which packets to
combine and how) relies on a group of virtual queues, maintained in the
transmitter, which are updated based on per-slot available receiver
ACK/NACKs. This queue-based coding concept has also been used in
\cite{SagEph09}, albeit for broadcast traffic with stochastic arrivals
where the stability region of the proposed algorithm becomes
asymptotically optimal as the erasure probability goes to 0, whereas
we consider systems with an arbitrarily fixed number of packets per
unicast session where the capacity is achieved for arbitrary values of
erasure probability.

During the preparation of this paper, we were informed that C.~Wang
has independently studied in \cite{Wang_Kuser} the same problem as
appears here and proposed coding algorithms that achieve capacity
under the same conditions as ours. Although the two works share common
ideas (namely, employing degraded channels to derive capacity outer
bounds and performing packet coding based on receiver feedback), the
proposed algorithms, the procedures for handling overhead, as well as
the methodology used for deriving their rate regions, are quite
different.

The paper is structured as follows. Section~\ref{sysmodel} describes
the exact model under investigation and provides the necessary
definitions in order to derive the capacity outer bound in
Section~\ref{outbound}.  The first coding algorithm is presented in
Section~\ref{algo1}, along with a discussion of the intuition behind
the algorithm and a detailed example. The main properties of the
algorithm are also presented.  The algorithm's optimal performance
under certain conditions on channel statistics and publicly available
feedback is established in Section~\ref{perf}. We also present a
variant of the algorithm that does not require public feedback due to
the incorporation of overhead and determine the corresponding
reduction. A modification of the algorithm that achieves capacity for
3 users under arbitrary channel statistics is presented in
Section~\ref{user3}, while Section~\ref{conclu} concludes the
paper. Appendices \ref{app1}--\ref{app7} contain most of the technical
proofs.

\section{System model and def\/initions} \label{sysmodel}

The system model is a direct extension to $N$ users of the
corresponding model in \cite{Leo_2user} but is nonetheless repeated
for completeness. We study a time-slotted system where a packet of
fixed length $L$ bits is transmitted in each slot. Without loss of
generality, we normalize to unity the actual time required to transmit
a single bit so that the time interval $[(l-1)L\;\, lL)$, for
$l=1,2,\ldots,$ corresponds to slot $l$. The communication system
consists of a single transmitter and a set
$\mc{N}\stackrel{\vartriangle}{=}\{1,2,\ldots,N\}$ of receivers/users
(we hereafter use these two terms interchangeably), while 
the channel is modeled as memoryless broadcast erasure (BE), so that
each broadcast packet is either received unaltered by a user or is
``erased'' (i.e.~the user does not receive the packet). The latter
case is equivalent to considering that the user receives the special
symbol $E$, which is distinct from any other possible transmitted
packet and does not actually map to a physical packet (since it models
an erasure). We hereafter use the term ``packet'' to refer to any
sequence of $L$ bits and the term ``symbol'' to refer to a packet or
an erasure $E$ (we retain however the standard nomenclature of ``input
symbol'' and ``output symbol'', although the former is a true packet
while the latter can also be an erasure).

In information-theoretic terms, the broadcast packet erasure channel is
described by the tuple $(\mc{X},(\mc{Y}_i: i\in
\mc{N}),p(\vec{Y}_l|X_l))$, where $\mc{X}$ is the input symbol
alphabet (we hereafter assume $\mc{X}=\mathbb{F}_q$, with
$\mathbb{F}_q$ a suitable field of size $q$),
$\mc{Y}_i=\mc{Y}=\mc{X}\cup\{E\}$ is the output symbol alphabet (where
$E\not \in \mc{X}$) for user $i$, and $p(\vec{Y}_l|X_l)$ is the
probability of having, at slot $l$, output
$\vec{Y}_l\stackrel{\vartriangle}{=} (Y_{i,l}, i\in \mc{N})$ for a
broadcast input symbol $X_l$.  The memoryless property implies that
$p(\vec{Y}_l|X_l)$ is independent of $l$, so that it is simply written
as $p(\vec{Y}|X)$. Since the transmitted symbols are packets of $L$
bits, we identify $\mathbb{F}_q$ with the set of $L$-bit sequences, so
that it holds $q=2^L$.

Define $Z_{i,l}\stackrel{\vartriangle}{=}\mathbb{I}[Y_{i,l}=E]$ as the
indicator function of an erasure occurring for user $i$ at slot $l$,
and consider the random vector $\vec{Z}_l=(Z_{i,l}, i\in \mc{N})$. The
sequence $\{\vec{Z}_l\}_{l=1}^\infty$ is assumed to consist of
temporally iid vectors (we denote with $\vec{Z}$ the random vector
with distribution equal to that of $\vec{Z}_l$), although, for a fixed
slot, arbitrary correlation between erasures for different users is
allowed. For any index set $\mc{I} \subseteq \mc{N}$, we define
$E_{\mc{I}} \stackrel{\vartriangle}{=}\{ Z_i=1,\; \forall\, i\in\mc{I}
\}= \cap_{i\in \mc{I}} \{ Z_i=1 \}$ as the event that an erasure
occurs for \textit{all} users in $\mc{I}$. We also use the convention
that an intersection over an empty index set yields the entire space
to define $E_{\varnothing}\stackrel{\vartriangle}{=}\Omega$ (the
sample space). We denote $\epsilon_{\mc{I}} \stackrel{\vartriangle}{=}
\Pr(E_{\mc{I}})$ (so that $\epsilon_\varnothing=1$) and, for
simplicity, write $\epsilon_i$ instead of $\epsilon_{\{i\}}$. In order
to avoid trivially degenerate cases, we henceforth assume
$\epsilon_i< 1$ for all $i\in \mc{N}$.

Using the introduced notation, when the transmitter at the beginning
of slot $l$ broadcasts symbol $X_l$, each user $i$ receives symbol
$Y_{i,l}= Z_{i,l} E + (1-Z_{i,l}) X_l$. At the end of each slot $l$,
all users inform the transmitter whether the symbol was received or
not, which is equivalent to each user $i$ sending the value of
$Z_{i,l}$ (essentially, a simple ACK/NACK) through an error-free
zero-delay control channel. 

A channel code, denoted as $(M_1,\ldots,M_N,n)$, for
the broadcast channel with feedback is now defined as the aggregate of
the following components (this is an extension of the standard
definition in \cite{Cover_info} to $N$ users):
\begin{itemize}
\item message sets $\mc{W}_i$ of size $\abs{\mc{W}_i}=M_i$ for each
  user $i\in \mc{N}$, where $\abs{\cdot}$ denotes set
  cardinality. Denote the message that needs to be communicated as
  $\vec{W}\stackrel{\vartriangle}{=}(W_i,i\in \mc{N}) \in \mc{W}$,
  where $\mc{W}\stackrel{\vartriangle}{=}\mc{W}_1\times \ldots \times
  \mc{W}_N$. It will also be helpful to interpret the message set
  $\mc{W}_i$ as follows: assume that user $i$ needs to decode a given
  set $\mc{K}_i$ of $L$-bit packets. Then, $\mc{W}_i$ is the set of
  all possible $\abs{\mc{K}_i} L$ bit sequences, so that it holds
  $\abs{\mc{W}_i}=M_i=2^{\abs{\mc{K}_i} L}$.

\item an encoder that transmits, at slot $l$, a symbol
  $X_l=f_l(\vec{W},\vec{Y}^{l-1})$ belonging to $\mathbb{F}_q$, based
  on the value of $\vec{W}$ and all previously gathered feedback
  $\vec{Y}^{l-1}\stackrel{\vartriangle}{=}(\vec{Y}_1,\ldots,\vec{Y}_{l-1})$. $X_1$
  is a function of $\vec{W}$ only. A total of $n$ symbols are
  transmitted for message $\vec{W}$.

\item $N$ decoders, one for each user $i\in \mc{N}$, represented by
  the decoding functions $g_i: \mc{Y}^n \to \mc{W}_i$, so that the
  reconstructed symbol is $\hat{W}_i=g(Y^n_i)$, where
  $Y^n_i\stackrel{\vartriangle} {=}(Y_{i,1},\ldots,Y_{i,n})$ is the
  sequence of symbols received by user $i$ (including any erasure
  symbols $E$) during the $n$ slots. Thus, the decoding performed by
  user $i$ depends only on packets received by $i$, i.e.~each user
  knows only its own feedback.
\end{itemize} 
Hence, a code \texttt{C} is fully specified by the tuple $\left(
M_1,\ldots,M_N,n,(f_l: l=1,\ldots,n),(g_i:i\in \mc{N}) \right)$,
which contains the message set size along with the encoding/decoding
functions; for brevity, we will simply write $(M_1,\ldots,M_N,n)$ to
denote \texttt{C}. The probability of erroneous decoding for message
$\vec{W}$ is $\lambda_n(\vec{W})=\Pr( \cup_{i\in\mc{N}} \{
g_i(Y^n_i)\neq W_i \} | \vec{W})$. The rate $\vec{R}$ for this code,
measured in information bits per transmitted symbol, is now defined as
the vector $\vec{R}=(R_i:i\in \mc{N})$ with $R_i=(\log_2
M_i)/n$. Hence, it holds $R_i=\abs{\mc{K}_i} L/n$.

Let $\frak{C}$ be a class of $(M_1,\ldots,M_N,n)$ codes. Then, a
vector rate $\vec{R}=(R_1,\ldots,R_N)$ is achievable \textit{under}
$\frak{C}$ if there exists a sequence of codes $(\lceil 2^{n R_1}
\rceil,\ldots, \lceil 2^{n R_N} \rceil,n)$ in $\frak{C}$ such that
$\frac{1}{\abs{\mc{W}}} \sum_{\vec{W}\in \mc{W}} \lambda_n(\vec{W})
\to 0$ as $n\to \infty$. Equivalently, we say that $\frak{C}$ achieves
rate $\vec{R}$. The closure of the set of rates $\vec{R}$ that are
achievable under $\frak{C}$ constitutes the rate region of
$\frak{C}$. We further define a rate $\vec{R}$ to be achievable if
there exists some class $\frak{C}$ of codes that achieves
$\vec{R}$. Finally, the capacity region of a channel is defined as the
closure of the set of all achievable rates, i.e.~the closure of the
union of rate regions of all possible classes of codes $\frak{C}$ for
this channel.

The following definition, introduced in \cite{Bergmans_first}, will be
useful in deriving the outer bound for the capacity of the broadcast
erasure channel.
\begin{definition} \label{def:phydeg}
A broadcast, not necessarily erasure, channel $\left( \mc{X},(\mc{Y}_i:
i\in \mc{N}),p(\vec{Y}|X)\right)$  with receiver set $\mc{N}$ is
physically degraded if there exists a permutation $\hat{\pi}$ 
on $\mc{N}$ such that the sequence $X\to Y_{\hat{\pi}(1)}\to \ldots \to 
Y_{\hat{\pi}(N)}$ forms a Markov chain.
\end{definition}
A generalization to $N$ users of the 2-user proof in
\cite{Gamal_feedback} provides the following result.
\begin{lemma} \label{lem:noinc}
Feedback does not increase the capacity region of a physically
degraded broadcast channel.
\end{lemma}
We now have all necessary tools to compute a capacity outer bound.

\section{Capacity outer bound} \label{outbound}

Our derivation of the capacity outer bound is based on a method
similar to the approaches in
\cite{Wang_Kuser}, \cite{Oz84}--\nocite{ViKa02}\cite{Liu09}. We initially
state a general result on the capacity of broadcast erasure channels
\textit{without feedback} \cite{DaHa05}.
\begin{lemma} \label{lem:cap_nf}
The capacity region (measured in information bits per transmitted
symbol) of a broadcast erasure channel with receiver set $\mc{N}$ and
no feedback is
\beq \label{cap_nf}
\mc{C}_{noFB}=\left\{ \vec{R}\geq \vec{0}: \sum_{i\in \mc{N}} \frac{R_i}{1-\epsilon_i}
  \leq L \right\} ,
\eeq
\end{lemma}
which implies that capacity can be achieved by a simple timesharing scheme.

We denote with $C$ the channel under consideration and, for an
arbitrary permutation $\pi$ on $\mc{N}$, introduce a new,
hypothetical, broadcast channel $\hat{C}_\pi$ with the same
input/output alphabets as $C$ and an erasure indicator function of
$\hat{Z}_{\pi(i),l}= \prod_{j=1}^i Z_{\pi(j),l}$. In other words, a
symbol at slot $l$ is erased by user $\pi(i)$ in $\hat{C}_\pi$ if and
only it is erased by \textit{all} users $\pi(j)$ in channel $C$, with
$j\leq i$, at slot $l$. This occurs with probability
$\hat{\epsilon}_{\pi(i)}\stackrel
{\vartriangle}{=}\epsilon_{\cup_{j=1}^i \{\pi(j)\}}$. A
straightforward calculation reveals that it holds $X\to Y_{\pi(N)}\to
\ldots \to Y_{\pi(1)}$. Hence, choosing the permutation $\hat{\pi}$ in
Definition~\ref{def:phydeg} such that $\hat{\pi}(i)=\pi(N-i+1)$, we
deduce that channel $\hat{C}_{\pi}$ is physically degraded.

In fact, channel $\hat{C}_{\pi}$ can be viewed as an augmentation of
the original channel $C$, where additional error-free virtual channels
are introduced between the receivers. Specifically, each user $\pi(i)$
in $\hat{C}_{\pi}$, for $1\leq i\leq N-1$, sends its output symbol to
user $\pi(i+1)$ through an error-free channel. Hence, any achievable
rate for channel $C$ can also be achieved for $\hat{C}_{\pi}$ using
the same code as in $C$ and ignoring any symbols transmitted through
the virtual channels. Denoting with $\mc{C}_{FB}$,
$\hat{\mc{C}}_{\pi,FB}$ the feedback capacity regions of channels $C$,
$\hat{C}_\pi$, respectively, we conclude that it holds $\mc{C}_{FB}
\subseteq \hat{\mc{C}}_{\pi,FB}$.

The above set inclusion already provides an outer bound to
$\mc{C}_{FB}$. In order to derive this bound, we note that the
previous results imply that the feedback capacity region of the
physically degraded channel $\hat{C}_\pi$ is identical, due to
Lemma~\ref{lem:noinc}, to the capacity region of $\hat{C}_\pi$ without
feedback. The latter is described, in general form, in
Lemma~\ref{lem:cap_nf} whence the following result follows.
\begin{lemma} \label{lem:bound2}
The feedback capacity region of $\hat{C}_\pi$ is given by
\beq
\hat{\mc{C}}_{\pi,FB}= \left\{ \vec{R}\geq \vec{0}: \sum_{i\in \mc{N}} 
  \frac{R_{\pi(i)}}{1-\hat{\epsilon}_{\pi(i)}} \leq L \right\} .
\eeq
\end{lemma}
The above analysis was based on a particular permutation
$\pi$. Considering all $N!$ permutations on $\mc{N}$ provides a
tighter general outer bound.
\begin{lemma} \label{lem:outer}
It holds $\mc{C}_{FB} \subseteq \mc{C}^{out} \stackrel{\vartriangle}{=}
\cap_{\pi\in \mc{P}} \hat{\mc{C}}_{\pi,FB}$, where $\mc{P}$ is the set of
all possible permutations on $\mc{N}$.
\end{lemma}

The outer bound $\mc{C}^{out}$ has been derived based on the decoding
rule in Section~\ref{sysmodel}, i.e.~each user in channel $C$ knows
only its own feedback at each slot (hereafter referred to as
``private'' feedback). This raises a question regarding whether this
bound is also valid for publicly available feedback (i.e.~when each
user in $C$ knows the feedback from all other users at each
slot). This question can be answered in the affirmative by extending
the bounding arguments in the recent work of \cite{CzPrDi12_ISIT},
which considered the case $N=3$ and public feedback (which corresponds
to a decoding function of the form $g_i(Y^n_i,\vec{Z}^n)$), to general
$N$. Since the use of public feedback simplifies the presentation of
the proposed algorithms, we initially assume that public feedback is
available and propose a coding algorithm named
$\mbox{\texttt{CODE1}}_{pub}$. We remove this assumption later in
Section~\ref{overhead} by proposing a simple overhead scheme on top of
the former algorithm, which leads to a new algorithm, named
$\mbox{\texttt{CODE1}}_{pri}$, that only requires private feedback.

\section{A class of codes} \label{algo1}

In this Section, we present a class of codes, collectively referred to
as algorithm $\mbox{\texttt{CODE1}}_{pub}$ (the index emphasizes the
assumption of public feedback), and describe the basic properties that
guarantee its correctness.

\subsection{The intuition behind the algorithm}

Before the algorithm's description, a brief discussion of its
underlying rationale will be useful. Since each user $i$ must decode
exactly the $\abs{\mc{K}_i}$ packets in its session and a packet is an
$L$-bit representation of an element in $\mathbb{F}_q$, the
transmitter transmits appropriate linear combinations of packets so
that each user $i$ eventually receives $\abs{\mc{K}_i}$
linearly independent combinations of the packets in $\mc{K}_i$. Hence,
all quantities appearing in subsequent expressions are elements of
$\mathbb{F}_q$ and all linear operations are performed in
$\mathbb{F}_q$.
 
The algorithm's operation can be summarized as follows: the
transmitter maintains a set of virtual queues $Q_{\mc{S}}$, indexed by
all non empty subsets $\mc{S}\subseteq \mc{N}$ and properly
initialized, as well as queues $Q_{D_i}$ for $i\in \mc{N}$. The queues
$Q_{D_i}$ contain copies of the packets that have been successfully
received by user $i\in \mc{N}$. The algorithm processes each queue
$Q_{\mc{S}}$ sequentially; during the processing of each queue
$Q_{\mc{S}}$, the packet $s$ to be transmitted next is selected as a
linear combination of all packets currently stored in $Q_{\mc{S}}$,
i.e.~$s=\sum_{p\in Q_{\mc{S}}} a_s(p) p$, where $a_s(p)$ are suitably
chosen coefficients in $\mathbb{F}_q$. Notice that, unless $a_s(p)$ is
non-zero for exactly one $p\in Q_{\mc{S}}$, the transmitted packet $s$
is not actually stored in $Q_{\mc{S}}$ but is created on-the-fly.

After transmitting $s$, the transmitter gets the ACK/NACKs for $s$
from the receivers and (depending on which users successfully received
$s$) potentially adds packet $s$ into a single queue
$Q_{\mc{S}^\prime}$, with $\mc{S}^\prime \supset \mc{S}$, and/or to
queues $Q_{D_i}$, for all users $i$ that received the packet. Some
additional bookkeeping, to be described in detail in Section
\ref{algdec}, is also performed. The algorithm terminates when all
queues $Q_{\mc{S}}$ have been processed, at which point each user $i$
can decode its original packets based on the packets contained in
$Q_{D_i}$.

A central concept in the proposed algorithm is the notion of ``token''
which is defined as follows.
\begin{definition} \label{tokeness}
A packet $s$ is a token for user $i$ iff $s$ can be written in the form
\beq \label{tok}
s=\sum_{p\in \mc{K}_i} b^{(i)}_s(p) p+c^{(i)}_s ,
\eeq
where $\vec{b}^{(i)}_s\stackrel{\vartriangle}{=} (b^{(i)}_s(p), \;
p\in \mc{K}_i)$, $c^{(i)}_s \in \mathbb{F}_q$ are known to user
$i$. We call $\vec{b}^{(i)}_s$ the ``coefficient vector'' of packet
$s$ for user $i$.
\end{definition}
In words, a token for user $i$ is any packet $s$ that allows $i$, upon
reception of $s$, to effectively construct a linear equation with the
packets in $\mc{K}_i$ as unknowns (since $\vec{b}^{(i)}_s$,
$c^{(i)}_s$ are known). For efficiency reasons, this equation should
ideally be linearly independent w.r.t.~all equations constructed by
user $i$ through the previously received packets (equivalently,
$\vec{b}^{(i)}_s\not\in span(\{ \vec{b}^{(i)}_{s^\prime}: s^\prime
\mbox{ received by } i \mbox { prior to } s \})$). In this case,
borrowing from network coding terminology, the packet is considered to
be an ``innovative'' token.

Hence, each user $i$ must receive $\abs{\mc{K}_i}$ innovative tokens
in order to decode its packets, at which point the algorithm
stops. Notice that it is possible, and actually very desirable for
throughput purposes, for a packet to simultaneously be a token (better
yet, innovative token) for multiple users. In the context of this
paper, we introduce the related, but not identical, notion of a
``\textit{Basis}'' token, rigorously defined in
Section~\ref{prop_code1}, which is needed for the proof of the
algorithm's correctness and its performance analysis. However, we will
still use the notion of ``innovative'' token to gain some intuition
into the algorithm.


An important remark that follows from the previous discussion is that
a token for user $i$ may not only be non-innovative for $i$, but it
may actually ``contain'' no packet intended for it,
i.e.~$b^{(i)}_s(p)=0$ for all $p\in \mc{K}_i$. For example, consider
the case where the transmitter sends a packet $s=p_1$, where $p_1 \in
\mc{K}_1$ and $s$ is received by user 2 only. Using the delta
Kronecker $\delta_{m,n}$ notation and setting $\vec{b}^{(1)}_s=
(\delta_{p,p_1}:p\in \mc{K}_1)$, $c^{(1)}_s=0$,
$\vec{b}^{(2)}_s=\vec{0}$ and $c^{(2)}_s=p_1$, it is easy to see that
$s$ is a token, according to Definition~\ref{tokeness}, for both users
1, 2 and none other. However, packet $s$ contains no packet intended
for 2, so that one could deduce that this slot was ``wasted''. Of
course, this is not actually the case (i.e.~the slot was not really
``wasted'') since user 2 gained some side information, so that the
question now becomes how to optimally exploit the side information
obtained through overhearing.

A distinctive characteristic of the proposed algorithm is that it
efficiently exploits such cases (where users receive packets that are
of no direct interest to them) by placing the packets into proper queues
instead of discarding them. This results in better opportunities for
efficient packet combinations in the future by creating simultaneous
innovative tokens for multiple users and essentially compensating for
previously ``wasted'' slots. The crux of the algorithm is in the careful
bookkeeping required to handle these cases in an efficient manner and
ensure that all users eventually receive the necessary number of
innovative tokens.

The following proposition, which establishes that any linear
combination of tokens is a new token (not necessarily innovative),
will be useful.
\begin{proposition} \label{keep_tok} 
  Consider a set of packets in a queue $Q$ and a set of users $\mc{S}$
  such that each packet $p\in Q$ is a token for all users $i\in
  \mc{S}$.  Then, any linear combination $s=\sum_{p\in Q} a_s(p) p$ of
  the packets in $Q$ is a token for all $i\in \mc{S}$, provided that
  $a_s(p)$ are known to all users $i\in \mc{S}$.
\end{proposition}
The above proposition is easily proved by noting that each packet/token 
$p\in Q$ for user $i\in\mc{S}$ can be written as $p=\sum_{u\in \mc{K}_i} 
b^{(i)}_p(u) u+ c^{(i)}_p$, whence it follows 
\beq \label{still_tok}
s=\sum_{u\in\mc{K}_i} \underbrace{ \left( \sum_{p\in Q} a_s(p) b^{(i)}_p(u)
   \right) }_{b^{(i)}_s(u)} u 
  + \underbrace{ \left( \sum_{p\in Q} a_s(p) c^{(i)}_p \right) }_{c^{(i)}_s} ,
\eeq
so that $s$ is still a token for each $i\in \mc{S}$.

\subsection{Description of algorithm $\mbox{\texttt{CODE1}}_{pub}$} \label{algdec}

\begin{figure}[t]
\centering
\begin{algorithmic}[1]
\Statex \hspace{-1.5em} \hrulefill {}
\Statex \vspace{-4pt} \hspace{-1.5em} \textbf{Algorithm} $\mbox{\texttt{CODE1}}_{pub}$
\Statex \vspace{-\baselineskip} \vspace{8pt} \hspace{-1.5em} \hrulefill {}
\State initialize $Q_{\mc{S}}$, $K^i_{\mc{S}}$ for all $\mc{S}\subseteq 
\mc{N}$ and $i\in\mc{S}$ and $Q_{D_i}$, $K_{D_i}$ for all $i\in \mc{N}$; 
\State $t\leftarrow 0$;
\For{$\ell \gets 1,\ldots, N$}
\ForAll{($Q_{\mc{S}}$ with $\abs{\mc{S}}=\ell$})  \Comment{arbitrary order of processing}
\While{($K^i_{\mc{S}}(t) >0$ for at least one $i\in\mc{S}$)}
\State compute \textit{suitable} coefficients $(a_s(p), p\in Q_{\mc{S}})$;
\State transmit packet $s=\sum_{p\in Q_{\mc{S}}} a_s(p) p$; 
\State apply procedure \texttt{ACTFB1} based on receiver feedback for $s$; 
\State $t\leftarrow t+1$;
\EndWhile   
\EndFor
\EndFor
\end{algorithmic}
\caption{Pseudocode for algorithm $\mbox{\texttt{CODE1}}_{pub}$.}
\label{CODE1}
\end{figure}

\begin{figure}[t]
\centering
\includegraphics[scale=0.4]{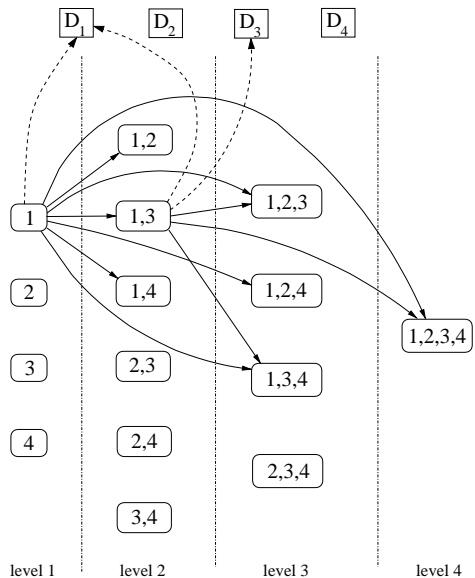}
\caption{Transmitter side virtual queue network for 4 users. Each oval
  box represents the queue indexed by the corresponding subset of
  $\{1,2,3,4\}$, while solid and dotted line arrows indicate potential
  packet movement between the queues.}
\label{4users}
\end{figure}

Algorithm $\mbox{\texttt{CODE1}}_{pub}$ is succinctly described in
pseudocode form in \fig{CODE1}. Specifically, the transmitter
maintains a network of virtual queues $Q_{\mc{S}}$, indexed by the
non-empty subsets $\mc{S}$ of $\mc{N}$, as well as $N$ queues denoted
as $Q_{D_i}$, for $i\in \mc{N}$. \fig{4users} provides an illustration
for 4 users, where an oval box represents $Q_{\mc{S}}$, for the
corresponding set $\mc{S}$ appearing as the box label, a square box
represents queue $Q_{D_i}$ and the vertical lines are used to classify
the queues into ``levels'', as will be explained below. The solid
(dotted) line arrows indicate potential packet movement into a queue
$Q_{\mc{S}}$ ($Q_{D_i}$). For graphical clarity, \fig{4users} only
shows the packet movements originating from queues $Q_{\{1\}}$,
$Q_{\{1,3\}}$; however, similar packet movements are allowed for the
other queues, as will be explained soon.
\subsubsection{Initialization}
All $Q_{D_i}$ queues are initially empty while $Q_{\mc{S}}$ are
initialized with the unicast packets as follows:
\beq \label{initQ}
Q_{\mc{S}}(0)= \left\{ \begin{array}{l@{\quad}l} \mc{K}_i & \mbox{if } 
  \mc{S}=\{i\}, \\ \varnothing & \mbox{otherwise.} \end{array} \right.
\eeq
The performed initialization guarantees that all packets placed in
queues $Q_{\{i\}}$ are tokens for user $i\in \mathcal{N}$ according to 
Definition~\ref{tokeness}.

Additionally, the algorithm keeps track of non-negative integer
indices $K_{D_i}$, $K^i_{\mc{S}}$. The former are associated to queues
$Q_{D_i}$, for all $i\in \mc{N}$, while the latter are associated to
queues $Q_{\mc{S}}$, for all $\mc{S}\subseteq \mc{N}$, $i\in
\mc{S}$. The indices $K_{D_i}$ are initialized to $0$ for all $i\in
\mc{N}$, while $K^i_{\mc{S}}$ are initialized as
\beq \label{initT}
K^i_{\mc{S}}(0)=\left\{ \begin{array}{l@{\quad}l} \abs{\mc{K}_i} &
  \mbox{if } \mc{S}=\{i\}, \\ 0 & \mbox{otherwise.} \end{array} \right.
\eeq
The entities $Q_{\mc{S}}$, $Q_{D_i}$, $K^i_{\mc{S}}$, $K_{D_i}$ will
be dynamically updated during the algorithm's execution (depending on
the exact ACK/NACKs reported by the users), which is why we placed an
explicit time dependence in \eq{initQ}, \eq{initT}. In fact, the
following note on notation will be useful: we write $K^i_{\mc{S}}$ to
refer to the index when the exact instant at which the index is
examined is unimportant (this is akin to using a variable name in a
programming language: although the contents of the variable may change
over time, we can always refer to the variable by name). We write
$K^i_{\mc{S}}(t)$ when we specifically refer to the value of the index
at time $t$. Furthermore, since the values of these indices depend on
the erasures that occur, $K^i_{\mc{S}}(t)$ is actually a random
variable. We will use the notation $\dot{K}^i_{\mc{S}}(t)$ (or
$\dot{K}^i_{\mc{S}}$ when time is unimportant) when we want to
emphasize the random nature of the indices.

For each user $i\in \mc{N}$, the algorithm also keeps track of
subsets, denoted as $\mc{B}^{(i)}_{\mc{S}}$ (for $\mc{S}\subseteq
\mc{N}$ with $i\in \mc{S}$, and $\mc{B}_{D_i}$, for $i\in \mc{N}$), of
coefficient vectors $\vec{b}^{(i)}_s$ of tokens $s$ for user $i$
stored in $Q_{\mc{S}}$, $Q_{D_i}$ (respectively). These coefficient
vector sets, which will be seen to have the important property that
they can be selected so that their union forms a basis for vector
space $\mathbb{F}^{\abs{\mc{K}_i}}_q$ for all $i\in \mathcal{N}$, are
initialized as $\mc{B}_{D_i}(0)=\varnothing$ for $i\in \mc{N}$ and
\beq \label{initB}
\mc{B}^{(i)}_{\mc{S}}(0)= \left\{ \begin{array}{l@{\quad}l} standard\_basis(
  \mathbb{F}^{\abs{\mc{K}_i}}_q) & \mbox{if } \mc{S}=\{i\}, \\ \varnothing & 
  \mbox{otherwise,} \end{array} \right. 
\eeq
where the $standard\_basis$ of an $\abs{\mc{K}_i}$-dimensional vector
space is the set of vectors $\vec{e}_i$ which have all components
equal to zero except for the $i$-th component, which is set to one.

\subsubsection{Encoding}

We define as ``level $\ell$'' the groups of all queues $Q_{\mc{S}}$
with $\abs{\mc{S}}=\ell$. The algorithm operates in $N$ phases so that
in phase $\ell$, with $1\leq \ell \leq N$, only transmissions of
linear combinations of packets in one of the queues in level $\ell$
occur. Specifically, at phase $\ell$, the transmitter orders the
queues in level $\ell$ according to a predetermined rule, known to all
users (say, according to lexicographic order of the index set
$\mc{S}$, which corresponds to the top-to-bottom ordering shown in
\fig{4users}). The transmitter then examines the first, according to
this order, queue $Q_{\mc{S}}$ and transmits a packet $s$ that is a
linear combination of all packets in $Q_{\mc{S}}$, i.e.
\beq \label{shuffle}
s=\sum_{p\in Q_{\mc{S}}} a_s(p) p .
\eeq
We slightly abuse parlance and say that ``$s$ is transmitted from
$Q_{\mc{S}}$'', although it is clear that $s$ is not actually stored
in $Q_{\mc{S}}$ but is created on-the-fly. Proposition~\ref{keep_tok}
guarantees that $s$ is a token for all users $i\in\mc{S}$, provided
that all packets $p\in Q_{\mc{S}}$ are also tokens for all $i\in
\mc{S}$.

The exact generation method for $a_s(p)$ is unimportant as long as two
general criteria are met.
\begin{criterion} \label{minass1}
  The procedure for generating $a_s(p)$ is known to all users, so that
  they can always reproduce the values of $a_s(p)$ even when they
  don't receive the packet $s$. This implies that the receivers also
  know the size of all queues $Q_{\mc{S}}$, $\mc{S}\subseteq \mc{N}$,
  at all times.
\end{criterion}
  
\begin{criterion} \label{minass2}
Assume that at the beginning of slot $t$, there exist (possibly empty)
sets of vectors $\mc{B}_{D_i}(t)\subseteq \{ \vec{b}^{(i)}_p: p\in
Q_{D_i}(t) \}$, for all $i\in \mc{N}$, and $\mc{B}^{(i)}_{\mc{I}}(t)
\subseteq \{ \vec{b}^{(i)}_p: p\in Q_{\mc{I}}(t) \}$, for all
$\mc{I}\subseteq \mc{N}$ and $i\in \mc{I}$, with the following
properties (note that for $t=0$, these properties automatically hold
by selecting $\mc{B}^{(i)}_{\mc{I}}(0)$, $\mc{B}_{D_i}(0)$ according
to \eq{initB}):
\beq \begin{split} \label{eq:prop1}
& \abs{\mc{B}^{(i)}_{\mc{I}}(t)}= K^i_{\mc{I}}(t) \mbox{ and }
  \abs{\mc{B}_{D_i}(t)}=K_{D_i}(t) , \\
& \mc{B}_{D_i}(t) \cup \bigcup_{\substack{\mc{I}: \mc{I} \subseteq \mc{N}\\ 
  K^i_{\mc{I}}(t)>0}} \mc{B}^{(i)}_{\mc{I}}(t) \mbox{ is a basis of }
  \mathbb{F}^{\abs{\mc{K}_i}}_q \mbox{ for all } i\in \mc{N} ,
\end{split} \eeq
and, for each $i\in \mc{S}$ with $K^i_{\mc{S}}(t)>0$, we pick an
\textit{arbitrary} $\hat{\vec{b}}_i\in
\mc{B}^{(i)}_{\mc{S}}(t)$. Then, the generating algorithm for $a_s(p)$
should return as output any $(a_s(p):p\in Q_{\mc{S}})$ such that the
transmitted packet $s=\sum_{p\in Q_{\mc{S}}} a_s(p) p$ has a
corresponding coefficient vector $\vec{b}^{(i)}_s$ with the property
\beq \label{eq:prop2}
\{ \vec{b}^{(i)}_s \} \cup \mc{B}_{D_i}(t) \cup \bigcup_{\substack{\mc{I}: 
  \mc{I} \subseteq \mc{N}\\ K^i_{\mc{I}}(t)>0}} \mc{B}^{(i)}_{\mc{I}}(t) - 
  \{ \hat{\vec{b}}_i \} \mbox{ is a basis of } \mathbb{F}^{\abs{\mc{K}_i}}_q 
  \mbox{ for all } i\in \mc{S} \mbox{ with } K^i_{\mc{S}}(t)>0 .
\eeq
\end{criterion}

The above Criteria should be interpreted as two tests that any
generating algorithm should pass, and conformance to these criteria is
what the term ``suitable coefficients'' appearing in line 6 of
\fig{CODE1} actually means. It is important to note that
Criterion~\ref{minass2} is essentially a conditional result: it
requires that the generator of $a_s(p)$ returns an output that
satisfies \eq{eq:prop2} \textit{provided} that there exist sets
$\mc{B}^{(i)}_{\mc{I}}(t)$, $\mc{B}_{D_i}(t)$ that satisfy
\eq{eq:prop1}, without making any claims about the \textit{actual}
existence of these sets in the first place. It will be shown later
(Lemmas~\ref{cruc}, \ref{trueind}) that there actually exist sets
$\mc{B}^{(i)}_{\mc{I}}(t)$, $\mc{B}_{D_i}(t)$ that satisfy
\eq{eq:prop1} and, furthermore, there always exist $(a_s(p): p\in
Q_{\mc{S}})$ that satisfy \eq{eq:prop2} for $L> \log_2 N$.

Of the two Criteria, the second one is clearly the more difficult to
satisfy. It will be shown that if coefficients $a_s(p)$ are selected
so as to satisfy both Criteria, all users in $\mc{N}$ will eventually
receive a sufficient number of packets to individually solve a linear
system that has a full rank matrix w.p.~1. Criterion 2 can be relaxed
so that the generator of $a_s(p)$ returns output that satisfies
\eq{eq:prop2} with probability arbitrarily close to 1; this choice
leads to a simple generator for $a_s(p)$ based on random
selection. Both variants of Criterion~\ref{minass2} can be satisfied
by choosing a sufficiently large field size $q$; however, for ease of
presentation, we only consider the case where \eq{eq:prop2} is true
w.p.~1.

\subsubsection{Feedback-based actions}

Once the linear combination $s$, in the form of \eq{shuffle}, is
transmitted from $Q_{\mc{S}}$ at slot $t$ and the transmitter receives
the corresponding feedback from all users, the following actions (or
steps), collectively referred to as \texttt{ACTFB1}, are taken (all 4
cases must be examined, since they are not mutually exclusive). We
denote with $\mc{G}$ the set of users that successfully received $s$
and omit the $t$ dependence from all $K^i_{\mc{S}}$ indices. \\
\underline{\texttt{ACTFB1} actions:}
\begin{enumerate}
\item if no user in $\mc{N}$ receives $s$, it is retransmitted. \label{retrans}

\item if it holds $\mc{G}\subseteq \mc{S}$ and $K^i_{\mc{G}}=0$ for
  all $i\in \mc{G}$, then $s$ is retransmitted.\label{nonewtok}

\item for each user $i\in \mc{S}$ that receives $s$ \textit{and}
  satisfies $K^i_{\mc{S}}>0$, $K^i_{\mc{S}}$ is decreased by 1 and
  $K_{D_i}$ is increased by 1. \label{gaintok}

\item if $s$ has been erased by at least one user $i\in \mc{S}$ \textit{and}
it holds $\mc{G}\cap (\mc{N}-\mc{S}) \neq \varnothing$, then
\label{movtok}
\begin{itemize}
\item packet $s$ is added to queue $Q_{\mc{S}\cup\mc{G}}$.

\item for each user $i\in \mc{S}$ that erased $s$ \textit{and}
satisfies $K^i_{\mc{S}}>0$, $K^i_{\mc{S}}$ is decreased by 1 and
$K^i_{\mc{S}\cup\mc{G}}$ is increased by 1.
\end{itemize}
\end{enumerate}
No new coefficients are produced for the retransmissions in steps 1,
2. \fig{4users} presents the permissible token movements from queues
$Q_{\{1\}}$, $Q_{\{1,3\}}$ that occur in step~\ref{movtok} of
\texttt{ACTFB1}, where, for graphical clarity, transitions from the
other queues are not shown (dashed lines correspond to
step~\ref{gaintok} of \texttt{ACTFB1}). Hence, a packet $s$
transmitted from $Q_{\mc{S}}$ can only be moved to a queue
$Q_{\mc{T}}$ with $\mc{T} \supset \mc{S}$ (a copy of the packet is
also added to queue $Q_{D_i}$ if $s$ was correctly received by user
$i$). 

\subsubsection{Algorithm termination and decoding procedure}
Processing of $Q_{\mc{S}}$ (i.e.~transmission of linear combinations
of packets from $Q_{\mc{S}}$) continues for as long as there exists at
least one $i\in \mc{S}$ with $K^i_{\mc{S}}>0$. When it holds
$K^i_{\mc{S}}=0$ for all $i\in\mc{S}$, the transmitter moves to the
next queue $Q_{\mc{S}^\prime}$ in level $\ell$ and repeats the above
procedure until it has processed all queues in level $\ell$. When this
occurs, phase $\ell$ is complete and the algorithm moves to phase
$\ell+1$, where it processes the queues in level $\ell+1$.

Since the session length $\vec{K}=(\abs{\mc{K}_i}: i\in \mc{N})$ and
the exact algorithm for generating coefficients $a_s(p)$ are known to
\textit{all} users before execution of $\mbox{\texttt{CODE1}}_{pub}$
begins, the presence of public feedback implies that, at the end of
each slot, \textit{all} users individually have exactly the same
feedback information as the transmitter. Hence, they can ``replay''
the execution of $\mbox{\texttt{CODE1}}_{pub}$ in real time and
iteratively compute $\vec{b}^{(i)}_s$, $c^i_s$ for each transmitted
packet through \eq{still_tok} so that, by the time
$\mbox{\texttt{CODE1}}_{pub}$ terminates at the end of phase $N$, each
user $i$ has received sufficiently many tokens (i.e.~packets stored in
$Q_{D_i}$) to solve the related system of equations and decode the
packets in $\mc{K}_i$.

\subsection{Properties and correctness of $\mbox{\texttt{CODE1}}_{pub}$} \label{prop_code1}

The following two Lemmas, proved in Appendices~\ref{app1}, \ref{app2}, 
respectively, contain all important properties of $\mbox{\texttt{CODE1}}_{pub}$, 
as they follow from its construction.
\begin{lemma} \label{cruc}
During the execution of $\mbox{\texttt{CODE1}}_{pub}$, the following statements are true:
\begin{enumerate}
\item Any packet $s$ that is stored in a queue $Q_{\mc{S}}$ at slot $t$, with
$\abs{\mc{S}}\geq 2$, is a linear combination of all packets in queue
$Q_{\mc{I}_s}$ (for some non-empty set $\mc{I}_s \subset \mc{S}$) that has
been transmitted at some prior slot $\tau<t$ and
received (at slot $\tau$) by all users in set
$\mc{S}-\mc{I}_s$ and erased by all users in set $\mc{N}-\mc{S}$. 

\item Any packet $s$ stored in queue $Q_{\mc{S}}$ can be decomposed as
  $s=\sum_{u\in \cup_{j\in \mc{S}} \mc{K}_j} \tilde{a}_s(u) u$,
  i.e.~packet $s$ is effectively a linear combination of packets
  destined for users in set $\mc{S}$ only.

\item Any packet $s$ stored in $Q_{\mc{S}}$ is a token for all $i\in
  \mc{S}$ (and only these $i\in \mc{S}$).

\item When transmitting a linear combination $s$ from $Q_{\mc{S}}$ at
  slot $t$, there always exist coefficients $a_{s}(p)$ that satisfy
  \eq{eq:prop2} of Criterion~\ref{minass2}, provided that there exist
  sets that satisfy \eq{eq:prop1} and it holds $L> \log_2 N$.
\end{enumerate}
\end{lemma}
The relation $L>\log_2 N$ will be assumed for the remainder of the
paper, so that all subsequent results (Theorems, Lemmas etc) are based
on this assumption. The following result essentially shows that each
user $i\in \mc{N}$ is able to decode its packets by the end of
$\mbox{\texttt{CODE1}}_{pub}$'s execution. The result is proved by
induction, using the algorithm's initialization and the fourth
statement in Lemma~\ref{cruc} to establish the crucial inductive step.
\begin{lemma} \label{trueind}
  Under the application of $\mbox{\texttt{CODE1}}_{pub}$, the following condition is
  true at the \textit{beginning} of each slot $t$: there exist vector
  sets $\mc{B}^{(i)}_{\mc{I}}(t) \subseteq \{ \vec{b}^{(i)}_p: p\in
  Q_{\mc{I}}(t) \}$, for all $\mc{I}\subseteq \mc{N}$ and $i\in
  \mc{I}$, and $\mc{B}_{D_i}(t) \subseteq \{ \vec{b}^{(i)}_p: p\in
  Q_{D_i}(t) \}$, for all $i\in \mathcal{N}$, such that
\begin{itemize}
\item $\abs{\mc{B}^{(i)}_{\mc{I}}(t)}= K^i_{\mc{I}}(t)$ and $\abs{\mc{B}_{D_i}(t)}=K_{D_i}(t)$.

\item $\mc{B}_{D_i}(t) \cup \bigcup_{\substack{\mc{I}:\mc{I}\subseteq \mc{N} 
\\ K^i_{\mc{I}}(t)>0 }} \mc{B}^{(i)}_{\mc{I}}(t)$ is a basis of $\mathbb{F}^
{\abs{\mc{K}_i}}_q$ for all $i\in \mc{N}$.
\end{itemize}
\end{lemma}
The existence of the above sets motivates the following definition.
\begin{definition} \label{definnov}
A packet $p$ is called a \textit{Basis} token for user $i\in \mc{N}$ at slot 
$t$ iff $\vec{b}^{(i)}_p \in \mc{B}_{D_i}(t) \cup \bigcup_{\substack{\mc{I}:\mc{I}
\subseteq \mc{N} \\ K^i_{\mc{I}}(t)>0}} \mc{B}^{(i)}_{\mc{I}}(t)$.
\end{definition}

Clearly, at the beginning of the slot $t_{end}$ immediately after the
completion of phase $N$, Lemma~\ref{trueind} implies (since
$K^i_{\mc{I}}(t_{end})=0$ for all $i$, $\mc{I}$) that
$\mc{B}_{D_i}(t_{end})$ is a basis of $\mathbb{F}^{\abs{\mc{K}_i}}_q$,
for all $i\in \mc{N}$. Hence, each user $i$ has received
$\abs{\mc{K}_i}$ linearly independent tokens (i.e.~\textit{Basis}
tokens) and can decode its packets on a one-shot manner by solving the
corresponding system of equations, using the \textit{Basis} tokens in
$\mc{B}_{D_i}$. Since this result holds for arbitrary channel
statistics, $\mbox{\texttt{CODE1}}_{pub}$ is, in principle,
universally applicable. In addition, no prior knowledge of channel
statistics is required for its execution.

\subsection{Some further intuitive remarks}

In retrospect, the combination of Lemmas~\ref{cruc}, \ref{trueind} and
their methods of proof give a very intuitive explanation to the
algorithm's operation, which we provide next. The sets
$\mc{B}^{(i)}_{\mc{I}}(t)$ contain the vectors that span the
\textit{subspace} to which the $\vec{b}^{(i)}_s$ vector of any packet
$s$ received by user $i$ from queue $Q_{\mc{I}}$ at slot $t$ must
belong in order to provide ``useful'' information to $i$ (i.e.~allow
$i$ to create an equation, w.r.t.~packets in set $\mc{K}_i$, from the
received $s$ that is linearly independent w.r.t~all previously created
equations by user $i$). This follows from the fact that, for all $i\in
\mc{N}$, any vector in $span(\mc{B}^{(i)}_{\mc{I}}(t))$ is linearly
independent w.r.t.~the vectors in $\mc{B}_{D_i}(t)$ (i.e.~the space
spanned by the coefficient vectors of the tokens already received by
user $i$), since the union of all these vector sets constitutes a
basis of $\mathbb{F}^{\abs{\mc{K}_i}}_q$. Similarly, $K^i_{\mc{I}}(t)$
is the \textit{number} of the elements of the basis of
$\mathbb{F}^{\abs{\mc{K}_i}}_q$ that belong to
$\mc{B}^{(i)}_{\mc{I}}(t)$.

Furthermore, by the algorithm's construction and
Proposition~\ref{keep_tok}, only the users $i\in \mc{S}$ can have
\textit{Basis} coefficient vectors corresponding to packets stored in
$Q_{\mc{S}}$. This is due to Lemma~\ref{cruc}, which states that any
linear combination of packets in $Q_{\mc{S}}$ contains packets that
are intended for users $i\in \mc{S}$ only. Similarly, Criterion~2 can
be intuitively summarized as follows: when the algorithm processes
queue $Q_{\mc{S}}$ and selects a packet $s=\sum_{p\in Q_{\mc{S}}}
a_s(p) p$ for transmission at slot $t$, we should select $a_s(p)$ such
that $s$ is an innovative token for \textit{all} $i\in \mc{S}$ with
$K^i_{\mc{S}}(t)>0$, \textit{provided} that there exist certain sets
with specific properties at slot $t$. The existence of these sets is
guaranteed again by Lemma~\ref{trueind}.

Regarding the rationale behind \texttt{ACTFB1}, step~\ref{gaintok} of
\texttt{ACTFB1} is equivalent to saying that when user $i$ receives a
``useful'' token at slot $t$ (meaning that $K^i_{\mc{S}}(t)>0$ so that
there remain \textit{Basis} tokens to receive) from $Q_{\mc{S}}$, this
token should be added to $\mc{B}_{D_i}$ (with a corresponding increase
to $K_{D_i}$), so that it becomes a \textit{Basis} token for user $i$
at slot $t+1$. If this is not the case and there exist users,
comprising set $\mc{G}\subseteq \mc{N}-\mc{S}$, who receive this
packet (step~\ref{movtok} of \texttt{ACTFB1}), then the packet has
become a token for users in $\mc{S}\cup\mc{G}$ and should be placed in
queue $Q_{\mc{S}\cup \mc{G}}$. This allows the token to be
simultaneously received by multiple users in the future and thus
compensate for the current loss. Additionally, since user $i$ can now
recover this token more efficiently from $Q_{\mc{S}\cup\mc{G}}$
instead of $Q_{\mc{S}}$, the indices $K^i_{\mc{S}}$,
$K^i_{\mc{S}\cup\mc{G}}$ should be modified accordingly to account for
the token transition. Step~\ref{nonewtok} of \texttt{ACTFB1} merely
states that the packet is retransmitted when it is only received by
users $i$ who have already recovered from the queue all innovative
tokens intended for them (i.e.~$\mc{B}^{(i)}_{\mc{S}}$ is empty).

If $K^i_{\mc{S}}$ becomes 0 at the end of some slot $\tilde{t}$,
queue $Q_{\mc{S}}$ is no longer useful for user $i$, since all
linearly independent combinations that could be created from
$Q_{\mc{S}}$ have either been received by $i$ or stored in higher
level queues (due to step~\ref{movtok} of \texttt{ACTFB1}) for future
recovery by $i$. Of course, the queue is still useful for any other
users $j\in \mc{S}$ with $K^j_{\mc{S}}(\tilde{t})>0$.

\subsection{An example of execution of $\mbox{\texttt{CODE1}}_{pub}$}

\input{example.tex}

\section{Performance analysis for $\mbox{\texttt{CODE1}}_{pub}$} \label{perf}

In this Section, we analyze the performance of
$\mbox{\texttt{CODE1}}_{pub}$ for arbitrary channel statistics and
conclude that $\mbox{\texttt{CODE1}}_{pub}$ achieves the capacity
outer bound of Lemma~\ref{lem:outer} (i.e.~achieves capacity),
provided that the users in $\mc{N}$ can be ordered according to a
specific relation that depends on channel statistics and the chosen
rates; this provision is shown to be true for the special case of
symmetric channels, i.e.~channels which satisfy the condition
$\epsilon_{\mc{I}}=\epsilon_{\mc{J}}$, for all $\mc{I},\mc{J}$ with
$\abs{\mc{I}}=\abs{\mc{J}}$ (i.e.~the probability that all users in
set $\mc{I}$ erase a packet is a function of $\abs{\mc{I}}$ only).

We also consider the case of spatially independent channels
(i.e.~$\epsilon_{\mc{I}}= \prod_{i\in \mc{I}} \epsilon_i$) with
(one-sided) fairness constraints, a notion first introduced in
\cite{Wang_Kuser}. To define this notion, we assume, without loss of
generality, that it holds $\epsilon_1\geq \ldots \geq \epsilon_N$ and
define a rate $\vec{R}$ to be (one-sided) fair iff it belongs to the
set $\mc{R}_{fair}\stackrel{\vartriangle} {=}\{ (R_1, \ldots,R_N)\geq
\vec{0}: \epsilon_1 R_1 \geq \ldots \geq \epsilon_N R_N \}$. We will
subsequently show that $\mbox{\texttt{CODE1}}_{pub}$ achieves any rate
$\vec{R}\in \mc{C}^{out}\cap \mc{R}_{fair}$,
i.e.~$\mbox{\texttt{CODE1}}_{pub}$ achieves all achievable fair rates
for the BPEC channel.

The complete performance analysis for $\mbox{\texttt{CODE1}}_{pub}$ is
quite lengthy so, for the reader's convenience, we present here the
main results.
\begin{theorem} \label{theo:genthru}
Denote $\hat{f}^i_{\mc{S}}\stackrel{\vartriangle}{=}\sum_{\mc{H}\subseteq 
\mc{S}-\{i\}} \frac{(-1)^{\abs{\mc{S}}-\abs{\mc{H}}-1}}{1-\epsilon_{\mc{N}-\mc{H}}}$ 
for all $\mc{S}\subseteq \mc{N}$ with $i\in \mc{S}$. For \textit{arbitrary} channel 
statistics, the rate region of $\mbox{\texttt{CODE1}}_{pub}$, in information bits per 
transmitted symbol, is given by
\beq \label{code1ach}
\mc{R}_{\mbox{\scriptsize\texttt{CODE1}}_{pub}}= \left\{ \vec{R}\geq \vec{0}: \sum_{\mc{S}\subseteq \mc{N}} 
  \max_{i\in \mc{S}}\; (\hat{f}^i_{\mc{S}} R_i) \leq L \right\} .
\eeq
\end{theorem}
\begin{IEEEproof}[Outline of proof]
We provide here an outline of the proof with complete details given in
Appendix~\ref{app3}. Since $\mbox{\texttt{CODE1}}_{pub}$, as described
in Section~\ref{algdec}, is a variable-length coding scheme (i.e.~the
total number of transmissions $\dot{T}^\ast$ required by the algorithm
is a random variable, hence unknown a priori), we propose the
following modification to make it compatible with a fixed blocklength
coding scheme that is required by the information-theoretic rate
definition of Section~\ref{sysmodel}. For a given rate vector
$\vec{R}$ and fixed $n$, we create, for each user $i\in \mc{N}$, a set
of packets $\mc{K}_i$, where $\abs{\mc{K}_i}=K_i(\vec{R})=\lceil n R_i
\rceil$, and consider $\mc{K}_i$ as the intended message for user
$i$. We then apply $\mbox{\texttt{CODE1}}_{pub}$ but stop at $n$
transmissions and declare an error if $\mbox{\texttt{CODE1}}_{pub}$
has not terminated yet (i.e.~an error is declared if
$\dot{T}^\ast>n$).
	
Hence, the modified fixed blocklength code has a probability of error
$p_n(e)= \Pr(\dot{T}^\ast> n)= \Pr( \dot{T}^\ast/n >1)$; furthermore,
using the SLLN, we can show that $\dot{T}^\ast/n$ tends to a
deterministic quantity $\bar{T}^\ast(\vec{R})$ (the $\vec{R}$
dependence is due to the fact that $\dot{T}^\ast$ implicitly depends
on $\vec{K} \stackrel{\vartriangle}{=} \lceil n\vec{R} \rceil$) w.p.~1
as $n\to \infty$. Hence, the information-theoretic rate region
achieved by $\mbox{\texttt{CODE1}}_{pub}$ is the set of rates
$\vec{R}$, measured in information symbols per transmission, for which
$p_n(e)\to 0$ as $n\to \infty$, which is intuitively equal to $\{
\vec{R}: \bar{T}^\ast(\vec{R}) \leq 1 \}$. To compute the rate region
in information bits per transmission, we use the fact that each symbol
contains $L$ bits and $\bar{T}^\ast(\vec{R})$ is a homogeneous
function of degree 1 with respect to its argument
(i.e.~$\bar{T}^\ast(\alpha \vec{R})= \alpha \bar{T}^\ast(\vec{R})$ for
any $\alpha>0$). Appendix~\ref{app3} provides a detailed calculation
of $\bar{T}^\ast(\vec{R})$ and makes the above argument rigorous.
\end{IEEEproof}
In order to provide a general optimality criterion for
$\mbox{\texttt{CODE1}}_{pub}$, we need to define the following set.
\beq \label{deford}
\mc{R}_{ord}\stackrel{\vartriangle}{=} \left\{ \vec{R}\geq \vec{0}: \exists\, 
  \mbox{permutation } \tilde{\pi} 
  \mbox{ s.t. } \forall\, \mc{S}\subseteq \mc{N} \mbox{ it holds } \argmax_{i\in \mc{S}} 
  \left( \hat{f}^i_{\mc{S}} R_i \right) = \argmin_{i\in\mc{S}} \; (\tilde{\pi}(i) )
  \right\} .
\eeq
Although the permutation $\tilde{\pi}$ in \eq{deford} may implicitly depend on $\vec{R}$ 
(as well as on channel statistics through $\hat{f}^i_{\mc{S}}$) and should actually be 
written as $\tilde{\pi}_{\vec{R}}$, we opt to simplify the notation by henceforth 
omitting this dependence. In words, $\mc{R}_{ord}$ contains all rates $\vec{R}$, whose 
indices can be rearranged according to $\tilde{\pi}$ so that the relation in \eq{deford} is 
satisfied. Notice that $\mc{R}_{ord}$ is a cone set, i.e.~$\vec{R}\in \mc{R}_{ord}$ 
implies $\alpha \vec{R}\in \mc{R}_{ord}$ for all $\alpha\geq 0$. Hence, as long as there
exists some non-zero $\vec{R}\in \mc{R}_{ord}$, the set $\mc{R}_{ord}$ intersects the 
boundary of $\mc{C}^{out}$.

Introducing the subset of $\mc{R}_{ord}$ 
\beq
\mc{D}\stackrel{\vartriangle}{=} \left\{ \vec{R}\in \mc{R}_{ord}: \sum_{i=1}^N 
  \frac{R_{\tilde{\pi}^{-1}(i)}}{1-\epsilon_{\{\tilde{\pi}^{-1}(1),\ldots,
  \tilde{\pi}^{-1}(i) \}}} \leq L \right\} ,
\eeq
where $\tilde{\pi}$ is the permutation corresponding to $\vec{R}\in \mc{R}_{ord}$ 
via \eq{deford}, we prove the following result in Appendix~\ref{app4}.
\begin{lemma} \label{lem:simp}
If $\vec{R}\in \mc{R}_{ord}$, it holds
\beq
\sum_{\mc{S}\subseteq \mc{N}} \max_{i\in \mc{S}} \left( \hat{f}^i_{\mc{S}} R_i
  \right)= \sum_{i=1}^N \frac{R_{\tilde{\pi}^{-1}(i)}}{1-\epsilon_{\{\tilde{\pi}^{-1}(1),
  \ldots,\tilde{\pi}^{-1}(i) \}}} .
\eeq
This implies, through Theorem~\ref{theo:genthru}, that 
$\mc{R}_{\mbox{\scriptsize\texttt{CODE1}}_{pub}} \cap \mc{R}_{ord}=\mc{D}$.
\end{lemma}

Theorem~\ref{theo:genthru} and Lemma~\ref{lem:simp} now lead to the main optimality criterion.
\begin{theorem} \label{theo:genopt}
The rate region of $\mbox{\texttt{CODE1}}_{pub}$ satisfies the
relation $\mc{R}_{\mbox{\scriptsize\texttt{CODE1}}_{pub}} \cap
\mc{R}_{ord}=\mc{C}^{out} \cap \mc{R}_{ord}=\mc{D}$
(i.e.~$\mbox{\texttt{CODE1}}_{pub}$ achieves any achievable rate in
$\mc{R}_{ord}$). Therefore, if it holds $\mc{R}_{ord} \supseteq
\mc{C}^{out}$, the rate region of $\mbox{\texttt{CODE1}}_{pub}$
satisfies the relation $\mc{R}_{\mbox{\scriptsize
    \texttt{CODE1}}_{pub}} =\mc{C}^{out}=\mc{D}$,
i.e.~$\mbox{\texttt{CODE1}}_{pub}$ achieves capacity.
\end{theorem}
More details are provided in Appendices~\ref{app3}, \ref{app4}. Theorem~\ref{theo:genopt} implies 
the following result (whose proof is given in Appendix~\ref{app5}) regarding the optimality of 
$\mbox{\texttt{CODE1}}_{pub}$. 
\begin{theorem} \label{optc1}
The set $\mc{R}_{ord}$ satisfies the following relations: 1) 
$\mc{R}_{ord}= \{ \vec{R}: \vec{R}\geq \vec{0} \}$, for symmetric channels and 
2) $\mc{R}_{ord} \supseteq \mc{R}_{fair}$ for spatially independent one-sided 
fair channels which satisfy the condition $\epsilon_1\geq \ldots \geq \epsilon_N$. 
Hence, $\mbox{\texttt{CODE1}}_{pub}$ achieves capacity for symmetric channels and 
also achieves all rates in $\mc{R}_{fair}\cap \mc{C}^{out}$ for spatially independent channels.
\end{theorem}

\subsection{Incorporation of overhead} \label{overhead}

The previous analysis rests on two assumptions: 1) \textit{public}
feedback is instantaneously available to \textit{all} users, and 2)
each user $i\in \mc{N}$ always knows the values of $\vec{b}^{(i)}_s$,
$c^{(i)}_s$ for any packet $s$ it receives. In order to remove the
former assumption (so that each user need only know its own feedback),
and still satisfy the latter requirement, the feedback information
must be conveyed to the receivers by the transmitter at the expense of
achievable rate (i.e.~incorporation of overhead). In fact, the second
requirement is equivalent to the requirement that all users know the
coefficients $a_s(p)$ of \textit{any} generated packet $s$, even if
they don't receive it. This follows from the fact that all
$\vec{b}^{(i)}_s$ are iteratively computed, through \eq{still_tok},
based on the selected $a_s(p)$. Hence, the second requirement is
satisfied if the algorithm for generating $a_s(p)$ (see the final
remarks in the proof of Lemma~\ref{cruc}) is available at each
receiver.  This eliminates the need for appending the coefficient
vector into the packet header as was originally proposed in
\cite{Chou_MS}. We next describe a simple, not necessarily optimal,
overhead scheme that can be applied to the original algorithm
$\mbox{\texttt{CODE1}}_{pub}$ (with or without the fixed blocklength
modification) and leads to a new algorithm, named
$\mbox{\texttt{CODE1}}_{pri}$, which does not require public
feedback. The latter algorithm consists of two stages, called ``pure
information transmission'' and ``feedback recovery'', as is explained
next.

During the pure information transmission stage, a single overhead bit
$h_1$ is reserved in each packet of length $L$. Hence, the information
payload contains $L-1$ bits and the linear combinations are performed
only over the information payload (i.e.~we treat the sequence of $L-1$
bits as an element of $\mathbb{F}_q$). The transmitter executes
$\mbox{\texttt{CODE1}}_{pub}$ normally\footnote{based on the
  algorithm's description in Section~\ref{algdec}, the reader will
  notice that the existence of public feedback may affect the exact
  decoding procedure at each user but does \textit{not} affect the
  transmitter's actions in any way, since the latter always has access
  to feedback from \textit{all} users.}, by setting $h_1=0$ in each
transmitted packet and taking the received feedback into account
according to \texttt{ACTFB1}. For each transmitted linear combination
$s$ (\textit{including} retransmissions due to steps 1, \ref{nonewtok}
of \texttt{ACTFB1}), the transmitter also creates an $N$-bit group
$(f_1,\ldots,f_N)$, where $f_i$ is 1 or 0, depending on whether or not
user $i$ received $s$, and stores it into a feedback log. Denoting
with $\dot{T}^\ast$ the (random) number of time slots required by
$\mbox{\texttt{CODE1}}_{pub}$ to process all queues, an equal number
of $N$-bit groups is created and added to the feedback log. Meanwhile,
each user stores the packets it receives in a single queue in a FIFO
manner since, at this point, it can do nothing more without additional
information on the other users' feedback.

In principle, if each user learns the exact feedback log, it will gain
the same information it would have in the case of public feedback;
hence, it can ``replay'' the algorithm as it was executed at the
transmitter side and deduce the values of $\vec{b}^{(i)}_s$,
$c^{(i)}_s$ for the packets $s$ it received. Hence, the objective now
becomes to multicast the feedback log to all $N$ users in a manner
that does not introduce significant overhead. This is performed in the
second stage of feedback recovery, in which 2 overhead bits $h_1$,
$h_2$ are reserved for each packet. When $\mbox{\texttt{CODE1}}_{pub}$
terminates (i.e.~phase $N$ is complete), the transmitter splits the
entire feedback log into packets of length $L$ (so that a total of
$\lceil N \dot{T}^\ast/(L-2) \rceil$ packets is required, considering
the 2 bit overhead per packet; we hereafter call these ``feedback''
packets) and broadcasts each feedback packet until it is received by
all users.\footnote{it is not necessary that any feedback packet is
  successfully received by all users simultaneously. During the
  transmission of the feedback log, the transmitter keeps track of
  which users receive a feedback packet, say by raising a flag
  whenever a user receives a packet. Hence, the transmitter need
  transmit a single feedback packet only until the flags for all users
  have been raised, at which point it starts transmitting the next
  feedback packet (resetting all flags).} Notice that a single
feedback packet actually contains the exact feedback that occurred in
a group of $\lfloor (L-2)/N \rfloor$ consecutive slots.

Each feedback packet has its $h_1$ bit set to 1, so any user that
receives it can distinguish it from ``pure'' information packets
(which had $h_1=0$) received during the previous phases of
$\mbox{\texttt{CODE1}}_{pub}$.  Furthermore, the transmitter applies
the following procedure for bit $h_2$. The first transmitted feedback
packet has $h_2=0$. The transmitter keeps sending this packet until
\textit{all} users receive it. When this occurs, the transmitter sends
the next feedback packet by flipping the $h_2$ bit.

The flipping of the $h_2$ bit is necessary to guard against the
following case: if a feedback packet is not received by all users upon
its first transmission, it is retransmitted so that it is possible
that a user may receive multiple copies of a ``single'' feedback
packet (meaning that all these packets contain feedback for the same
group of slots). Without any additional provisioning, this user cannot
distinguish this case from the case of multiple feedback packets that
occurred in contiguous groups of slots and happened to experience
exactly the same erasures. This problem is solved by enforcing the
rule of flipping $h_2$ between transmission of feedback packets that
correspond to \textit{different} groups of slots during the $N$ phases
of $\mbox{\texttt{CODE1}}_{pub}$.

\begin{figure}[htbp]
\centering
\psfrag{dotT}[][][0.9]{$\dot{T}^\ast$}
\psfrag{f1}[][][0.8]{$\vec{f}_1$}
\psfrag{f2}[][][0.8]{$\vec{f}_2$}
\psfrag{f3}[][][0.8]{$\vec{f}_{\dot{T}^\ast}$}
\includegraphics[scale=0.4]{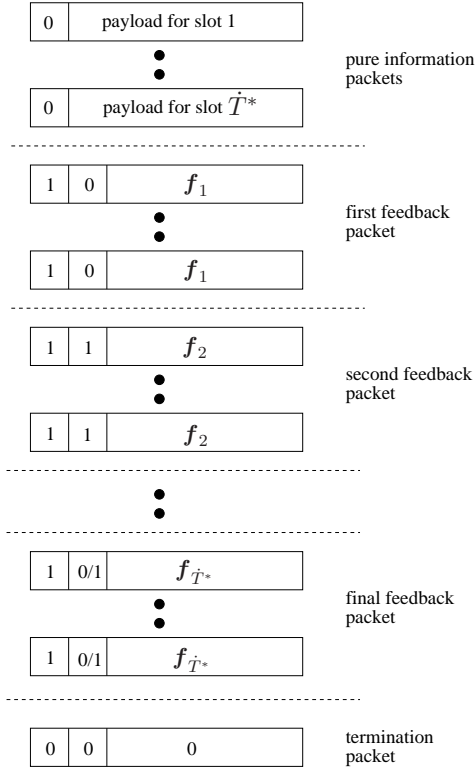}
\caption{Distinguishing packets at the receivers based on overhead bits.}
\label{fig:packets}
\end{figure}

After all log packets have been successfully received, the transmitter
broadcasts a final packet with all bits (including $h_1$, $h_2$) set
to 0 until it is also received by all users. This packet, which can be
easily distinguished by previous feedback log packets since it differs
in the $h_1$ bit, informs the receivers that transmission of all
relevant information is complete. The entire overhead scheme is
pictorially demonstrated in \fig{fig:packets}.

Assuming the order of processing $Q_{\mc{S}}$ to be known a priori,
each receiver can actually ``replay'' the execution of
$\mbox{\texttt{CODE1}}_{pub}$, up to the point for which it has
received the corresponding part of the feedback log, since it can
reproduce the coefficients $a_s(p)$ using the same coefficient
generation procedure and linear independence checking procedure (see
discussion at the end of the proof of Lemma~\ref{cruc} in the
Appendix) as the transmitter. Hence, the receiver can create local
copies of the transmitter side queues $Q_{\mc{S}}$ and counters
$K^i_{\mc{S}}$ and use \eq{still_tok} to iteratively compute the
$\vec{b}^{(i)}_s$, $c^{(i)}_s$ values of each transmitted packet
$s$. The FIFO manner of storing packets at the receiver is crucial,
since it associates each received packet to the correct ACK/NACK
group. The following result now follows from
Theorem~\ref{theo:genthru}.
\begin{theorem}
Under the overhead scheme described above, the rate region of
$\mbox{\texttt{CODE1}}_{pri}$, measured in information bits per transmission,
for arbitrary channel statistics satisfies the following relation
\beq \label{theo:ach_prv}
\mc{R}_{\mbox{\scriptsize \texttt{CODE1}}_{pri}} \supseteq \left\{ \vec{R}: 
  \sum_{\mc{S}\subseteq\mc{N}} \max_{i\in\mc{S}} \; (R_i \hat{f}^i_{\mc{S}}) \leq 
  \frac{L-1}{1+\frac{N^2}{(L-2)(1-\epsilon_{max})}} \right\} ,
\eeq
where $\epsi_{max}= \max_{i\in \mc{N}} \epsi_i$.
\end{theorem}
$\mc{R}_{\mbox{\scriptsize \texttt{CODE1}}_{pri}}$ approximates
$\mc{R}_{\mbox{\scriptsize \texttt{CODE1}}_{pub}}$ within 1 bit as
$L\to \infty$, so that the overhead-induced rate loss is minimal. An
an example, for $N=10$ and $\epsilon_{max}= 0.5$ (the latter
represents very poor channel conditions; $\epsilon_{max}$ is typically
much smaller), a length of $L=8000$ bits leads to a rate loss of 2.5\%
w.r.t.~$\mc{R}_{\mbox{\scriptsize \texttt{CODE1}}_{pub}}$.

\begin{IEEEproof}
The proof is similar to the proof of Theorem~\ref{theo:genthru} for
the case of public feedback, with the important difference that we
must now also take into account the number of slots required for the
transmission of the feedback log to all users. Based on the
description of $\mbox{\texttt{CODE1}}_{pri}$, the total number of
slots $\dot{T}^{\ast\ast}$ needed by this algorithm is
\beq \label{dotTT}
\dot{T}^{\ast\ast}= \dot{T}^\ast+ \sum_{l=1}^{1+\lceil N \dot{T}^\ast/(L-2) \rceil} \dot{N}_l ,
\eeq
where the first part in the above sum (i.e.~$\dot{T}^\ast$) is the
number of slots required by $\mbox{\texttt{CODE1}}_{pub}$ and the
second part is the total number of slots required to transmit the
packetized feedback log (i.e.~$1+\lceil N \dot{T}^\ast/(L-2) \rceil$
packets, including the termination packet), where
$\dot{N}_l\stackrel{\vartriangle}{=} \max_{i\in \mc{N}}
\dot{N}_{i,l}$, with $\dot{N}_{i,l}$ the (random) number of
transmissions required until the $l$-th feedback packet is received by
user $i$. It is clear that $\dot{N}_{i,l}$ are geometrically
distributed with $\Pr(\dot{N}_{i,l}=\nu) = \epsi^{\nu-1}_i
(1-\epsi_i)$ while $\dot{N}_l$ are (temporally) iid random
variables. The following relations will also be useful.
\beq \begin{split} \label{bound_max}
\Pr( \dot{N}_l \geq \nu ) &= \Pr \left( \bigcup_{i\in \mc{N}} \{ \dot{N}_{i,l} \geq \nu \} \right) 
  \leq \sum_{i\in \mc{N}} \Pr(\dot{N}_{i,l}\geq \nu)= \sum_{i\in \mc{N}} \epsi^{\nu-1}_i \leq N 
  \epsi^{\nu-1}_{max} , \\
\mathbb{E}[\dot{N}_l] &= \sum_{\nu=1}^{\infty} \Pr( \dot{N}_l \geq \nu) \leq \sum_{\nu=1}^\infty N 
  \epsi^{\nu-1}_{max} = \frac{N}{1-\epsi_{max}} .
\end{split} \eeq

Rewriting \eq{dotTT} as
\beq
\frac{\dot{T}^{\ast\ast}}{n}= \frac{\dot{T}^\ast}{n} + 
  \frac{1+\lceil N \dot{T}^\ast/(L-2) \rceil}{n} \, \left[ \frac{1}{1+\lceil N 
  \dot{T}^\ast/(L-2) \rceil} \sum_{l=1}^{1+\lceil N \dot{T}^\ast/(L-2) \rceil} \dot{N}_l \right] ,
\eeq
and using \eq{auxme} of Appendix~\ref{app3} for the asymptotic
behavior of $\dot{T}^\ast/n$ as $n\to \infty$, and the fact that
$\dot{T}^\ast \to \infty$ w.p.~1 as $n\to \infty$, so that we can
invoke the SLLN for the term inside brackets, we conclude that
\beq \label{bound_TT}
\bar{T}^{\ast\ast}(\vec{R}) \stackrel{\vartriangle}{=} \lim_{n\to \infty} \frac{\dot{T}^{\ast\ast}}{n} = 
  \left[ 1+\frac{N}{L-2}\, \mathbb{E}[\dot{N}_l] \right] \lim_{n\to \infty} \frac{\dot{T}^\ast}{n} \leq \left[
  1+\frac{N^2}{(L-2) (1-\epsi_{max})} \right] \sum_{\varnothing\neq \mc{S}\subseteq \mc{N}} \max_{i\in \mc{S}}
  \; (\hat{f}^i_{\mc{S}} R_i) ,
\eeq  
where we used \eq{bound_max} in the last inequality of the above
expression. We can now apply verbatim the argument used in
Appendix~\ref{app3} (Section~\ref{app:genthru}) to show that the
achievable rate region of $\mbox{\texttt{CODE1}}_{pri}$, in
information symbols per transmission, is
\beq
\mc{R}_{\mbox{\scriptsize \texttt{CODE1}}_{pri}}= \left\{ \vec{R} \geq \vec{0}: \dot{T}^{\ast\ast}(\vec{R})
  \leq 1 \right\} \supseteq \left\{ \vec{R}\geq \vec{0}: \left[ 1+\frac{N^2}{(L-2) (1-\epsi_{\max})} \right]
  \sum_{\varnothing \neq \mc{S}\subseteq \mc{N}} \max_{i\in \mc{S}} \; (\hat{f}^i_{\mc{S}} R_i) \leq 1
  \right\} ,
\eeq
where the last set inequality is due to
\eq{bound_TT}. Eq.~\eq{theo:ach_prv} follows immediately by noting
that each transmitted packet in the pure information transmission
phase (the feedback packets, although necessary for decoding, only
carry feedback information that is independent from the actual
message) has an information payload of $L-1$ bits.
\end{IEEEproof}

\section{Achieving capacity for 3 users and arbitrary channel statistics} \label{user3}

Although $\mbox{\texttt{CODE1}}_{pub}$ achieves the capacity
outer bound of Lemma~\ref{lem:outer} for some channel statistics
(namely, those that satisfy condition $\mc{R}_{ord}\supseteq
\mc{C}^{out}$ in Theorem~\ref{theo:genopt}), this is not always true, 
i.e.~for certain channel statistics there exist rates $\vec{R}\in
\mc{C}^{out}$ that are \textit{not} achievable by
$\mbox{\texttt{CODE1}}_{pub}$. This is easily verified for 3 users as
follows: consider the case of equal rates, i.e.~$R_i=R$ for all
$i\in\{1,2,3\}$ (which implies that $\abs{\mc{K}_i}=K$ for all $i$),
and assume that it holds
\beq \begin{split} \label{ass2}
& \epsilon_1=\epsilon_2=\epsilon_3 , \\
& \epsilon_{\{1,2\}}> \epsilon_{\{1,3\}}> \epsilon_{\{2,3\}} .
\end{split} \eeq
Considering all possible permutations on $\{1,2,3\}$ and applying 
Lemma~\ref{lem:outer} yields the following bound
\beq \label{eqrate}
\mc{C}^{out}_{eq}= \left\{ R \vec{1}: R \left( \frac{1}{1-\epsilon_1}
  +\frac{1}{1-\epsilon_{\{1,2\}}} +\frac{1}{1-\epsilon_{\{1,2,3\}}}
  \right) \leq L \right\} .
\eeq
Applying \eq{code1ach} of Theorem~\ref{theo:genthru} to the case of
equal rates and using \eq{ass2} produces, after some algebra,
\beq
\mc{R}_{eq,\texttt{CODE1}_{pub}}= \left\{ R\vec{1}: R\left( \frac{1}{1-\epsilon_1}+\frac{2}
  {1-\epsilon_{\{1,2\}}}-\frac{1}{1-\epsilon_{\{2,3\}}}+\frac{1}{1-\epsilon_{\{1,2,3\}}}
  \right) \leq L \right\} ,
\eeq
which implies, since $\frac{1}{1-\epsilon_{\{1,2\}}}> \frac{1}{1-\epsilon_{\{2,3\}}}$, 
that $\mc{R}_{eq,\texttt{CODE1}_{pub}} \subset \mc{C}^{out}_{eq}$. This demonstrates the 
suboptimality of $\mbox{\texttt{CODE1}}_{pub}$.

A more intuitive explanation for the suboptimal performance of
$\mbox{\texttt{CODE1}}_{pub}$ under asymmetric channel statistics for
the 3-receiver case can also be given through the following argument
(note that, for $N=3$, the network corresponding to \fig{4users}
contains only queues for sets $\mc{S}\in \{
\{1\},\{2\},\{3\},\{1,2\},\{1,3\}, \{2,3\}$, $\{1,2,3\} \}$, in addition
to $Q_{D_1}$, $Q_{D_2}$, $Q_{D_3}$). Assume that in phase 2 of
$\mbox{\texttt{CODE1}}_{pub}$, the order in which the queues are
processed is $\{1,2\}, \{1,3\}, \{2,3\}$. When the transmitter sends
linear combinations of packets from $Q_{\{1,2\}}$, it is quite
possible that the indices $K^1_{\{1,2\}}$, $K^2_{\{1,2\}}$ do not
become zero simultaneously. Say it happens that, at some slot $t$, it
holds $K^1_{\{1,2\}}(t)=0$ and $K^2_{\{1,2\}}(t)>0$. By construction,
$\mbox{\texttt{CODE1}}_{pub}$ will continue to transmit linear
combinations from $Q_{\{1,2\}}$ until $K^2_{\{1,2\}}$ also becomes
$0$.  However, this introduces a degree of inefficiency, as evidenced
in step~\ref{nonewtok} of \texttt{ACTFB1}.

Specifically, if a transmitted packet $s$ is only received by user 1,
step~\ref{nonewtok} will force $s$ to be retransmitted until some user
other than 1 receives it, essentially ``wasting'' this slot. We claim
that there exists potential for improvement at this point, by mixing
the packets in $Q_{\{1,2\}}$ with the packets in
$Q_{\{1,2,3\}}$. Clearly, the first two statements in Lemma~\ref{cruc}
are still true, so that each packet stored in either $Q_{\{1,2\}}$ or
$Q_{\{1,2,3\}}$ is a token for both users 1,2. Combining this fact
with Proposition~\ref{keep_tok}, any linear combination $s$ of the
packets in $Q_{\{1,2\}}$, $Q_{\{1,2,3\}}$ is a token. In fact, since
it will be later shown that it is still possible to define sets
$\mc{B}^{(i)}_{\mc{I}}(t)$, $\mc{B}_{D_i}(t)$ so that
Lemma~\ref{trueind} holds, a proper selection of $a_s(p)$ allows $s$
to become a \textit{Basis} token, in the next slot, for both 1,2
(provided that it holds $K^1_{\{1,2,3\}}>0$). Hence, even if the
packet is received only by 1, the slot is not wasted, since 1 recovers
a \textit{Basis} token.

Unfortunately, the previous reasoning implies that the rule of always
combining packets from a single queue must be discarded if the
objective is to achieve capacity. For $N>3$, it is not even clear what
structure a capacity achieving algorithm should have. However, for
$N=3$, we present the following algorithm, named
$\mbox{\texttt{CODE2}}_{pub}$, which achieves capacity for
arbitrary channels, assuming public feedback is available.

\begin{figure}[t]
\centering
\includegraphics[scale=0.4]{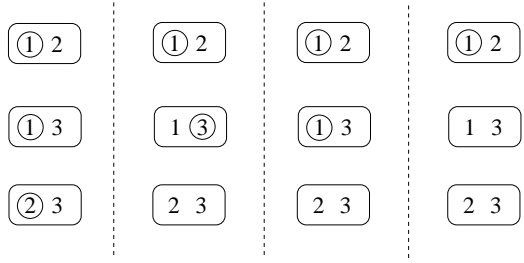}
\caption{Possible states of innovative token indices $\dot{K}^i_{\mc{S}}$ for 
the level 2 queues at epoch $t^\ast_2$.}
\label{remcases}
\end{figure}

$\mbox{\texttt{CODE2}}_{pub}$ operates in phases as follows. Phase 1
of $\mbox{\texttt{CODE2}}_{pub}$ is identical to phase 1 of
$\mbox{\texttt{CODE1}}_{pub}$, with the transmitter acting according
to the rules in \texttt{ACTFB1} (note that step~\ref{nonewtok} of
\texttt{ACTFB1} cannot occur in this phase of
$\mbox{\texttt{CODE2}}_{pub}$). In phase 2 of
$\mbox{\texttt{CODE2}}_{pub}$, the transmitter orders the level 2
queues $Q_{\mc{S}}$ according to an arbitrary rule and sequentially
processes each $Q_{\mc{S}}$ by transmitting linear combinations from
$Q_{\mc{S}}$ until it holds $K^i_{\mc{S}}=0$ for \textit{at least one}
user $i\in \mc{S}$. When this occurs, the transmitter moves to the
next level 2 queue. Again, the steps in \texttt{ACTFB1} are
applied. When all level 2 queues have been processed, each such queue
$Q_{\mc{S}}$ has at most one surviving user index (meaning some $i\in
\mc{S}$ with $K^i_{\mc{S}}>0$). For convenience, we denote this time
instant with $t^\ast_2$ and define the survival number $\dot{S}u(i)$
of index $i\in \{1,2,3\}$ as
$\dot{S}u(i)\stackrel{\vartriangle}{=}\abs{ \{ \mc{S}: \abs{\mc{S}}=2,
  \; \dot{K}^i_{\mc{S}}(t^\ast_2)>0\} }$. In words, $\dot{S}u(i)$ is
equal to the number of level 2 queues which contain unrecovered
\textit{Basis} tokens for user $i$ at time $t^\ast_2$. Clearly,
$\dot{S}u(i)$ is a random variable that depends on the prior erasure
events (hence, the dot accent) and satisfies $0\leq \dot{S}u(i)\leq 2$
for all $i\in \{1,2,3\}$. The transmitter now distinguishes cases as
follows:
\begin{enumerate}
\item if it holds $\dot{S}u(i)=0$ for all $i\in \{1,2,3\}$,
  $\mbox{\texttt{CODE2}}_{pub}$ reverts to
  $\mbox{\texttt{CODE1}}_{pub}$, starting at phase 3. \label{case1}

\item if it holds $\dot{S}u(i)=1$ for all $i\in \{1,2,3\}$,
  $\mbox{\texttt{CODE2}}_{pub}$ reverts to
  $\mbox{\texttt{CODE1}}_{pub}$ and continues processing each
  $Q_{\mc{S}}$ queue in level 2 until all $K^i_{\mc{S}}$ become
  zero. \label{case2}

\item otherwise, there exists at least one pair of users $i,j$ such
  that $\dot{S}u(i)=0$, $\dot{S}u(j)>0$. In this case, simple
  enumeration reveals that all possible configurations of
  $\dot{S}u(l)$ for $l\in \{1,2,3\}$ fall in exactly one of the
  following 4 categories:
\begin{enumerate}
\item there exist distinct users $i^\ast,j^\ast,k^\ast\in \{1,2,3\}$ 
such that $\dot{S}u(i^\ast)=0$, $\dot{S}u(j^\ast)=1$, $\dot{S}u(k^\ast)=2$.

\item there exist distinct users $i^\ast,j^\ast,k^\ast\in 
\{1,2,3\}$ such that $\dot{S}u(i^\ast)=0$, $\dot{S}u(j^\ast)=\dot{S}u(k^\ast)=1$.

\item there exist distinct users $i^\ast,j^\ast,k^\ast\in \{1,2,3\}$ 
such that $\dot{S}u(i^\ast)=\dot{S}u(j^\ast)=0$ and $\dot{S}u(k^\ast)=2$.

\item there exist distinct users $i^\ast,j^\ast,k^\ast\in \{1,2,3\}$ 
such that $\dot{S}u(i^\ast)=\dot{S}u(j^\ast)=0$ and $\dot{S}u(k^\ast)=1$.
\end{enumerate}
To provide some concrete examples, \fig{remcases} contains 4 possible
configurations (each belonging, from left to right, to one of the
above categories), where circles are used to denote surviving
indices. The values $(i^\ast,j^\ast,k^\ast)$ for each configuration
are $(3,2,1)$, $(2,1,3)$, $(3,2,1)$, $(3,2,1)$, respectively. 
\end{enumerate}

We hereafter concentrate on case 3 of the above list, since cases 1, 2
revert to $\mbox{\texttt{CODE1}}_{pub}$. The transmitter now
constructs the set $\mc{Q}_{\dot{S}u}= \{ Q_{\{i^\ast,j\}}:
\dot{S}u(i^\ast)=0,\; \dot{K}^j_{\{i^\ast,j\}}(t^\ast_2)>0 \}$
consisting of all level 2 queues that contain a surviving index $j$
and an index $i^\ast$ with $\dot{S}u(i^\ast)=0$. Relative order within
$\mc{Q}_{\dot{S}u}$ is unimportant. A subphase, called 2.1, is now
initiated, in which the following actions are performed:
\begin{itemize}
\item the transmitter processes each queue $Q_{\{i^\ast,j\}}$ in
$\mc{Q}_{\dot{S}u}$ and transmits a packet $s$ which is a linear combination
of all packets in queues $Q_{\{i^\ast,j\}}$ \textit{and}
$Q_{\{1,2,3\}}$ (``and'' denotes grouping in this context and
should not be interpreted in the Boolean sense). The coefficients $a_s(p)$ 
are selected such that $s$ is a \textit{Basis} token for $j$ as well as 
$i^\ast$ (for the latter case, this is true if it holds 
$K^{i^\ast}_{\{1,2,3\}}>0$). It will be proved in Appendix~\ref{app6} 
that this selection is always possible. Depending on the received 
feedback, the following actions, collectively referred to as \texttt{ACTFB2}, 
are taken.\\
\underline{\texttt{ACTFB2} actions:}
\begin{enumerate}
\item if $s$ is erased by all users, $s$ is retransmitted.

\item if $s$ is received only by $i^\ast$ when it holds 
$K^{i^\ast}_{\{1,2,3\}}=0$, $s$ is retransmitted.\label{nonewtok2}

\item if $j$ receives $s$, $K^j_{\{i^\ast,j\}}$ is
decreased by 1 and $K_{D_j}$ is increased by 1.

\item if $i^\ast$ receives $s$ \textit{and} it holds
  $K^{i^\ast}_{\{1,2,3\}}>0$, $K^{i^\ast}_{\{1,2,3\}}$ is decreased by
  1 and $K_{D_i^\ast}$ is increased by 1. \label{dectrip}

\item if $j$ erases $s$ and $k\in \mc\{1,2,3\}-\{i^\ast,j\}$ receives
it, $s$ is added to $Q_{\{1,2,3\}}$, $K^j_{\{i^\ast,j\}}$ is decreased
by 1 and $K^j_{\{1,2,3\}}$ is increased by 1. \label{inctrip}
\end{enumerate}
Notice that, apart from step~\ref{dectrip}) in the above list,
\texttt{ACTFB2} is similar to \texttt{ACTFB1}. The above procedure is
repeated until it holds $K^j_{\{i^\ast,j\}}=0$, at which point the
next queue in $\mc{Q}_{\dot{S}u}$ is processed. The above procedure is
repeated until all queues in $\mc{Q}_{\dot{S}u}$ have been processed.

\item once all queues in $\mc{Q}_{\dot{S}u}$ have been processed, the
  transmitter computes the new values of $\dot{S}u(i)$ for
  $i\in\{1,2,3\}$ and constructs $\mc{Q}_{\dot{S}u}$ from scratch. If
  $\mc{Q}_{\dot{S}u}=\varnothing$, $\mbox{\texttt{CODE2}}_{pub}$
  reverts to $\mbox{\texttt{CODE1}}_{pub}$ starting at phase 3,
  otherwise it repeats the above procedure verbatim for the new
  $\mc{Q}_{\dot{S}u}$. It can be easily verified that at most 2
  iterations of this procedure will be performed until it holds
  $\mc{Q}_{\dot{S}u}=\varnothing$.
\end{itemize}

As a final comment, step~\ref{nonewtok2} of \texttt{ACTFB2} is similar
to step~\ref{nonewtok} of \texttt{ACTFB1} so one could argue that
$\mbox{\texttt{CODE2}}_{pub}$ still performs inefficiently.
However, by construction of $\mc{Q}_{\dot{S}u}$, it is easy to verify that
if, during the combination of $Q_{\{i^\ast,j\}}\in \mc{Q}_{\dot{S}u}$ with
$Q_{\{1,2,3\}}$, $K^{i^\ast}_{\{1,2,3\}}$ becomes $0$ before
$K^j_{\{i^\ast,j\}}$ does, then $i^\ast$ has no more \textit{Basis}
tokens to recover (i.e.~it holds $K^{i^\ast}_{\mc{S}}=0$ for all
$\mc{S}\subseteq \mc{N}$). Hence, $i^\ast$ cannot gain any more
linearly independent tokens by combining $Q_{\{i^\ast,j\}}$ with
$Q_{\{1,2,3\}}$ and no efficiency is lost.

To provide a concrete example for the last statement, consider the
application of subphase 2.1 to the leftmost configuration in
\fig{remcases}. It holds $\mc{Q}_{\dot{S}u}=\{ Q_{\{1,3\}},
Q_{\{2,3\}}\}$ and the transmitter starts combining $Q_{\{1,3\}}$ with
$Q_{\{1,2,3\}}$ until $K^2_{\{2,3\}}$ becomes $0$. If it happens that
$K^3_{\{1,2,3\}}$ becomes $0$ before $K^2_{\{2,3\}}$, then 3 has
indeed recovered all \textit{Basis} tokens so that, even if
step~\ref{nonewtok} occurs, no efficiency gain is possible. The same
conclusion is reached by examining the 3 other categories shown in
\fig{remcases}. Hence, at the end of subphase 2.1, it holds
$K^i_{\mc{S}}=0$ for all $i\in \mc{S}$ with $\abs{\mc{S}}=2$ and
$\mbox{\texttt{CODE2}}_{pub}$ reverts to $\mbox{\texttt{CODE1}}_{pub}$
starting at phase 3.

The properties and achievable rate region of
$\mbox{\texttt{CODE2}}_{pub}$ can be determined by an approach similar
to that of $\mbox{\texttt{CODE1}}_{pub}$. Specifically, the
correctness of $\mbox{\texttt{CODE2}}_{pub}$ is proved in
Appendix~\ref{app6}, where a slight modification of Lemma~\ref{cruc}
is used to show that Lemma~\ref{trueind} is still true for
$\mbox{\texttt{CODE2}}_{pub}$. This guarantees that at the end of
$\mbox{\texttt{CODE2}}_{pub}$, all 3 users have received the required
number of linearly independent tokens and can decode their
packets. The performance analysis for $\mbox{\texttt{CODE2}}_{pub}$ is
identical to $\mbox{\texttt{CODE1}}_{pub}$, up to time
$t^\ast_2$. From this point on, the number of tokens produced during
the combination of the queues in $\mc{Q}_{\dot{S}u}$ with
$Q_{\{1,2,3\}}$ must be carefully computed. The computation is
relatively straightforward but lengthy, and is deferred to
Appendix~\ref{app7}. The final result is:
\begin{theorem} \label{opt3}
$\mbox{\texttt{CODE2}}_{pub}$ achieves the capacity outer bound of
$\mc{C}^{out}$, for $L \geq 2$. In case only private feedback is
available, we can construct algorithm $\mbox{\texttt{CODE2}}_{pri}$, based
on $\mbox{\texttt{CODE2}}_{pub}$, using the overhead scheme employed
in $\mbox{\texttt{CODE1}}_{pri}$. The final result is that the rate
region of $\mbox{\texttt{CODE2}}_{pri}$ asymptotically differs from
the capacity outer bound by 1 bit as $L\to \infty$.
\end{theorem}

\section{Conclusions} \label{conclu}

This paper presented 2 coding algorithms,
$\mbox{\texttt{CODE1}}_{pub}$ and $\mbox{\texttt{CODE2}}_{pub}$, which
achieve (assuming public feedback) an outer bound of the feedback
capacity region of the $N$-user broadcast erasure channel with $N$
unicast sessions for the following cases, respectively: 1) arbitrary
$N$ and channel statistics that satisfy the general condition in
Theorem~\ref{theo:genopt} (this includes symmetric channels as a
special case), and 2) arbitrary channel statistics, for $N=3$. If
public feedback is not available, a simple overhead scheme can be
applied on top of each algorithm, leading to a rate region that
asymptotically differs from the outer bound by 1 bit as $L\to
\infty$. The main characteristic of the algorithms is the introduction
of virtual queues to store packets, depending on received feedback,
and the appropriate mixing of the packets, without requiring any
knowledge of channel statistics, to allow for simultaneous reception
of innovative packets by multiple users.

Since only an outer bound to the capacity region is known for $N\geq
4$ and arbitrary channels, the search for capacity achieving
algorithms for $N\geq 4$ is an obvious future research topic. It is
expected that such algorithms cannot be constructed through minor
modifications of $\mbox{\texttt{CODE1}}_{pub}$, as was the case with
$\mbox{\texttt{CODE2}}_{pub}$, and may possibly require complete
knowledge of channel statistics. If this is the case, adaptive
algorithms that essentially ``learn'' the relevant statistics may be
appropriate. Suboptimal algorithms with guaranteed performance bounds
in the spirit of \cite{SagEph09} may also be of interest.

\appendices

\section{Proof of Lemma~\ref{cruc}} \label{app1}
 
By construction of $\mbox{\texttt{CODE1}}_{pub}$, the only way a
packet $s$ can be stored in queue $Q_{\mc{S}}$, with $\abs{\mc{S}}\geq
2$, is during step~\ref{movtok} of \texttt{ACTFB1} (since, excluding
packets that are received by a user $i\in \mc{S}$ and moved to queue
$Q_{D_i}$, no packets are moved between queues $Q_{\mc{S}}$ in the
other steps of \texttt{ACTFB1}). Thus, the execution of
step~\ref{movtok} implies that $s$ is a linear combination of packets
in some queue $Q_{\mc{I}_s}$, with $\varnothing\neq \mc{I}_s\subset
\mc{S}$, and $s$ is received by \textit{all} users in
$\mc{S}-\mc{I}_s$ and erased by \textit{all} users in
$\mc{N}-\mc{S}$. This completes the proof of the first statement.

For the second statement of the Lemma, we note that the algorithm's
operation implies that any transmitted packet $s$ is
decomposed as $s=\sum_{u\in \cup_{j\in \mc{N}} \mc{K}_j}
\tilde{a}_s(u) u$ (the algorithm essentially sends linear combinations
of linear combinations etc.). Furthermore, we can combine the
initialization of $\mbox{\texttt{CODE1}}_{pub}$ (for queues
$Q_{\mc{S}}$ with $\abs{\mc{S}}=1$) with the first statement in
Lemma~\ref{cruc} (proved in the previous paragraph) to show, via
strong induction on $\abs{\mc{S}}=2,\ldots,N$, that, for all
$\mc{S}\subseteq \mc{N}$ and any packet $s$ stored in $Q_{\mc{S}}$, it
holds $\tilde{a}_s(u)=0$ for all $u\in \mc{K}_j$ with $j\not \in
\mc{S}$. Specifically, any $s\in Q_{\mc{S}}$ must have entered
$Q_{\mc{S}}$ during step~\ref{movtok} of \texttt{ACTFB1}, so that it
holds $s=\sum_{p\in Q_{\mc{I}_s}} a_s(p) p$, where $\mc{I}_s\subset
\mc{S}$. Using the strong induction hypothesis for $\mc{I}_s$, we know
that any $p\in Q_{\mc{I}_s}$ is written as $p=\sum_{u\in \cup_{j\in
    \mc{I}_s}} \tilde{a}_p(u) u$. Combining the last two expressions,
we conclude that any packet $s$ stored in $Q_{\mc{S}}$ can be written
as
\beq \label{justS}
s= \sum_{u\in \cup_{j\in \mc{S}} \mc{K}_j} \tilde{a}_s(u) u ,
\eeq
for suitable $\tilde{a}_s(u)$, and the second statement is also
proved.

To prove the third statement of the Lemma, we apply strong induction
on $\abs{\mc{S}}$, starting with $\abs{\mc{S}}=1$. Due to the
initialization of $\mbox{\texttt{CODE1}}_{pub}$, any packet $s$
stored in $Q_{\{i\}}$ belongs to set $\mc{K}_i$, so that $s$ is a
(trivial) token for user $i$ and no other user. We now consider any
$s$ stored in queue $Q_{\mc{S}}$ with $\abs{\mc{S}}>1$, and use the
first statement of the Lemma to write $s=\sum_{p\in Q_{\mc{I}_s}}
a_s(p) p$, where $\mc{I}_s \subset \mc{S}$. This also implies that $s$
was received by all users in set $\mc{S}-\mc{I}_s$, so that $s$ is a
token for all users in the set $\mc{S}-\mc{I}_s$. Combining the
inductive hypothesis for set $\mc{I}_s$ with
Proposition~\ref{keep_tok}, we conclude that $s$ is a token for all
$i\in \mc{I}_s$ as well, so that $s$ is a token for all $i\in
\mc{S}$. To show that $s$ is not a token for any $i\not \in \mc{S}$,
we combine the fact that $s$ is a linear combination of packets
destined for users in set $\cup_{j\in \mc{I}_s} \mc{K}_j$ only (second
statement of the Lemma) with the fact that $s$ was erased by all users
$i\not \in \mc{S}$ (first statement of Lemma). Hence, $i$ cannot be a
token for any $i\not \in \mc{S}$.

Before we prove the fourth statement in Lemma~\ref{cruc}, we need to
establish some intermediate results. The following Proposition is easily
proved by considering the union bound for the probabilities of the complementary
events.
\begin{proposition} \label{prop_bound}
For any events $A_j$, with $j=1,\ldots,m$, it holds
\beq \nonumber
\Pr( \cap_{j=1}^m A_j) \geq \sum_{j=1}^m \Pr(A_j) -m+1 .
\eeq
\end{proposition}
The following result will be crucial in proving Lemma~\ref{cruc}.
\begin{lemma} \label{change_base}
Let $\{ \vec{v}_1,\ldots,\vec{v}_M \}$ be a basis set of the vector space 
$\mathbb{F}^M_q$ and consider a subspace $\mc{U}$ with dimension $l\geq 1$, 
which contains the set $\{\vec{v}_1,\ldots,\vec{v}_K\}$, with $1\leq K\leq l$. 
Then, the subspace $\mc{U}\cap span(\{\vec{v}_2,\ldots,\vec{v}_M\})$ has dimension 
at most $l-1$, and $\mc{U}-span(\{ \vec{v}_2,\ldots,\vec{v}_M\}$ is a non-empty set. 
Additionally, for any vector $\vec{u}\in \mc{U}-span(\{\vec{v}_2,\ldots,\vec{v}_M\}$, 
the set $\{\vec{u},\vec{v}_2,\ldots,\vec{v}_M\}$ is a basis of $\mathbb{F}^M_q$.
\end{lemma}

\begin{IEEEproof}
We use contradiction to show that $\dim(\mc{U}\cap span(\{ \vec{v}_2,\ldots,
\vec{v}_M\}) \leq l-1$. Specifically, assume that $\dim(\mc{U}\cap span(\{ \vec{v}_2, 
\ldots,\vec{v}_M \})=l$. Then there exists a set $\{\tilde{\vec{v}}_1,
\ldots, \tilde{\vec{v}}_l\}$ which forms a basis of $\mc{U}\cap span(\{\vec{v}_2,\ldots,
\vec{v}_M\})$. Therefore, $\{\tilde{\vec{v}}_1,\ldots,\tilde{\vec{v}}_l\} \subseteq 
\mc{U}\cap span(\{\vec{v}_2,\ldots,\vec{v}_M\})$ is a basis of $\mc{U}$ as well, since 
it is a linearly independent set of cardinality $l$ that is contained in the subspace 
$\mc{U}$ of dimension $l$. The basis property for $\mc{U}$ now implies that $\vec{v}_1
\in span(\{ \tilde{\vec{v}}_1,\ldots,\tilde{\vec{v}}_l\})$ and, since $\tilde{\vec{v}}_i
\in span (\{\vec{v}_2,\ldots,\vec{v}_M\})$ for $1\leq i\leq l$, it also holds $\vec{v}_1
\in span(\{\vec{v}_2,\ldots,\vec{v}_M\})$. This contradicts the assumption that 
$\{\vec{v}_1,\ldots,\vec{v}_M\}$ are linearly independent and proves the desired result. 
Additionally, since $\vec{v}_1\not\in span(\{\vec{v}_2,\ldots,\vec{v}_M\})$, it also 
holds $\lambda \vec{v}_1 \in \mc{U}-span(\{ \vec{v}_2,\ldots,\vec{v}_M\})$ for all 
$\lambda\in \mathbb{F}_q-\{0\}$, so that $\mc{U}-span(\{\vec{v}_2,\ldots,\vec{v}_M\}) 
\neq \varnothing$.

In order to show that $\{\vec{u},\vec{v}_2,\ldots,\vec{v}_M\}$ is a basis of $\mathbb{F}^M_q$
for \textit{any} $\vec{u}\in \mc{U}-span(\{ \vec{v}_2,\ldots,\vec{v}_M\})$, it suffices 
to show that $\{ \vec{u},\vec{v}_2,\ldots,\vec{v}_M\}$ is a linearly independent set. Indeed,
pick any $\vec{u}\in \mc{U}-span(\{ \vec{v}_2,\ldots,\vec{v}_M\})$ and assume that there 
exist $\alpha,c_i\in \mathbb{F}_q$ such that 
\beq \nonumber
\alpha \vec{u}+ \sum_{i=2}^M c_i \vec{v}_i =0 .
\eeq
Then it must hold $\alpha=0$, since the case $\alpha\neq 0$ implies that $\vec{u}\in span(\{
\vec{v}_2,\ldots,\vec{v}_M\})$, which is impossible by the selection of $\vec{u}$. 
The condition $\alpha=0$ now implies $c_i=0$, due to the linear independence of $\{\vec{v}_2,
\ldots,\vec{v}_M\}$, so that $\{ \vec{u},\vec{v}_2,\ldots,\vec{v}_M\}$ is also linearly
independent and the proof is complete.
\end{IEEEproof}

The last intermediate result we need before proving the fourth
statement in Lemma~\ref{cruc} is provided below.
\begin{lemma} \label{linalg} 
Let $\vec{v}_j$, with $j=1,\ldots,k$, be vectors in $\mathbb{F}^M_q$. 
Denote $\mc{V}=span( \{\vec{v}_j\}, j=1,\ldots,k)$ and $l=\dim(\mc{V})$, 
with $l\geq 1$. Let $\alpha_j$, with $j=1,\ldots,k$, be independent 
random variables uniformly distributed in $\mathbb{F}_q$ and construct 
the random vector $\vec{v}=\sum_{j=1}^k \alpha_j \vec{v}_j$. Then, 
$\vec{v}$ is uniformly distributed in $\mc{V}$, i.e.
\beq \nonumber
\Pr(\vec{v}=\vec{e})= \frac{1}{q^l} \quad \forall\, \vec{e}\in \mc{V} .
\eeq

Additionally, let $\{\vec{b}_1,\ldots,\vec{b}_M\}$ be a basis of 
$\mathbb{F}^M_q$ and assume that $\{ \vec{b}_1,\ldots,\vec{b}_K\}
\subseteq \mc{V}$ for $1\leq K\leq M$. It then holds
\beq \nonumber
\Pr\left( \{\vec{v},\vec{b}_2,\ldots,\vec{b}_M\} \mbox{ is basis of } 
  \mathbb{F}^M_q \right) \geq 1-\frac{1}{q} \; .
\eeq
\end{lemma}

\begin{IEEEproof}
Since $\mc{V}$ has dimension $l$, we can pick $l$ vectors $\vec{v}_i$ 
(out of the $k$ available) as a basis for $\mc{V}$; without loss of 
generality, we can permute vector indices so that the basis set is 
$\{\vec{v}_1,\ldots,\vec{v}_l\}$. Hence, $\vec{v}$ can be written as
$\vec{v}=\sum_{j=1}^l \alpha_j \vec{v}_j + \vec{g}$, where
$\vec{g}\stackrel{\vartriangle}{=}\sum_{j=l+1}^k \alpha_j \vec{v}_j$
is a random vector independent from $\sum_{j=1}^l \alpha_j
\vec{v}_j$. Furthermore, any vector $\vec{e}\in \mc{V}$ can be written
uniquely, through the basis set, as $\vec{e}=\sum_{j=1}^l e_j
\vec{v}_j$. It now holds
\beq \begin{split}
\Pr(\vec{v}=\vec{e}) &= \sum_{\vec{r}\in\mc{V}} \Pr \left( \left. \sum_{j=1}^l \alpha_j 
  \vec{v}_j +\vec{g}= \vec{e} \right| \vec{g}=\vec{r} \right) \Pr(\vec{g}=\vec{r}) \\
&= \sum_{\vec{r}\in\mc{V}} \Pr\left( \left. \sum_{j=1}^l \alpha_j \vec{v}_j= \vec{e}
  -\vec{r} \right| \vec{g}=\vec{r} \right) \Pr(\vec{g}=\vec{r}) \\
&= \sum_{\vec{r}\in\mc{V}} \Pr\left( \sum_{j=1}^l \alpha_j \vec{v}_j= \sum_{j=1}^l 
  (e_j-r_j) \vec{v}_j \right) \Pr(\vec{g}=\vec{r}) \\
&= \sum_{\vec{r}\in\mc{V}} \Pr( \cap_{j=1}^l \{ \alpha_j=e_j-r_j \}) \Pr(\vec{g}=\vec{r})  
  =\sum_{\vec{r}\in\mc{V}} \frac{1}{q^l} \Pr(\vec{g}=\vec{r})= \frac{1}{q^l} ,
\end{split} \eeq
where we used the independence of $\sum_{j=1}^l \alpha_j \vec{v}_j$
from $\vec{g}$ to remove the conditional probability and exploited the
facts that $\{\vec{v}_1,\ldots,\vec{v}_l\}$ is a basis set for
$\mc{V}$ and $\alpha_j$ are independent and uniformly distributed in $\mathbb{F}_q$.

To prove the second part, we note that
\beq
\Pr\left( \{\vec{v},\vec{b}_2,\ldots,\vec{b}_M\} \mbox{ basis of } 
  \mathbb{F}^M_q \right)= 1-\Pr(\vec{v}\in span(\{ \vec{b}_2,\ldots,
  \vec{b}_M\}) ) .
\eeq
For notational convenience, denote $\mc{M}=span(\{ \vec{b}_2,\ldots,\vec{b}_M\})$.
It now holds
\beq \label{comp1}
\Pr(\vec{v}\in \mc{M})= \sum_{\vec{r}\in \mc{M}\cap \mc{V}} \Pr(\vec{v}=\vec{r})= 
  \frac{1}{q^l} \abs{\mc{M}\cap \mc{V}} ,
\eeq
where the last equality is due to the uniform distribution of
$\vec{v}$ in $\mc{V}$.  For all vector spaces over a finite field, it
also holds $\abs{\mc{M}\cap \mc{V}}= q^{\dim(\mc{M}\cap \mc{V})} \leq
q^{\dim(\mc{V})-1}=q^{l-1}$, where the inequality is due to
Lemma~\ref{change_base}. Inserting this inequality into \eq{comp1}
produces $\Pr(\vec{v}\in \mc{M})\leq 1/q$, whence the desired result
follows immediately.
\end{IEEEproof}

We are now in position to prove the fourth statement of
Lemma~\ref{cruc}.  Specifically, recalling the notation of
Criterion~\ref{minass2}, we assume that there exist sets
$\mc{B}^{(i)}_{\mc{I}}(t)$, $\mc{B}_{D_i}(t)$ such that
$\mathring{\mc{B}}\stackrel{\vartriangle}{=} \mc{B}_{D_i}(t)\cup
\bigcup_{\substack{\mc{I}:\mc{I}\subseteq \mc{N}
    \\ K^i_{\mc{I}}(t)>0}} \mc{B}^{(i)}_{\mc{I}}(t)$ is a basis of
$\mathbb{F}^{\abs{\mc{K}_i}}_q$ for all $i\in \mc{N}$. Assuming that
$\mbox{\texttt{CODE1}}_{pub}$ is currently processing $Q_{\mc{S}}$,
define the set $\mc{R}_{\mc{S}}(t) \stackrel{\vartriangle}{=}\left\{ i
\in \mc{S}: K^i_{\mc{S}}(t)>0 \right\}$. We need to show that if, for
each $i\in \mc{R}_{\mc{S}}(t)$, we pick an arbitrary vector
$\hat{\vec{b}}_i\in \mc{B}^{(i)}_{\mc{S}}(t)$, then there exists a
coefficient vector $\vec{a}_s= (a_s(p), p\in Q_{\mc{S}})$ such that
the vectors $\vec{b}^{(i)}_{s}= \sum_{p\in Q_{\mc{S}}} a_{s}(p)
\vec{b}^{(i)}_p$, corresponding to the combination $s=\sum_{p\in
  Q_{\mc{S}}} a_{s}(p) p$, satisfy the following condition
\beq \label{condi}
\{ \vec{b}^{(i)}_s \} \cup \mc{B}_{D_i(t)} \cup \bigcup_{\substack{\mc{I}:\mc{I} \subseteq 
  \mc{N}\\ K^i_{\mc{I}}(t)>0}} \mc{B}^{(i)}_{\mc{S}}(t) -\{ \hat{\vec{b}}_i \} 
  \mbox{ is basis of } \mathbb{F}^{\abs{\mc{K}_i}}_q \quad \forall\, i\in \mc{R}_{\mc{S}}(t) .
\eeq

The proof is via a standard probabilistic argument. Specifically, consider the case
where coefficients $\vec{a}_{s}$ are iid randomly generated according to a 
uniform distribution in $\mathbb{F}_q$. For a given user $i\in \mc{R}_{\mc{S}}(t)$, 
define the event $A_i\stackrel{\vartriangle}{=} \left\{ \{\vec{b}^{(i)}_s \} \cup 
\mathring{\mc{B}}-\{ \hat{\vec{b}}_i \} \mbox{ is basis of } 
\mathbb{F}^{\abs{\mc{K}_i}}_q \right\}$, whence it follows from Lemma~\ref{linalg} 
that $\Pr(A_i) \geq 1-1/q$. Applying Proposition~\ref{prop_bound} to the event 
$\cap_{i\in \mc{R}_{\mc{S}}(t)} A_i$ yields
\beq \label{nbound}
\Pr\left( \cap_{i\in \mc{R}_{\mc{S}}(t)} A_i \right) \geq \abs{\mc{R}_{\mc{S}}(t)}
  \left(1-\frac{1}{q}\right)-\abs{\mc{R}_{\mc{S}}(t)}+1 \geq 
  1-\frac{\abs{\mc{R}_{\mc{S}}(t)}}{q} \geq 1-\frac{N}{q} .
\eeq
Selecting $q>N$ (since $q$ can be as large as $2^L$, the condition $q>N$ can be satisfied if
$L> \log_2 N$) results in a strictly positive probability, 
which implies that there exist some vectors $\vec{b}^{(i)}_s$ that simultaneously satisfy 
\eq{condi} for all $i\in \mc{R}_{\mc{S}}(t)$. This completes the proof of the fourth 
statement in Lemma~\ref{cruc}.

The previous analysis suggests the following alternative approach to
an exhaustive search for generating $a_s(p)$ in accordance with
Criterion~\ref{minass2}. If the sets $\mc{B}^{(i)}_{\mc{I}}$,
$\mc{B}_{D_i}$ are actually stored at the transmitter and receivers,
and since Lemma~\ref{cruc} ensures that, for $q>N$, there exist
coefficients $a_s(p)$ which satisfy \eq{eq:prop2} of
Criterion~\ref{minass2}, then $a_s(p)$ can be generated randomly and
uniformly in $\mathbb{F}_q$ (so that \eq{nbound} holds) followed by an
explicit check by the transmitter whether the generated vectors
$\vec{b}^{(i)}_s$ indeed satisfy \eq{condi}. If \eq{condi} is violated
for at least one $i\in \mc{R}_{\mc{S}}(t)$, new coefficients are
repeatedly created until the condition is satisfied for all $i\in
\mc{R}_{\mc{S}}(t)$. Only then is the packet $s$ actually transmitted,
using the most recent coefficients $a_s(p)$ . The average number of
trials required to find the suitable coefficients is easily computed
as $(1-N/q)^{-1}$.

\section{Proof of Lemma~\ref{trueind}} \label{app2}

Proof is by induction on $t$. At the beginning of slot $t=0$, we can
satisfy all conditions by choosing for each $i\in \mc{N}$ as follows:
$\mc{B}_{D_i}(0)= \varnothing$, $\mc{B}^{(i)}_{\mc{I}}(0)=\varnothing$
for $\mc{I}\neq \{i\}$ and
$\mc{B}^{(i)}_{\{i\}}(0)=standard\_basis\left(
\mathbb{F}^{\abs{\mc{K}_i}}_q \right)$. We now assume that the
inductive hypothesis is true at the beginning of slot $t$ and the
queue currently being processed is $Q_{\mc{S}}$. We construct
$\mc{R}_{\mc{S}}(t)=\{ i\in \mc{S}: K^i_{\mc{S}}(t)>0 \}$ and further
assume w.l.o.g.~that $\mc{R}_{\mc{S}}(t)\neq \varnothing$ since, in
the opposite case, $\mbox{\texttt{CODE1}}_{pub}$ will skip processing
$Q_{\mc{S}}$ and continue to the next queue. Lemma~\ref{cruc} now
guarantees that, due to the validity of the hypothesis (i.e.~the
existence of $\mc{B}^{(i)}_{D_i}(t)$, $\mc{B}^{(i)}_{\mc{I}}$) at the
beginning of slot $t$, we can select vectors $\hat{\vec{b}}_i \in
\mc{B}^{(i)}_{\mc{S}}(t)$, for each $i\in \mc{R}_{\mc{S}}(t)$, and
coefficients $a_s(p)$ for the next packet $s$ to be transmitted from
$Q_{\mc{S}}$ so that \eq{eq:prop2} of Criterion~\ref{minass2} is
satisfied for all $i\in \mc{R}_{\mc{S}}(t)$.

For each $i\in \mc{N}-\mc{R}_{\mc{S}}(t)$, it either holds $i\not\in
\mc{S}$ or $i\in \mc{S}, \; K^i_{\mc{S}}(t)=0$. In both cases, by
construction of \texttt{ACTFB1}, the transmission of $s$ does not
change any of the $K^i_{\mc{I}}$, $K_{D_i}$ indices. Hence, at the beginning of
slot $t+1$, we can select $\mc{B}^{(i)}_{\mc{I}}(t+1)=
\mc{B}^{(i)}_{\mc{I}}(t)$, for all $\mc{I}\subseteq \mc{N}$, and
$\mc{B}_{D_i}(t+1)=\mc{B}_{D_i}(t)$ so that, for all $i\in \mc{N}-
\mc{R}_{\mc{S}}(t)$, the inductive hypothesis holds for $t+1$ as
well. We now concentrate on $i\in \mc{R}_{\mc{S}}(t)$ and consider the
following mutually exclusive cases:
\begin{itemize}
\item if $i$ receives $s$, \texttt{ACTFB1} forces $s$ to be added to
  $Q_{D_i}$ and $K^i_{\mc{S}}$ to be decreased by one, while $K_{D_i}$
  is increased by one. Accordingly, we select
  $\mc{B}^{(i)}_{\mc{S}}(t+1)= \mc{B}^{(i)}_{\mc{S}}(t)-\{
  \hat{\vec{b}}_i \}$ and $\mc{B}_{D_i}(t+1)=\mc{B}_{D_i}(t) \cup
  \{\vec{b}^{(i)}_s \}$, while all other sets $\mc{B}^{(i)}_{\mc{I}}$
  remain unaffected. Lemma~\ref{cruc} now implies that the union of
  the new sets at slot $t+1$ form a basis of
  $\mathbb{F}^{\abs{\mc{K}_i}}_q$.

\item if $i$ erases $s$ and all users in a maximal set $\mc{G}\subseteq \mc{N}-\mc{S}$ 
receive $s$, then $K^i_{\mc{S}}$ is decreased by one and $K^i_{\mc{S}\cup \mc{G}}$ is increased
by one, according to \texttt{ACTFB1}. We now select $\mc{B}^{(i)}_{\mc{S}}(t+1)= 
\mc{B}^{(i)}_{\mc{S}}(t)-\{ \hat{\vec{b}}_i \}$ and $\mc{B}^{(i)}_{\mc{S}\cup\mc{G}}(t+1)=
\mc{B}^{(i)}_{\mc{S}\cup\mc{G}}(t)\cup \{ \vec{b}^{(i)}_s \}$ while all other sets remain 
unchanged. Lemma~\ref{cruc} again implies that the new sets form a basis of $\mathbb{F}^
{\abs{\mc{K}_i}}_q$ at $t+1$.

\item if $i$ erases $s$ and the only users that receive $s$ belong to
  a set $\mc{G}\subseteq \mc{S}$, no $K^i_{\mc{I}}$, $K_{D_i}$ indices
  are affected so that no sets need be changed. In this case, the
  inductive hypothesis holds trivially at $t+1$.
\end{itemize}
 
Since the above list contains all possible cases, we conclude that the hypothesis is true at the 
beginning of slot $t+1$ and the proof is complete.

\section{Proof of Theorem~\ref{theo:genthru}} \label{app3}

\subsection{Some auxiliary results}

We first need to establish some additional notation and intermediate 
results. Denote with $R_{\mc{G}}\stackrel{\vartriangle}{=}\{Z_i=0,\;\forall\, 
i\in\mc{G}\}$ the event that \textit{all} users in set $\mc{G}$ receive the 
transmitted packet, whence it follows from De Morgan's law that
\beq \label{sst}
R^c_{\mc{G}}= \biguplus_{\mc{H}\neq \varnothing: \mc{H}\subseteq \mc{G}}
  \left( E_{\mc{H}}\cap R_{\mc{G}-\mc{H}} \right) ,
\eeq
where $^c$ stands for set complement and $\biguplus$ denotes a union
of disjoint sets. For completeness, we define $E_{\varnothing}=
R_{\varnothing}=\Omega$ (the sample space). Introducing the quantity
$p_{\mc{S},\mc{G}} \stackrel{\vartriangle}{=} \Pr(E_{\mc{S}} \cap
R_{\mc{G}})$ for all disjoint $\mc{S}, \mc{G}\subseteq \mc{N}$, we can
use \eq{sst} to convert the expression $\Pr(E_{\mc{S}})=\Pr(E_{\mc{S}}
\cap R_{\mc{G}})+\Pr(E_{\mc{S}}\cap R^c_{\mc{G}})$ into
\beq \label{aux1}
\Pr(E_{\mc{S}})= \Pr(E_{\mc{S}}\cap R_{\mc{G}})+ \sum_{\mc{H}\neq 
  \varnothing: \mc{H}\subseteq \mc{G}} \Pr( E_{\mc{S}\cup\mc{H}} \cap 
  R_{\mc{G}-\mc{H}} )
\Leftrightarrow p_{\mc{S},\mc{G}}= \epsilon_{\mc{S}}-\sum_{\mc{H}\neq \varnothing:
  \mc{H}\subseteq \mc{G}} p_{\mc{S}\cup\mc{H},\mc{G}-\mc{H}} .
\eeq
Evaluating the last relation for arbitrary $\mc{S}$ and $\mc{G}=\{j\}$, with $j\not\in
\mc{S}$, yields
\beq \label{G1}
p_{\mc{S},\{j\}}= \epsilon_{\mc{S}}-\epsilon_{\mc{S}\cup \{j\}} .
\eeq
The following result provides a general expression for $p_{\mc{S},\mc{G}}$.
\begin{lemma} \label{lem:genp}
For any non-empty disjoint sets $\mc{S},\mc{G}\subseteq \mc{N}$, it holds
\beq \label{indu1}
p_{\mc{S},\mc{G}}= \sum_{\mc{H}\subseteq \mc{G}} (-1)^{\abs{\mc{H}}} 
  \epsilon_{\mc{S}\cup\mc{H}} .
\eeq
\end{lemma}
\begin{IEEEproof}
Proof is by strong induction on $\abs{\mc{G}}$. Specifically, for arbitrary
$\mc{S}$ and $\abs{\mc{G}}=1$ (say, $\mc{G}=\{j\}$), \eq{indu1} becomes
\beq
p_{\mc{S},\{j\}}= \sum_{\mc{H}\subseteq \{j\}} (-1)^{\abs{\mc{H}}} \epsilon_{\mc{S}
  \cup\mc{H}}= (-1)^0 \epsilon_{\mc{S}\cup\varnothing}+(-1)^1 \epsilon_{\mc{S}\cup\{j\}} ,
\eeq
which is identical to \eq{G1}. We now assume that \eq{indu1} is true
for \textit{all} $\mc{S}$ and \textit{all} $\mc{G}$ with
$\abs{\mc{G}}=1,\ldots,l$ and show that \eq{indu1} is still
true for \textit{all} $\mc{S}$ and \textit{all} $\hat{\mc{G}}$ with
$\abs{\hat{\mc{G}}}=l+1$. Specifically, we can write
$\hat{\mc{G}}=\{i\}\cup \mc{G}$ where $i\not\in \mc{G}$ and
$\abs{\mc{G}}=l$, so that we only need to show
\beq \label{des1}
p_{\mc{S},\hat{\mc{G}}}=p_{\mc{S},\mc{G}\cup\{i\}} \stackrel{?}{=} \sum_{\mc{H}
  \subseteq \mc{G}\cup \{i\}} (-1)^{\abs{\mc{H}}} \epsilon_{\mc{S}\cup\mc{H}} .
\eeq
Since any subset $\mc{H}$ of $\mc{G}\cup\{i\}$ is either a subset of $\mc{G}$ 
(and therefore does not contain $i$) or (exclusive or) $\mc{H}$ contains $i$ and a,
possibly empty, subset $\tilde{\mc{H}}$ of $\mc{G}$, the sum in \eq{des1} can be 
written as
\beq \begin{split} \label{decomp2}
p_{\mc{S},\mc{G}\cup\{i\}}& \stackrel{?}{=} \sum_{\mc{H}\subseteq\mc{G}} 
  (-1)^{\abs{\mc{H}}} \epsilon_{\mc{S}\cup\mc{H}}+\sum_{\tilde{\mc{H}}\subseteq\mc{G}} 
  (-1)^{\abs{\tilde{\mc{H}}}+1} \epsilon_{\mc{S}\cup\{i\}\cup \tilde{\mc{H}}} \\
&= \sum_{\mc{H}\subseteq\mc{H}} (-1)^{\abs{\mc{H}}} \left[ \epsilon_{\mc{S}\cup\mc{H}}
  -\epsilon_{\mc{S}\cup\{i\}\cup\mc{H}} \right] .
\end{split} \eeq

However, it also holds
\beq \begin{split} \label{tr1}
p_{\mc{S},\mc{G}\cup \{i\}} &= \Pr(E_{\mc{S}}\cap R_{\{i\}}\cap R_{\mc{G}})=
  \Pr(E_{\mc{S}}\cap R_{\mc{G}})-\Pr(E_{\mc{S}}\cap E_{\{i\}} \cap R_{\mc{G}}) \\
&= \Pr(E_{\mc{S}}\cap R_{\mc{G}}) - \Pr(E_{\mc{S}\cup\{i\}} \cap R_{\mc{G}})= 
  p_{\mc{S},\mc{G}}-p_{\mc{S}\cup\{i\},\mc{G}} ,
\end{split} \eeq
Since $\abs{\mc{G}}\leq l$, the inductive hypothesis holds for 
$p_{\mc{S},\mc{G}}$, $p_{\mc{S}\cup\{i\},\mc{G}}$, whence we conclude that
\beq \begin{split} \label{assu}
p_{\mc{S},\mc{G}} &= \sum_{\mc{H}\subseteq\mc{G}} (-1)^{\abs{\mc{H}}}
  \epsilon_{\mc{S}\cup\mc{H}} , \\
p_{\mc{S}\cup\{i\},\mc{G}} &= \sum_{\mc{H}\subseteq\mc{G}} (-1)^{\abs{\mc{H}}}
  \epsilon_{\mc{S}\cup\{i\}\cup \mc{H}} .
\end{split} \eeq
Inserting \eq{assu} in \eq{tr1} immediately produces the RHS of \eq{decomp2}
and the proof is complete.
\end{IEEEproof}
An immediate consequence of Lemma~\ref{lem:genp} is the following result.
\begin{corollary} \label{onlyS}
For any $\mc{S}\subseteq \mc{N}$ with $i\in \mc{S}$, the probability that a 
transmitted packet is received \textit{exactly} by all users in $\mc{S}-\{i\}$ 
(and none other) is given by 
\beq \nonumber
p_{\mc{N}-(\mc{S}-\{i\}),\mc{S}-\{i\}}= \sum_{\mc{H}\subseteq \mc{S}-\{i\}} 
  (-1)^{\abs{\mc{H}}} \epsilon_{(\mc{N}-(\mc{S}-\{i\}))\cup \mc{H}} .
\eeq
\end{corollary}

For the next auxiliary result, we need to introduce some further
notation. Consider some given $n$, $\vec{R}> \vec{0}$ and the
application of the original $\mbox{\texttt{CODE1}}_{pub}$
(i.e.~without the fixed blocklength modification) for $\vec{K}=\lceil
n \vec{R} \rceil$ packets. We hereafter use consistently a dot accent
to explicitly denote a random variable. We denote with
$\dot{T}^\ast_{i,\mc{S}}$ the number of slots (viewed as a random
variable due to the random erasures) it takes under
$\mbox{\texttt{CODE1}}_{pub}$ for index $K^i_{\mc{S}}$ to become 0
during the processing of queue $Q_{\mc{S}}$, while
$\dot{T}^\ast_{\mc{S}}$ (resp.~$\dot{T}^\ast$) denotes the number of
slots it takes under $\mbox{\texttt{CODE1}}_{pub}$ to process queue
$Q_{\mc{S}}$ (resp.~\textit{all} queues). Hence, it holds
\beq \begin{split} \label{def_dotT}
\dot{T}^\ast_{\mc{S}} &= \max_{i\in \mc{S}} \; \dot{T}^\ast_{i,\mc{S}} , \\
\dot{T}^\ast &= \sum_{\varnothing \neq \mc{S}\subseteq \mc{N}} \dot{T}^\ast_{\mc{S}} . 
\end{split} \eeq

Due to the random erasures, the time-varying index
$\dot{K}^i_{\mc{S}}(t)$ is a random process. We denote with
$\tilde{t}_{\mc{S}}$ the time when processing of queue $Q_{\mc{S}}$
begins and define the random variable $\dot{k}^i_{\mc{S}}=
\dot{K}^i_{\mc{S}}(\tilde{t}_{\mc{S}})$ so that, by the algorithm's
initialization, it holds $\dot{k}^i_{\{i\}}= K_i$ w.p.~1. By
construction of $\mbox{\texttt{CODE1}}_{pub}$, when queue $Q_{\mc{S}}$
is processed at slot $t$ by transmitting a linear combination $s$,
index $\dot{K}^i_{\mc{S}}(t)$ is reduced by one (assuming that
$\dot{K}^i_{\mc{S}}(t)>0$) only if $s$ is received by at least one
user in set $\mc{N}-(\mc{S}-\{i\})$ (i.e.~received by either $i$ or at
least one user in $\mc{N}-\mc{S}$). Denoting with
$\dot{N}^i_{\mc{S},l}$ the number of slots in the time interval
between the $(l-1)$-th and the $l$-th reduction of index
$K^i_{\mc{S}}$ during the processing of $Q_{\mc{S}}$, it clearly
follows that\footnote{for consistency, we assume that the $0$-th
  reduction of $K^i_{\mc{S}}$ occurs at $\tilde{t}_{\mc{S}}$,
  i.e.~when processing of $Q_{\mc{S}}$ begins.}
\beq \label{geom}
\dot{T}^\ast_{i,\mc{S}}= \sum_{l=1}^{\dot{k}^i_{\mc{S}}} \dot{N}^i_{\mc{S},l} ,
\eeq
where $\dot{N}^i_{\mc{S},l}$ are iid geometric random variables with
$\Pr(\dot{N}^i_{\mc{S},l}=\nu)= (a^i_{\mc{S}})^{\nu-1}
(1-a^i_{\mc{S}})$, where $a^i_{\mc{S}}=\epsi_{\mc{N}-(\mc{S}-\{i\})}$.

Assuming that packet $s$ is transmitted from $Q_{\mc{S}}$ at slot $t$
and $\dot{K}^i_{\mc{S}}(t)$ is reduced by 1 at the end of the slot,
exactly one of the following two mutually exclusive events occurs:
either $s$ is successfully received by $i$
(w.p.~$\frac{1-\epsi_i}{1-a^i_{\mc{S}}}$) or $s$ is not received by
$i$ but is received by all users in set $\mc{T}-\mc{S}$ (and erased by
all users in $\mc{N}-\mc{T}$), so that it is placed in queue
$Q_{\mc{T}}$, with $\mc{T}\supset \mc{S}$, due to step~\ref{movtok} of
\texttt{ACTFB1}. The latter case occurs with probability
$\frac{p^i_{\mc{S}\to \mc{T}}}{1-a^i_{\mc{S}}}$, where
\beq \label{pars}
p^i_{\mc{S} \to \mc{T}}= p_{\mc{N}-(\mc{T}-\{i\}),\mc{T}-\mc{S}} .
\eeq
Note that the above events occur \textit{provided} that
$\dot{K}^i_{\mc{S}}(t)>0$ is actually decreased by 1, so that the
corresponding probabilities are actually conditional
probabilities. This is the reason for the appearance of the term
$(1-a^i_{\mc{S}})$ in the denominator of both probabilities.

We denote with $\dot{\mc{D}}^i_{\mc{S},l} \supset \mc{S}$ the index
set of the queue to which the transmitted packet $s$ is moved after
the $l$-th reduction of index $K^i_{\mc{S}}$, with $1\leq l\leq
\dot{k}^i_{\mc{S}}$, during the processing of $Q_{\mc{S}}$. Obviously,
this is a random variable (hence, the dot) that depends on the exact
erasures that occurred during the slot of the $l$-th reduction. From
the previous discussion, it holds $\Pr(\dot{\mc{D}}^i_{\mc{S},l}=
\mc{T})= \frac{p^i_{\mc{S}\to \mc{T}}}{1-a^i_{\mc{S}}}$ for all
$\mc{T}\supset \mc{S}$ and the total number of tokens for user $i$
that were moved into $Q_{\mc{T}}$ during the processing of
$Q_{\mc{S}}$ is
\beq
\dot{k}^i_{\mc{S}\to \mc{T}}= \sum_{l=1}^{\dot{k}^i_{\mc{S}}} \mathbb{I}[\dot{\mc{D}}^i_{\mc{S},l}=\mc{T}]
\eeq
where
\beq \label{indic}
\mathbb{I} \left[ \dot{\mc{D}}^i_{\mc{S},l}= \mc{T} \right]= \left\{
  \begin{array}{l@{\quad}l} 1 & \mbox{w.p. }\frac{p^i_{\mc{S}\to \mc{T}}}{1-a^i_{\mc{S}}} , \\
  0 & \mbox{w.p. } 1-\frac{p^i_{\mc{S}\to \mc{T}}}{1-a^i_{\mc{S}}} . \end{array} \right.
\eeq

Step~\ref{movtok} of \texttt{ACTFB1} now implies the following recursion for all $\mc{S}$ with
$\abs{\mc{S}}\geq 2$
\beq \label{rv_rec}
\dot{k}^i_{\mc{S}}= \sum_{\substack{\varnothing\neq \mc{I}\subset \mc{S}\\ i\in \mc{S}}} 
  k^i_{\mc{I}\to \mc{S}}= \sum_{\substack{\varnothing\neq \mc{I}\subset \mc{S}\\ i\in \mc{S}}} 
  \sum_{l=1}^{\dot{k}^i_{\mc{I}}} \mathbb{I} \left[ \dot{\mc{D}}^i_{\mc{I},l}=\mc{S} \right] ,
\eeq
which captures the property that $\dot{k}^i_{\mc{S}}$ (i.e.~the value
of $K^i_{\mc{S}}$ at the beginning of processing $Q_{\mc{S}}$) is
equal to the cumulative number of tokens for user $i$ that were moved
to $Q_{\mc{S}}$ during the prior processing of queues $Q_{\mc{I}}$,
for $\mc{I}\subset \mc{S}$. %
Rewriting \eq{rv_rec} as
\beq \label{rv_overn}
\frac{\dot{k}^i_{\mc{S}}}{n}= \sum_{\substack{\varnothing\neq \mc{I}\subset \mc{S}\\ i\in \mc{S}}}  
  \frac{\dot{k}^i_{\mc{I}}}{n} \, \frac{1}{\dot{k}^i_{\mc{I}}} \sum_{l=1}^{\dot{k}^i_{\mc{I}}} 
  \mathbb{I} \left[ \dot{\mc{D}}^i_{\mc{I},l}= \mc{S} \right] ,
\eeq
we now state the next result.
\begin{lemma} \label{just_recur}
Under the application of $\mbox{\texttt{CODE1}}_{pub}$ for
$\vec{K}=\lceil n\vec{R} \rceil$, with $\vec{R}> \vec{0}$, it holds
for all $\mc{S}\subseteq \mc{N}$ and $i\in \mc{S}$
\beq \begin{split} \label{asym}
& \lim_{n\to \infty} \frac{\dot{k}^i_{\mc{S}}}{n}= k^i_{\mc{S}} \quad a.e. \\
& \lim_{n\to \infty} \frac{\dot{T}^\ast_{i,\mc{S}}}{\dot{k}^i_{\mc{S}}}= \frac{1}{1-\alpha^i_{\mc{S}}} \quad a.e.
\end{split} \eeq
where $k^i_{\mc{S}}>0$ are deterministic quantities defined through the recursive relation
\beq \label{rec_det}
k^i_{\mc{S}}= \sum_{\substack{\varnothing \neq I \subset \mc{S}\\ i\in \mc{I}}} \frac{k^i_{\mc{I}}}
  {1-\epsi_{\mc{N}-(\mc{I}-\{i\})}}\, p_{\mc{N}-(\mc{S}-\{i\}),\mc{S}-\mc{I}} \quad \forall\, \mc{S}: 
  \abs{\mc{S}} \geq 2 ,
\eeq
and the initial condition $k^i_{\{i\}}= R_i$.
\end{lemma}

\begin{IEEEproof}
Proof is by strong induction on $\abs{\mc{S}}$. For $\abs{\mc{S}}=1$,
the initialization of the algorithm implies that
$\dot{k}^i_{\{i\}}=K_i= \lceil n R_i \rceil$, whence we conclude that
$\dot{k}^i_{\{i\}}/n \to R_i$ a.e.~as $n\to \infty$. Additionally, it
holds $\dot{T}^\ast_{i,\{i\}}= \sum_{l=1}^{\dot{k}^i_{\{i\}}}
\dot{N}^i_{\{i\},l}$ so that the SLLN yields
\beq
\frac{1}{\dot{k}^i_{\{i\}}} \, \dot{T}^\ast_{i,\{i\}}= \frac{1}{\dot{k}^i_{\{i\}}}\, 
  \sum_{l=1}^{\dot{k}^i_{\{i\}}} \dot{N}^i_{\{i\},l} \to \mathbb{E}[\dot{N}^i_{\{i\}}]=
  \frac{1}{1-a^i_{\{i\}}} \quad a.e. \mbox{ as } n\to \infty ,
\eeq
since $\dot{k}^i_{\{i\}}\to \infty$ a.e. as $n\to \infty$.

We now assume that \eq{asym} is true for all $\mc{S}$ with
$\abs{\mc{S}}\leq m$. Applying \eq{rv_overn} to any $\mc{S}$ with
$\abs{\mc{S}}=m+1$, taking a limit as $n\to \infty$ and using the
inductive hypothesis for all $\mc{I}\subset \mc{S}$ (since it holds
$\abs{\mc{I}}\leq m$) and the SLLN (since $\dot{k}^i_{\mc{I}}\to
\infty$ a.e. as $n\to \infty$ and the indicator functions are iid
random variables), we arrive at
\beq \label{ind_p1}
\lim_{n\to \infty} \frac{\dot{k}^i_{\mc{S}}}{n} = \sum_{\substack{\varnothing \neq \mc{I} \subset 
  \mc{S}\\ i\in \mc{S}}} \left( \lim_{n\to \infty} \frac{\dot{k}^i_{\mc{I}}}{n} \right) \mathbb{E} 
  \left[ \mathbb{I} \left[ \dot{\mc{D}}^i_{\mc{I},l}= \mc{S} \right] 
  \right]= \sum_{\substack{\varnothing \neq \mc{I} \subset \mc{S}\\ i\in \mc{S}}} k^i_{\mc{I}} \, 
  \frac{p^i_{\mc{I}\to \mc{S}}}{1-a^i_{\mc{I}}} = k^i_{\mc{S}} .
\eeq  
Using \eq{pars} to substitute for $p^i_{\mc{I}\to \mc{S}}$,
$a^i_{\mc{S}}$, \eq{ind_p1} reduces to \eq{rec_det} for all $\mc{S}$
with $\abs{\mc{S}}=m+1$, so that the induction is complete for the
first equation in \eq{asym}. To prove the second equation in \eq{asym}
for all $\mc{S}$ with $\abs{\mc{S}}=m+1$, we follow a procedure
similar to the case of $\abs{\mc{S}}=1$ so that
\beq
\dot{T}^\ast_{i,\mc{S}}= \sum_{l=1}^{\dot{k}^i_{\mc{S}}} \dot{N}^i_{\mc{S},l} \Rightarrow \frac{\dot{T}^\ast_{i,\mc{S}}}
  {\dot{k}^i_{\mc{S}}}= \frac{1}{\dot{k}^i_{\mc{S}}} \, \sum_{l=1}^{\dot{k}^i_{\mc{S}}} \dot{N}^i_{\mc{S},l} \to 
  \mathbb{E}[\dot{N}^i_{\mc{S}}]= \frac{1}{1-a^i_{\mc{S}}} \quad a.e. \mbox{ as } n\to \infty ,
\eeq
This proves the second equation in \eq{asym} and completes the proof.
\end{IEEEproof}

Using Lemma~\ref{just_recur} and rewriting \eq{def_dotT} as
\beq \begin{split}
& \frac{\dot{T}^\ast_{\mc{S}}}{n}= \max_{i\in \mc{S}} \left( \frac{\dot{T}^\ast_{i,\mc{S}}}{n} 
  \right)= \max_{i\in \mc{S}} \left( \frac{\dot{T}^\ast_{i,\mc{S}}}{\dot{k}^i_{\mc{S}}}\, 
  \frac{\dot{k}^i_{\mc{S}}}{n} \right) , \\
& \frac{\dot{T}^\ast}{n}= \sum_{\varnothing \neq \mc{S}\subseteq \mc{N}} \frac{\dot{T}^\ast_{\mc{S}}}{n} ,
\end{split} \eeq
we can take a limit as $n\to \infty$, use \eq{asym} and exploit the
continuity of $\max$ to pass the limit through it and arrive at the
following Corollary.
\begin{corollary} \label{limits}
Under the application of $\mbox{\texttt{CODE1}}_{pub}$, it holds
\beq \begin{split} \label{avgs}
& \lim_{n\to \infty} \frac{\dot{T}^\ast_{\mc{S}}}{n}= \max_{i\in \mc{S}} \left( \frac{k^i_{\mc{S}}}
  {1-a^i_{\mc{S}}} \right) \quad a.e. \\
& \lim_{n\to \infty} \frac{\dot{T}^\ast}{n}= \sum_{\varnothing \neq \mc{S} \subseteq \mc{N}} 
  \max_{i\in \mc{S}}\, \left( \frac{k^i_{\mc{S}}}{1-a^i_{\mc{S}}} \right) \quad a.e.
\end{split} \eeq
\end{corollary}

The last auxiliary result is an explicit solution of \eq{rec_det}
(along with the initial condition $k^i_{\{i\}}= R_i$) which,
introducing the variable
\beq \label{def_f}
f^i_{\mc{S}} \stackrel{\vartriangle}{=} \frac{k^i_{\mc{S}}}{ R_i (1-\epsi_{\mc{N}-(\mc{S}-\{i\})})},
\eeq
is cast into the more convenient form
\beq \label{recur_f}
f^i_{\mc{S}}= \frac{1}{1-\epsilon_{\mc{N}-(\mc{S}-\{i\})}}\, 
  \sum_{\substack{\varnothing\neq \mc{I}\subset \mc{S}\\ i\in\mc{I}}} f^i_{\mc{I}}
  \, p_{\mc{N}-(\mc{S}-\{i\}),\mc{S}-\mc{I}} \quad \forall\, \mc{S}: \abs{\mc{S}} \geq 2 ,
\eeq
with an initial condition of
$f^i_{\{i\}}=\frac{1}{1-\epsilon_{\mc{N}}}$.  The following Lemma
provides an explicit representation of $f^i_{\mc{S}}$ and shows that
$f^i_{\mc{S}}$ is identical to the quantity $\hat{f}^i_{\mc{S}}$
introduced in Theorem~\ref{theo:genthru}.
\begin{lemma} \label{lem:genf}
For any set $\mc{S}\subseteq \mc{N}$ with $i\in \mc{S}$, it holds
\beq \label{expf}
f^i_{\mc{S}}= \sum_{\mc{H}\subseteq \mc{S}-\{i\}} \frac{(-1)^{\abs{\mc{H}}}}{1-
  \epsilon_{(\mc{N}-(\mc{S}-\{i\}))\cup \mc{H}}} .
\eeq
\end{lemma}

\begin{IEEEproof}
The following equivalent expression can be derived from Lemma~\ref{lem:genp}.
\beq \begin{split} \label{indu2}
p_{\mc{S},\mc{G}} &= \sum_{\mc{H}\subseteq\mc{G}} (-1)^{\abs{\mc{H}}} \epsilon_{\mc{S}\cup\mc{H}}
  = \sum_{\mc{H}\subseteq\mc{G}} (-1)^{\abs{\mc{H}}} \left( 1-(1-\epsilon_{\mc{S}\cup\mc{H}})
  \right) \\
&= \sum_{\mc{H}\subseteq\mc{G}} (-1)^{\abs{\mc{H}}}+ \sum_{\mc{H}\subseteq\mc{G}}
  (-1)^{\abs{\mc{H}}+1} (1-\epsilon_{\mc{S}\cup\mc{H}})= \sum_{\mc{H}\subseteq\mc{G}}
  (-1)^{\abs{\mc{H}}+1} (1-\epsilon_{\mc{S}\cup\mc{H}}) ,
\end{split} \eeq
where we used the binomial theorem to compute $\sum_{\mc{H}\subseteq \mc{G}} (-1)^{\abs{\mc{H}}}= 
\sum_{r=0}^{\abs{\mc{G}}} {\abs{\mc{G}} \choose r} (-1)^r= (1-1)^{\abs{\mc{G}}}=0$. 

We initially manipulate \eq{recur_f} by substituting for $p_{\mc{N}-(\mc{S}-\{i\}),
\mc{S}-\mc{I}}$ through \eq{indu2}, which yields
\beq
f^i_{\mc{S}}= \frac{1}{1-\epsilon_{\mc{N}-(\mc{S}-\{i\})}} \sum_{\{i\}\subseteq 
  \mc{I}\subset \mc{S}} f^i_{\mc{I}} \sum_{\mc{H}\subseteq \mc{S}-\mc{I}} 
  (-1)^{\abs{\mc{H}}+1} (1-\epsilon_{(\mc{N}-(\mc{S}-\{i\}))\cup \mc{H}}) .
\eeq
Extracting the $\mc{H}=\varnothing$ term from the summation over $\mc{H}$ yields
\beq \begin{split}
f^i_{\mc{S}} &= \frac{1}{1-\epsilon_{\mc{N}-(\mc{S}-\{i\})}} \sum_{\{i\}\subseteq 
  \mc{I}\subset \mc{S}} f^i_{\mc{I}} \left[ -(1-\epsilon_{\mc{N}-(\mc{S}-\{i\})})
  +\sum_{\varnothing\neq \mc{H}\subseteq \mc{S}-\mc{I}} (-1)^{\abs{\mc{H}}+1}
  (1-\epsilon_{(\mc{N}-(\mc{S}-\{i\}))\cup \mc{H}}) \right] \\
&= -\sum_{\{i\}\subseteq\mc{I}\subset\mc{S}} f^i_{\mc{I}} +
  \frac{1}{1-\epsilon_{\mc{N}-(\mc{S}-\{i\})}} \sum_{\varnothing\neq \mc{H}\subseteq
  \mc{S}-\{i\}} \sum_{\{i\}\subseteq \mc{I}\subseteq \mc{S}-\mc{H}} f^i_{\mc{I}}
  (-1)^{\abs{\mc{H}}+1} (1-\epsilon_{(\mc{N}-(\mc{S}-\{i\}))\cup \mc{H}}) ,
\end{split} \eeq
where we changed the order of summation in the second sum of the last line. Moving the 
first sum in the RHS of the last expression to the LHS produces
\beq \label{recur_sum}
\sum_{\{i\}\subseteq \mc{I}\subseteq \mc{S}} f^i_{\mc{I}}=  
  \frac{1}{1-\epsilon_{\mc{N}-(\mc{S}-\{i\})}} \sum_{\varnothing\neq \mc{H}\subseteq
  \mc{S}-\{i\}} (-1)^{\abs{\mc{H}}+1} (1-\epsilon_{(\mc{N}-(\mc{S}-\{i\}))\cup \mc{H}})
  \left[ \sum_{\{i\}\subseteq \mc{I}\subseteq \mc{S}-\mc{H}} f^i_{\mc{I}} \right] ,
\eeq 
which provides a new recursion w.r.t.~the term $\sum_{\{i\}\subseteq \mc{I}
\subseteq \mc{S}} f^i_{\mc{I}}$. 

For a fixed $i$, we can use induction on $\abs{\mc{S}}$ to show the following 
relation
\beq \label{indu3}
\sum_{\{i\}\subseteq \mc{I}\subseteq \mc{S}} f^i_{\mc{I}}= \frac{1}{1-
  \epsilon_{\mc{N}-(\mc{S}-\{i\})}}, \quad \forall \, \mc{S}, \; \forall\, i\in \mc{S} .
\eeq
Indeed, for $\abs{\mc{S}}=1$, which implies $\mc{S}=\{i\}$, \eq{indu3} yields
$f^i_{\{i\}}=\frac{1}{1-\epsilon_{\mc{N}}}$, which is identical to the initial 
condition of \eq{recur_f}. We now assume that \eq{indu3} is true for all $\mc{S}$
with $\abs{\mc{S}}\leq l$ and show that it is also true for all $\mc{S}$ with
$\abs{\mc{S}}=l+1$. Specifically, for any $\mc{S}$ with $\abs{\mc{S}}=l+1$, 
\eq{recur_sum} becomes
\beq \begin{split} 
\sum_{\{i\}\subseteq \mc{I}\subseteq \mc{S}} f^i_{\mc{I}} &= \frac{1}{1-
  \epsilon_{\mc{N}-(\mc{S}-\{i\})}} \sum_{\varnothing\neq \mc{H}
  \subseteq \mc{S}-\{i\}} (-1)^{\abs{\mc{H}}+1} (1-\epsilon_{(\mc{N}-(\mc{S}-\{i\}))
  \cup\mc{H}})\, \frac{1}{1-\epsilon_{(\mc{N}-(\mc{S}-\{i\})) \cup\mc{H}}} \\
&= \frac{1}{1-\epsilon_{\mc{N}-(\mc{S}-\{i\})}} ,
\end{split} \eeq
where we used the inductive hypothesis for the terms $\sum_{\{i\}\subseteq \mc{I}
\subseteq \mc{S}-\mc{H}} f^i_{\mc{I}}$, since $\abs{\mc{I}}\leq \abs{\mc{S}-\mc{H}}
\leq l$ when $\mc{H}\neq \varnothing$, and applied the binomial theorem. This 
completes the induction and proves \eq{indu3}.   

We denote with $\hat{p}_{\mc{I}}\stackrel{\vartriangle}{=}p_{\mc{N}-\mc{I},\mc{I}}$ 
the probability that a packet is received by exactly the users in $\mc{I}$ (and none 
other), whence we deduce the following relation
\beq \label{rr}
\sum_{\mc{I}\subseteq \mc{S}} \hat{p}_{\mc{I}}= \Pr\left( \bigcup_{\mc{I}\subseteq \mc{S}} 
  (E_{\mc{N}-\mc{I}} \cap R_{\mc{I}}) \right)= \Pr(E_{\mc{N}-\mc{S}})= \epsilon_{\mc{N}-\mc{S}} , 
\eeq
which is true for \textit{any} $\mc{S}\subseteq \mc{N}$. Hence, it also holds 
$\sum_{\mc{I}\subseteq \mc{S}-\{i\}} \hat{p}_{\mc{I}}=\epsilon_{\mc{N}-(\mc{S}-\{i\})}$, 
so that the following is true for all $\mc{S}$ and $i\in \mc{S}$
\beq \begin{split} \label{conne}
& \sum_{\mc{I}\subseteq \mc{S}-\{i\}} \hat{p}_{\mc{I}}= \epsilon_{\mc{N}-(\mc{S}-\{i\})} ,\\
& \sum_{\mc{I}\subseteq \mc{S}-\{i\}} f^i_{\mc{I}\cup\{i\}}= \frac{1}
  {1-\epsilon_{\mc{N}-(\mc{S}-\{i\})}} ,
\end{split} \eeq
along with the initial conditions $\hat{p}_{\varnothing}=\epsilon_{\mc{N}}$, $f^i_{\{i\}}=
1/(1-\epsilon_{\mc{N}})$. The second equation in \eq{conne} is essentially a rewrite of
\eq{indu3}.

We now make the crucial observation that \eq{conne} allows for a separate recursive 
computation of $f^i_{\mc{S}}$, $\hat{p}_{\mc{S}}$ based on the 
corresponding initial condition. Since the only difference between the two recursions 
is the RHS term (the recursion for $\hat{p}_{\mc{I}}$, $f^i_{\mc{I}\cup\{i\}}$ uses 
$\epsilon_{\mc{N}-(\mc{S}-\{i\})}$, $(1-\epsilon_{\mc{N}-(\mc{S}-\{i\})})^{-1}$, 
respectively), we conclude that any relation that holds for $\hat{p}_{\mc{I}}$ also holds 
for $f^i_{\mc{I}\cup\{i\}}$ via a substitution $\epsilon_{\mc{N}-(\mc{S}-\{i\})} \to 
(1-\epsilon_{\mc{N}-(\mc{S}-\{i\})})^{-1}$. Combining the last statement with 
Corollary~\ref{onlyS} (which provides an expression for $\hat{p}_{\mc{S}-\{i\}}$), yields
\beq
f^i_{\mc{S}}= f^i_{(\mc{S}-\{i\})\cup\{i\}}= \sum_{\mc{H}\subseteq \mc{S}-\{i\}}
  \frac{(-1)^{\abs{\mc{H}}}}{1-\epsilon_{(\mc{N}-(\mc{S}-\{i\}))\cup \mc{H}}} ,
\eeq
which completes the proof.
\end{IEEEproof}

\subsection{Proof of Theorem~\ref{theo:genthru}} \label{app:genthru}

We are now in position to finally prove
Theorem~\ref{theo:genthru}. Through a change of variable
$\mc{H}^\prime=(\mc{S}-\{i\})-\mc{H}$, \eq{expf} can also be written
as $f^i_{\mc{S}}=\sum_{\mc{H}^\prime\subseteq \mc{S}-\{i\}}
\frac{(-1)^{\abs{\mc{S}}- \abs{\mc{H}^\prime}-1}}{1-\epsilon_{\mc{N}-\mc{H}^\prime}}=
\hat{f}^i_{\mc{S}}$. Additionally, using Lemma~\ref{lem:genf} and
\eq{def_f} to substitute for $k^i_{\mc{S}}$ in \eq{avgs} yields
\beq \begin{split} \label{auxme}
& \bar{T}^\ast_{\mc{S}}(\vec{R}) \stackrel{\vartriangle}{=} 
  \lim_{n\to \infty} \frac{\dot{T}^\ast_{\mc{S}}}{n} = \max_{i\in \mc{S}} \;
  \left( \frac{\hat{f}^i_{\mc{S}} (1-\epsi_{\mc{N}-(\mc{S}-\{i\})})}{1-a^i_{\mc{S}}} \right)= 
  \max_{i\in \mc{S}} \; ( \hat{f}^i_{\mc{S}} R_i ) , \\
& \bar{T}^\ast(\vec{R}) \stackrel{\vartriangle}{=} \lim_{n\to \infty} \frac{\dot{T}^\ast}{n}= 
  \sum_{\varnothing \neq \mc{S} \subseteq \mc{N}} \bar{T}^\ast_{\mc{S}}(\vec{R})=  \sum_{\varnothing 
  \neq \mc{S} \subseteq \mc{N}} \max_{i\in \mc{S}} \; (\hat{f}^i_{\mc{S}} R_i) ,
\end{split} \eeq
where we also used \eq{pars} to substitute for $a^i_{\mc{S}}$. We now show that the
achievable region of $\mbox{\texttt{CODE1}}_{pub}$, in information symbols per
transmission, is given by
\beq \label{conn}
\mc{R}_{\mbox{\scriptsize \texttt{CODE1}}_{pub}}= \left\{ \vec{R}: \bar{T}^\ast(\vec{R}) \leq 1 \right\} .
\eeq 
The reader can verify that \eq{conn} readily yields \eq{code1ach}
through \eq{auxme}, considering the fact that each symbol/packet
contains $L$ bits. Hence, it remains to prove \eq{conn}, which is
equivalent to proving the following statements: 1) any $\vec{R}$ such
that $\bar{T}^\ast(\vec{R})< 1$ is achievable by
$\mbox{\texttt{CODE1}}_{pub}$, and 2) no $\vec{R}$ with
$\bar{T}^\ast(\vec{R})>1$ is achievable by
$\mbox{\texttt{CODE1}}_{pub}$.

To prove the first part of \eq{conn}, consider any $\vec{R}\neq
\vec{0}$ with $\bar{T}^\ast(\vec{R})< 1$ and apply the fixed
blocklength version of $\mbox{\texttt{CODE1}}_{pub}$ (i.e.~stop after
$n$ transmissions), with $\vec{K}=\lceil n\vec{R} \rceil$. By
construction of the modified $\mbox{\texttt{CODE1}}_{pub}$, an error
occurs iff $\dot{T}^\ast > n$. Hence, the probability of error for the
modified $\mbox{\texttt{CODE1}}_{pub}$ is
\beq \label{perr}
p_n(e)= \Pr (\dot{T}^\ast> n)= \Pr \left( \frac{\dot{T}^\ast}{n} > 1 \right)= \Pr\left( \frac{\dot{T}^\ast}{n}
  -\bar{T}^\ast(\vec{R}) > 1- \bar{T}^\ast(\vec{R}) \right) .
\eeq
Letting $n\to \infty$, the relation $\bar{T}^\ast(\vec{R})< 1$
implies, through \eq{perr}, that $p_n(e)\to 0$, since the LHS of the
inequality in the last event in \eq{perr} goes to $0$ as $n\to
\infty$, while the RHS is strictly positive. This proves the first
part of \eq{conn}. A similar argument can be used to show that
$T^\ast(\vec{r})> 1$ implies $\lim_{n\to \infty} p_n(e)=1$, which
proves the second part of \eq{conn}.

\section{Proof of Lemma~\ref{lem:simp} and Theorem~\ref{theo:genopt}} \label{app4}

Consider an arbitrary $\vec{R}\in \mc{R}_{ord}$ and define the set
\beq \label{defPhi}
\Phi_{\tilde{\pi}}(j) \stackrel{\vartriangle}{=} \{ k\in \mc{N}: 
  \tilde{\pi}(k) \geq j\} ,
\eeq
where $\tilde{\pi}$ is the permutation corresponding to $\vec{R}$ via \eq{deford}. 
Additionally, there exists the functional inverse $\tilde{\pi}^{-1}$ of $\tilde{\pi}$ 
(since $\tilde{\pi}$ is a bijection on $\mc{N}$), which is a permutation on $\mc{N}$ as
well. In fact, the introduction of $\tilde{\pi}^{-1}$ allows us to rewrite 
\eq{defPhi} as
\beq
\Phi_{\tilde{\pi}}(j)= \{ \tilde{\pi}^{-1}(j),\tilde{\pi}^{-1}(j+1),\ldots,
  \tilde{\pi}^{-1}(N) \} ,
\eeq
which can be proved by standard bidirectional set inclusion. It now holds
\beq \label{fsum}
\sum_{\mc{S}\subseteq \mc{N}} \max_{i\in \mc{S}} \left( \hat{f}^i_{\mc{S}} R_i \right)= \sum_{l=1}^N
  \sum_{\mc{S}:l=\argmax_{i\in\mc{S}} (\hat{f}^i_{\mc{S}} R_i)} \hat{f}^l_{\mc{S}} R_l
  = \sum_{j=1}^N R_{\tilde{\pi}^{-1}(j)} \sum_{\mc{S}: \tilde{\pi}^{-1}(j)= \argmax_{i\in \mc{S}}
  (\hat{f}^i_{\mc{S}} R_i )} \hat{f}^{\tilde{\pi}^{-1}(j)}_{\mc{S}} ,
\eeq
where the last equality follows from the substitution $l=\tilde{\pi}^{-1}(j)$. Since 
$\vec{R}\in \mc{R}_{ord}$, \eq{deford} now implies
\beq
\left\{ \mc{S}: \argmax_{i\in \mc{S}} \left( \hat{f}^i_{\mc{S}} R_i \right)= \tilde{\pi}^{-1}(j)
  \right\}= \left\{ \mc{S}: \tilde{\pi}^{-1}(j)=\argmin_{i\in \mc{S}} (\tilde{\pi}(i)) 
  \right\} ,
\eeq
so that the inner sum in the RHS of \eq{fsum} becomes
\beq \label{trick4}
\sum_{\mc{S}: \tilde{\pi}^{-1}(j)= \argmin_{i\in\mc{S}} (\tilde{\pi}(i))}
  \hat{f}^{\tilde{\pi}^{-1}(j)}_{\mc{S}} = \sum_{\mc{S}: \{ \tilde{\pi}^{-1}(j)\}
  \subseteq \mc{S} \subseteq \Phi_{\tilde{\pi}}(j)} \hat{f}^{\tilde{\pi}^{-1}(j)}_{\mc{S}}
  = \frac{1}{1-\epsilon_{\mc{N}-(\Phi_{\tilde{\pi}(j)}-\{\tilde{\pi}^{-1}(j) \})}} ,
\eeq
where the first equality follows from the fact that, by construction, 
all sets $\mc{S}$ appearing in the summation of \eq{trick4} satisfy the relation
\beq
\left\{ \tilde{\pi}^{-1}(j) \right\} \subseteq \mc{S} \subseteq \left\{ k\in \mc{N}: 
  \tilde{\pi}(k) \geq \tilde{\pi}(\tilde{\pi}^{-1}(j)) \right\} = \Phi_{\tilde{\pi}}(j) ,
\eeq
and the second equality follows from \eq{indu3}.

The definition of $\Phi_{\tilde{\pi}}(j)$ now implies
\beq \label{nset}
\mc{N}-( \Phi_{\tilde{\pi}}(j)- \{ \tilde{\pi}^{-1}(j)\})= \{k\in \mc{N}: \tilde{\pi}(k)\leq j \}
  = \{ \tilde{\pi}^{-1}(1),\ldots, \tilde{\pi}^{-1}(j) \} ,
\eeq
which can again be proved by bidirectional set inclusion. Inserting \eq{nset} into
\eq{trick4} and \eq{fsum} finally yields
\beq \label{achcode}
\sum_{\mc{S}\subseteq \mc{N}} \max_{i\in \mc{S}} \left( \hat{f}^i_{\mc{S}} R_i \right) = 
  \sum_{j=1}^N \frac{R_{\tilde{\pi}^{-1}(j)}}{1-\epsilon_{\{\tilde{\pi}^{-1}(1),\ldots,
  \tilde{\pi}^{-1}(j)\}}} ,
\eeq
which completes the proof of Lemma~\ref{lem:simp}.

Regarding Theorem~\ref{theo:genopt}, we can prove that $\mc{R}_{\mbox{\scriptsize\texttt{CODE1}}_{pub}}
\cap \mc{R}_{ord}= \mc{C}^{out}\cap \mc{R}_{ord}$ by showing that $\mc{R}_{\mbox{\scriptsize\texttt{CODE1}}_{pub}} 
\cap \mc{R}_{ord} \supseteq \mc{C}^{out}\cap \mc{R}_{ord}$ (the inclusion in the other direction
follows trivially from the fact $\mc{R}_{\mbox{\scriptsize\texttt{CODE1}}_{pub}}\subseteq \mc{C}^{out}$). 
Indeed, pick any $\vec{R}\in \mc{C}^{out}\cap \mc{R}_{ord}$. Since $\vec{R}\in \mc{C}^{out}$, 
Lemma~\ref{lem:outer} implies that it holds
\beq \label{tp1}
\max_{\pi\in \mc{P}} \left( \sum_{i=1}^N \frac{R_{\pi(i)}}{1-\epsilon_{\{\pi(1),
  \ldots,\pi(i)\}}} \leq L \right) ,
\eeq
where $\mc{P}$ is the set of all possible permutations on $\mc{N}$, so that $\mc{P}$ includes 
both $\tilde{\pi}$ and $\tilde{\pi}^{-1}$. Hence, \eq{tp1} also holds for the specific 
permutation $\tilde{\pi}$ (corresponding to the chosen $\vec{R}$), which implies through
\eq{achcode} and Theorem~\ref{theo:genthru} that $\vec{R}\in \mc{R}_{\mbox{\scriptsize\texttt{CODE1}}_{pub}}$. Since 
$\vec{R}$ also belongs to $\mc{R}_{ord}$, it follows that $\mc{R}_{\mbox{\scriptsize\texttt{CODE1}}_{pub}} \cap 
\mc{R}_{ord} \supseteq \mc{C}^{out} \cap \mc{R}_{ord}$. This completes the proof of the first 
statement in Theorem~\ref{theo:genopt}.

The second statement of Theorem~\ref{theo:genopt} now follows from the fact that the assumption 
$\mc{R}_{ord} \supseteq \mc{C}^{out}$ (which also implies $\mc{R}_{ord} \supseteq 
\mc{R}_{\mbox{\scriptsize\texttt{CODE1}}_{pub}}$) transforms the established relation
$\mc{R}_{\mbox{\scriptsize\texttt{CODE1}}_{pub}}\cap \mc{R}_{ord}= \mc{C}^{out}\cap \mc{R}_{ord}=\mc{D}$ into 
$\mc{R}_{\mbox{\scriptsize\texttt{CODE1}}_{pub}}=\mc{C}^{out}=\mc{D}$. Hence, 
$\mbox{\texttt{CODE1}}_{pub}$ achieves capacity in this case.

\section{Proof of Theorem~\ref{optc1}} \label{app5}

For symmetric channels, we introduce the notation $\tilde{\epsilon}_{\abs{\mc{I}}}=
\epsilon_{\mc{I}}$ for all $\mc{I}\subseteq \mc{N}$ with a given $\abs{\mc{I}}$. It then 
holds $\tilde{\epsilon}_1 \geq \ldots \geq \tilde{\epsilon}_N$, which in turn implies 
$\frac{1}{1-\tilde{\epsilon}_1} \geq \ldots \geq \frac{1}{1-\tilde{\epsilon}_N}$. 
A simple index exchange argument in Lemma~\ref{lem:outer} reveals that 
$\mc{C}^{out}$ can be written as
\beq \label{rewsym}
\mc{C}^{out}=\left\{ \vec{R}\geq \vec{0}: 
  \sum_{i=1}^N \frac{R_{\mathring{\pi}^{-1}(i)}}{1-\tilde{\epsilon}_i} \leq L \right\} ,
\eeq
where $\mathring{\pi}$ is the permutation on $\mc{N}$ that rearranges 
$\vec{R}$ in non-decreasing order, i.e.~$R_{\mathring{\pi}^{-1}(1)} \geq  \ldots \geq 
R_{\mathring{\pi}^{-1}(N)}$.

By definition of symmetric channels, it also holds 
\beq \nonumber
f^i_{\mc{S}}= \sum_{\mc{H}\subseteq \mc{S}-\{i\}} \frac{(-1)^{\abs{\mc{S}}-
  \abs{\mc{H}}-1}}{1-\epsilon_{\mc{N}-\mc{H}}}= \sum_{m=0}^{\abs{\mc{S}}-1}
  {{\abs{\mc{S}}-1} \choose m} \frac{(-1)^{\abs{\mc{S}}-m-1}}{1-\tilde{\epsilon}_{N-m}} ,
\eeq
where we used the fact that there exist ${{\abs{\mc{S}}-1} \choose m}$ subsets $\mc{H}$ 
of $\mc{S}-\{i\}$ with cardinality $m$. Hence, $f^i_{\mc{S}}$ is independent of $i$, so 
that for all $\vec{R}\geq \vec{0}$ it holds
\beq
\argmax_{i\in \mc{S}} \;( f^i_{\mc{S}} R_i)=\argmax_{i\in \mc{S}} \;(R_i)= 
  \argmin_{i\in \mc{S}} \; (\mathring{\pi}(i)) ,
\eeq
where the last equality is due to the definition of
$\mathring{\pi}$. Hence, it holds $\mc{R}_{ord}=\{\vec{R}: \vec{R}\geq
\vec{0} \}$ since we can select, for each $\vec{R}\geq \vec{0}$, the
permutation $\tilde{\pi}=\mathring{\pi}$ to satisfy \eq{deford}.
Since $\mc{R}_{ord} \supseteq \mc{C}^{out}$,
$\mbox{\texttt{CODE1}}_{pub}$ achieves capacity for symmetric channels
and its rate region is given by \eq{rewsym}.

In the case of one-sided fair spatially independent channels, we must show that 
any vector $\vec{R}\in \mc{R}_{fair}$, i.e.~any vector which satisfies
\beq \begin{split} \label{ass3}
& \epsilon_1\geq \ldots \geq \epsilon_N, \\
& \epsilon_1 R_1 \geq \ldots \geq \epsilon_N R_N ,
\end{split} \eeq 
also belongs to $\mc{R}_{ord}$, i.e.~there exists a permutation $\tilde{\pi}$ 
such that it holds $\argmax_{i\in \mc{S}} (\hat{f}^i_{\mc{S}} R_i)=\argmin_{i\in\mc{S}} 
(\tilde{\pi}(i))$ for all $\mc{S}\subseteq \mc{N}$. In fact, we will show that the 
required permutation $\tilde{\pi}$ is the identity permutation; in other words, we will 
prove that \eq{ass3} implies $f^i_{\mc{S}} R_i \geq f^j_{\mc{S}} R_j$ for all $i,j\in
\mc{S}$ with $i<j$.

Consider an arbitrary set $\mc{S}\subseteq \mc{N}$ and let $i,j\in \mc{S}$. Using 
Lemma~\ref{lem:genf} and exploiting the spatial independence, we compute $f^i_{\mc{S}}$ 
as
\beq \begin{split} \label{ss1}
f^i_{\mc{S}} &= \sum_{\mc{H}\subseteq \mc{S}-\{i\}} \frac{(-1)^{\abs{\mc{H}}}}
  {1-\epsilon_{(\mc{N}-(\mc{S}-\{i\}))\cup \mc{H}}}= \sum_{\substack{\mc{H}\subseteq 
  \mc{S}-\{i\}\\ j\not\in\mc{H}}} \frac{(-1)^{\abs{\mc{H}}}}
  {1-\epsilon_{\mc{N}-\mc{S}} \epsilon_i \epsilon_{\mc{H}}} + \sum_{\substack{\mc{H}
  \subseteq \mc{S}-\{i\}\\ j\in\mc{H}}} \frac{(-1)^{\abs{\mc{H}}}}
  {1-\epsilon_{\mc{N}-\mc{S}} \epsilon_i \epsilon_j \epsilon_{\mc{H}-\{j\}}} \\
&= \sum_{\mc{H}\subseteq \mc{S}-\{i,j\}} \frac{(-1)^{\abs{\mc{H}}}}
  {1-\epsilon_{\mc{N}-\mc{S}} \epsilon_i \epsilon_{\mc{H}}}+ \sum_{\mc{H} \subseteq 
  \mc{S}-\{i,j\}} \frac{(-1)^{\abs{\mc{H}}+1}}{1-\epsilon_{\mc{N}-\mc{S}}\epsilon_i 
  \epsilon_j \epsilon_{\mc{H}}} .
\end{split} \eeq
For an arbitrary set $\mc{S}$, define $m_{\mc{S}}=\min\;\{k\in \mc{S} \}$, so that 
it suffices to show $f^{m_{\mc{S}}}_{\mc{S}} R_{m_{\mc{S}}} \geq f^i_{\mc{S}} R_i$ 
for all $\mc{S}$ and $i\in\mc{S}$. Since it holds, by \eq{ass3}, $\frac{R_{m_{\mc{S}}}}
{R_i}\geq \frac{\epsilon_i}{\epsilon_{m_{\mc{S}}}}$, we will prove the desired inequality
$\frac{R_{m_{\mc{S}}}}{R_i} \geq \frac{f^i_{\mc{S}}}{f^{m_{\mc{S}}}_{\mc{S}}}$ by proving 
the stronger inequality $\frac{\epsilon_i}{\epsilon_{m_{\mc{S}}}} \stackrel{?}{\geq} 
\frac{f^i_{\mc{S}}} {f^{m_{\mc{S}}}_{\mc{S}}}$, or equivalently 
\beq \label{proveit}
\epsilon_i f^{m_{\mc{S}}}_{\mc{S}} \stackrel{?}{\geq} \epsilon_{m_{\mc{S}}} 
  f^i_{\mc{S}}, \quad \forall\, \mc{S}, \; i\in\mc{S} . 
\eeq

We now concentrate on \eq{proveit} and manipulate it through \eq{ss1} to produce
the equivalent relation 
\beq \begin{split}
& \sum_{\mc{H}\subseteq \mc{S}-\{i,m_{\mc{S}}\} } (-1)^{\abs{\mc{H}}} 
  \frac{\epsilon_{m_{\mc{S}}}-\epsilon_{i}}{1-\epsilon_{\mc{N}-\mc{S}}
  \epsilon_{m_{\mc{S}}} \epsilon_{i} \epsilon_{\mc{H}}} \stackrel{?}{\geq} 
  \sum_{\mc{H}\subseteq \mc{S}-\{i,m_{\mc{S}}\} } (-1)^{\abs{\mc{H}}} \left( 
  \frac{\epsilon_{m_{\mc{S}}}}{1-\epsilon_{\mc{N}-\mc{S}} \epsilon_{i} 
  \epsilon_{\mc{H}}} -\frac{\epsilon_{i}}{1-\epsilon_{\mc{N}-\mc{S}}
  \epsilon_{m_{\mc{S}}}\epsilon_{\mc{H}}} \right) \\
\Leftrightarrow & \sum_{\mc{H}\subseteq \mc{S}-\{i,m_{\mc{S}}\} }
  (-1)^{\abs{\mc{H}}} \frac{\epsilon_{m_{\mc{S}}}-\epsilon_{i}}{1-\epsilon_{\mc{N}
  -\mc{S}} \epsilon_{m_{\mc{S}}} \epsilon_{i} \epsilon_{\mc{H}}} \stackrel{?}{\geq}
  \sum_{\mc{H}\subseteq \mc{S}-\{i,m_{\mc{S}}\} } (-1)^{\abs{\mc{H}}}
  \frac{ (\epsilon_{m_{\mc{S}}}-\epsilon_{i}) (1-\epsilon_{\mc{N}-\mc{S}} \epsilon_{i}
  \epsilon_{\mc{H}}-\epsilon_{\mc{N}-\mc{S}} \epsilon_{m_{\mc{S}}} 
  \epsilon_{\mc{H}})}{ (1-\epsilon_{\mc{N}-\mc{S}} \epsilon_{i} \epsilon_{\mc{H}}) 
  (1-\epsilon_{\mc{N}-\mc{S}} \epsilon_{m_{\mc{S}}} \epsilon_{\mc{H}})} . \label{eq2}
\end{split} \eeq
Using the fact that $\epsilon_{m_{\mc{S}}}\geq \epsilon_i$ and the following equality
\beq
\frac{1-\epsilon_{\mc{N}-\mc{S}}\epsilon_{i} \epsilon_{\mc{H}}-\epsilon_{\mc{N}-\mc{S}}
  \epsilon_{m_{\mc{S}}} \epsilon_{\mc{H}}} {(1-\epsilon_{N-S}\epsilon_{i}\epsilon_{H}) 
  (1-\epsilon_{N-S}\epsilon_{m}\epsilon_{H}} =1-\frac{\epsilon^2_{\mc{N}-\mc{S}} 
  \epsilon^2_{\mc{H}} \epsilon_{i} \epsilon_{m_{\mc{S}}}} {(1-\epsilon_{\mc{N}-\mc{S}} 
  \epsilon_{i} \epsilon_{\mc{H}}) (1-\epsilon_{\mc{N}-\mc{S}} \epsilon_{m_{\mc{S}}} 
  \epsilon_{\mc{H}})} ,
\eeq
we can write an equivalent expression to \eq{eq2} as
\beq \label{eq3}
\sum_{\mc{H}\subseteq \mc{S}-\{i,m_{\mc{S}}\} } \frac{(-1)^{\abs{\mc{H}}}}
  {1-\epsilon_{\mc{N}-\mc{S}} \epsilon_{m_{\mc{S}}} \epsilon_{i} \epsilon_{\mc{H}}}
  +\sum_{\mc{H}\subseteq \mc{S}-\{i,m_{\mc{S}}\} } (-1)^{\abs{\mc{H}}} \,
  \frac{\epsilon^2_{\mc{N}-\mc{S}} \epsilon^2_{\mc{H}} \epsilon_{i} \epsilon_{m_{\mc{S}}}}
  {(1-\epsilon_{\mc{N}-\mc{S}}\epsilon_{i} \epsilon_{\mc{H}}) (1-\epsilon_{\mc{N}-\mc{S}}
  \epsilon_{m_{\mc{S}}} \epsilon_{H})} \stackrel{?}{\geq} 0 ,
\eeq
where we also used the identity $\sum_{\mc{H}\subseteq \mc{S}-\{i,m_{\mc{S}}\}} 
(-1)^{\abs{\mc{H}}}=0$.

We now observe that the first term of \eq{eq3} is equal to the non-negative quantity
$f^i_{\mc{S}-\{m_{\mc{S}}\}}$ so that, in order to prove \eq{eq3}, it suffices to prove
the second term in \eq{eq3} to be non-negative, namely
\beq \label{eq4}
\sum_{\mc{H}\subseteq \mc{S}-\{i,m_{\mc{S}}\} } (-1)^{\abs{\mc{H}}} \,
  \frac{\epsilon^2_{\mc{N}-\mc{S}} \epsilon^2_{\mc{H}} \epsilon_{i} \epsilon_{m_{\mc{S}}}}
  {(1-\epsilon_{\mc{N}-\mc{S}}\epsilon_{i} \epsilon_{\mc{H}}) (1-\epsilon_{\mc{N}-\mc{S}}
  \epsilon_{m_{\mc{S}}} \epsilon_{H})} \stackrel{?}{\geq} 0 .
\eeq
Eq.~\eq{eq4} is now a special case of the following general result
\begin{lemma} \label{trick2}
For any $0\leq \alpha_1,\alpha_2< 1$, it holds
\beq \label{prov2}
\sum_{\mc{H}\subseteq \mc{S}} (-1)^{\abs{\mc{H}}}\, \frac{\prod_{i\in \mc{H}} 
  \epsilon^2_i} {\left(1-\alpha_1 \prod_{i\in \mc{H}} \epsilon_i\right) 
  \left(1-\alpha_2 \prod_{i\in \mc{H}} \epsilon_i \right)} \geq 0 .
\eeq
\end{lemma}

\begin{IEEEproof}
Using the geometric series $\sum_{l=0}^\infty z^l=1/(1-z)$, for all $0\leq z< 1$, and 
setting $z=\alpha_1 \prod_{i\in \mc{H}} \epsilon_i$ and $z=\alpha_2 \prod_{i\in \mc{H}}
\epsilon_i$, yields
\beq \begin{split} \label{trick3}
\frac{\prod_{i\in \mc{H}} \epsilon^2_i} {\left(1-\alpha_1 \prod_{i\in \mc{H}}
  \epsilon_i \right) \left( 1-\alpha_2 \prod_{i\in \mc{H}} \epsilon_i \right)} &=  
  \sum_{l=0}^{\infty} \sum_{k=0}^{\infty} \left( \alpha_1 \prod_{i\in \mc{H}}
  \epsilon_i \right)^l \left(\alpha_2 \prod_{i\in \mc{H}} \epsilon_i\right)^k 
  \left(\prod_{i\in \mc{H}} \epsilon_i\right)^2 \\
&= \sum_{l=0}^{\infty} \sum_{k=0}^{\infty} \alpha^l_1 \alpha^k_2 \prod_{i\in \mc{H}} 
   \epsilon_{i}^{l+k+2} .
\end{split} \eeq
Multiplying \eq{trick3} with $(-1)^{\abs{\mc{H}}}$, summing over all $\mc{H}\subseteq 
\mc{S}$ and using the identity $\prod_{i\in\mc{S}} (1-x_i)= \sum_{\mc{H}\subseteq 
\mc{S}} (-1)^{\abs{\mc{H}}} \prod_{i\in \mc{H}} x_i$ (which is easily proved 
by induction on $\abs{\mc{S}}$) now produces
\beq
\mbox{LHS of \eq{prov2}}= \sum_{l=0}^\infty \sum_{k=0}^\infty \alpha^l_1 \alpha^k_2
  \sum_{\mc{H}\subseteq\mc{S}} (-1)^{\abs{\mc{H}}} \prod_{i\in\mc{H}} \epsilon^{l+k+2}_i
  =\sum_{l=0}^\infty \sum_{k=0}^\infty \alpha^l_1 \alpha^k_2 \prod_{i\in \mc{S}} 
  (1-\epsilon^{l+k+2}_i) \geq 0 ,
\eeq
which is the desired result.
\end{IEEEproof}

\section{Correctness of $\mbox{\texttt{CODE2}}_{pub}$} \label{app6}

The following result is a close analogue to Lemma~\ref{cruc}.
\begin{lemma} \label{modp}
Consider a slot $t$ in subphase 2.1 of $\mbox{\texttt{CODE2}}_{pub}$,
when queues $Q_{\{i^\ast,j\}} \in \mc{Q}_{\dot{S}u}$ and $Q_{\{1,2,3\}}$ are
combined, and a packet $s=\sum_{p\in Q_{\{i^\ast,j\}} \cup
  Q_{\{1,2,3\}}} a_s(p) p$ is transmitted. Assume that at the
beginning of the slot (i.e.~before any packet transmission), there
exist sets $\mc{B}^{(l)}_{\mc{I}}(t)\subseteq \{ \vec{b}^{(l)}_p: p\in
Q_{\mc{I}} \}$, for all $\mc{I}\subseteq \mc{N}$ and $l\in \mc{I}$,
and $\mc{B}_{D_l}(t)=\{ \vec{b}^{(l)}_p: p\in Q_{D_l} \}$ such that
$\mc{B}_{D_l}(t) \cup \bigcup_{\substack{\mc{I}: \mc{I} \subseteq
    \mc{N} \\ K^l_{\mc{I}}(t)>0}} \mc{B}^{(l)}_{\mc{I}}$ is a basis of
$\mathbb{F}^{\abs{\mc{K}_l}}_q$ for all $l\in \mc{N}$. Define
$\mc{R}_{\{i^\ast,j\}} (t)\stackrel{\vartriangle}{=} \left\{ l\in
\{i^\ast,j\}: K^l_{\{i^\ast,j\}}(t) >0 \vee K^l_{\{1,2,3\}}(t) >0
\right\}$ and, for each $l\in \mc{R}_{\{i^\ast,j\}}(t)$, pick a vector
$\hat{\vec{b}}_l$ as follows
\beq \nonumber
\hat{\vec{b}}_l= \left\{ \begin{array}{l@{\quad}l} \mbox{arbitrary } \vec{b}^{(l)}_p\in 
  \mc{B}^{(l)}_{\{1,2,3\}}(t) & \mbox{if } \dot{S}u(l)=0, \\ \mbox{arbitrary } \vec{b}^{(l)}_p
  \in \mc{B}^{(l)}_{\{i^\ast,j\}}(t) & \mbox{otherwise}. \end{array} \right. 
\eeq
Then there exist coefficients $a_s(p)$ such that the set $ \{ \vec{b}^{(l)}_s \}
\cup \mc{B}_{D_l(t)} \cup \bigcup_{\substack{\mc{I}: \mc{I}\subseteq \mc{N}\\ 
K^l_{\mc{I}}(t)>0}} \mc{B}^{(l)}_{\mc{I}}(t)$ is a basis of 
$\mathbb{F}^{\abs{\mc{K}_l}}_q$ for all $l\in \mc{R}_{\{i^\ast,j\}}(t)$.
\end{lemma}

\begin{IEEEproof}
The proof is essentially a repetition of the proof of Lemma~\ref{cruc}, the main 
ingredients being the application of Lemma~\ref{linalg} to show that
\beq \nonumber
\Pr\left( \{ \vec{b}^{(l)}_s \} \cup \mc{B}_{D_l}(t) \cup \bigcup_{\substack{
  \mc{I}: \mc{I}\subseteq \mc{N}\\ K^l_{\mc{I}}(t)>0}} \mc{B}^{(l)}_{\mc{I}}(t)
  -\{ \hat{\vec{b}}_l \} \mbox{ is basis of } \mathbb{F}^{\abs{\mc{K}_i}}_q
  \right) \geq 1-\frac{1}{q} ,
\eeq
for all $l\in \mc{R}_{\{i^\ast,j\}}(t)$, and a standard probabilistic argument 
where $a_s(p)$ are selected iid uniformly in $\mathbb{F}_q$.
\end{IEEEproof}

Lemma~\ref{modp} can now be used to show that Lemma~\ref{trueind} is
also true for $\mbox{\texttt{CODE2}}_{pub}$. This is again proved by
induction on each slot $t$. In fact, since
$\mbox{\texttt{CODE2}}_{pub}$ is \textit{identical} to
$\mbox{\texttt{CODE1}}_{pub}$ up to $t^\ast_2$ (the time where each
level 2 queue has at most one surviving index), it follows that the
inductive hypothesis is true for all slots $t\leq t^\ast_2$, so we
only need to apply induction for $t> t^\ast_2$. Due to the queue
mixing in subphase 2.1, the proof of Lemma~\ref{trueind} must be
modified as follows.

\begin{IEEEproof}[Proof of Lemma~\ref{trueind} for $\mbox{\texttt{CODE2}}_{pub}$]
Assume that the inductive hypothesis holds at the beginning of slot
$t>t^\ast_2$ and we are currently combining $Q_{\{i^\ast,j\}}\in
\mc{Q}_{\dot{S}u}$ with $Q_{\{1,2,3\}}$.  We pick the coefficients for
the packet $s$ to be transmitted at slot $t$ according to
Lemma~\ref{modp} and distinguish the following mutually exclusive
cases for each $l\in \mc{R}_{\{i^\ast,j\}}(t)$ (for $l\not\in
\mc{R}_{\{i^\ast,j\}}(t)$, the hypothesis holds for $t+1$ without
changing any $\mc{B}^{(l)}_{\mc{I}}$, i.e.~we simply select
$\mc{B}^{(l)}_{\mc{I}}(t+1)=\mc{B}^{(l)}_{\mc{I}}(t)$)
\begin{itemize}
\item if $l$ receives $s$ and it holds $\dot{S}u(l)=0$,
  \texttt{ACTFB2} requires that $K^l_{\{1,2,3\}}$ is decreased by 1
  and $K_{D_l}$ is increased by 1. We set
  $\mc{B}^{(l)}_{\{1,2,3\}}(t+1)= \mc{B}^{(l)}_{\{1,2,3\}}(t)- \{
  \hat{\vec{b}}_l \}$ and $\mc{B}_{D_l}(t+1)= \mc{B}_{D_l}(t)\cup \{
  \vec{b}^{(l)}_s \}$ while all other sets remain unchanged.
  Lemma~\ref{modp} implies that the new sets form a basis of
  $\mathbb{F}^ {\abs{\mc{K}_l}}_q$ at slot $t+1$.

\item if $l$ receives $s$ and it holds $\dot{S}u(l)>0$, then,
  according to \texttt{ACTFB2}, $K^l_{\{i^\ast,j\}}$ is decreased by
  one and $K_{D_l}$ increased by 1. The hypothesis still holds for
  user $l$ and slot $t+1$ by setting
  $\mc{B}_{D_l}(t+1)=\mc{B}_{D_l}(t)\cup \{ \vec{b}^{(l)}_s \}$ and
  $\mc{B}^{(l)}_{\{i^\ast,j\}}(t+1)= \mc{B}^{(l)}_{\{i^\ast,j\}}(t)-
  \{ \hat{\vec{b}}_l \}$, while all other sets remain unchanged.

\item if $l$ erases $s$ and $k\in \{1,2,3\}-\{i^\ast,j\}$ receives it,
  $\mbox{\texttt{CODE2}}_{pub}$ requires $K^l_{\{i^\ast,j\}}$ to be
  decreased by 1 and $K^l_{\{1,2,3\}}$ increased by one. The inductive
  hypothesis at $t+1$ is still true by setting
  $\mc{B}^{(l)}_{\{i^\ast,j\}}(t+1)= \mc{B}^{(l)}_{\{i^\ast,j\}}(t)-
  \{ \vec{b}^{(l)}_s \}$ and
  $\mc{B}^{(l)}_{\{1,2,3\}}(t+1)=\mc{B}^{(l)}_{\{1,2,3\}}(t) \cup \{
  \vec{b}^{(l)}_s \}$.

\item in all other cases, no $K^l_{\mc{I}}$, $K_{D_l}$ indices change,
  so that sets $\mc{B}^{(l)}_{\mc{I}}$, $\mc{B}_{D_l}$ remain the same
  as in slot $t$, and the hypothesis is trivially true at slot $t+1$.
\end{itemize}
Since the above list contains all possible cases, the inductive hypothesis always 
holds for all $l\in \mc{N}$ in slot $t+1$ and the proof is complete.
\end{IEEEproof}

\section{Proof of Theorem~\ref{opt3}} \label{app7} 

Consider a vector $\vec{R}$ and assume without loss of generality that
$\vec{R}> \vec{0}$.  As in the analysis of
$\mbox{\texttt{CODE1}}_{pub}$, we consider a modified version with a
fixed blocklength $n$ where the transmitter creates sets of packets
$\mc{K}_i$ with $\abs{\mc{K}_i}= K_i(\vec{R})= \lceil n R_i \rceil$,
for $i\in \{1,2,3\}$, and transmits $n$ symbols. An error is declared
if $\mbox{\texttt{CODE2}}_{pub}$ has not terminated by the $n$-th
transmission. The proof is similar to that of
Theorem~\ref{theo:genthru}, in the sense that the total number of
slots $\dot{T}^\ast$ required by $\mbox{\texttt{CODE2}}_{pub}$ is
computed as a random variable and it is seen that $\dot{T}^\ast/n$
tends to a deterministic quantity $\bar{T}^\ast(\vec{R})$ w.p.~1 as
$n\to \infty$, so that the achievable region of
$\mbox{\texttt{CODE2}}_{pub}$ is $\{ \vec{R}:
\bar{T}^\ast(\vec{R})\leq 1\}$. Having found an exact expression for
$\bar{T}^\ast(\vec{R})$, simple algebra reveals the latter region to
be identical to the outer bound of Lemma~\ref{lem:outer}.

We denote $\mc{N}=\{1,2,3\}$ while $\dot{T}^\ast_{\mc{S}}$ is the (random) number of time 
slots it takes $\mbox{\texttt{CODE2}}_{pub}$ to process queue $Q_{\mc{S}}$, so that 
$\dot{T}^\ast= \sum_{\varnothing \neq \mc{S}\subseteq \mc{N}} \dot{T}^\ast_{\mc{S}}$.
Since $\mbox{\texttt{CODE2}}_{pub}$ is identical to $\mbox{\texttt{CODE1}}_{pub}$ until 
the end of phase 2 (i.e.~when each level 2 queue has at most one non-zero $K$ index), 
we conclude that all level 1 queues are processed identically to 
$\mbox{\texttt{CODE1}}_{pub}$, so that Corollary~\ref{limits} implies, through the
appropriate substitutions
\beq
\lim_{n\to \infty} \sum_{\mc{S}: \abs{\mc{S}}=1} \frac{\dot{T}^\ast_{\mc{S}}}{n}=
   \sum_{i\in \mc{N}} \hat{f}^i_{\{i\}} R_i= \frac{R_1+R_2+R_3}{1-\epsi_{\mc{N}}} \quad a.e.
\eeq

We now make the following crucial observation regarding the decision
taken by $\mbox{\texttt{CODE2}}_{pub}$ at the end of phase 2 (denoted
as $t^\ast_2$). Depending on the exact values of $\dot{S}u(i)$, the
following cases exist:
\begin{itemize}
\item if $\dot{S}u(i)=0$ for all $i\in \mc{N}$, or $\dot{S}u(i)=1$ for
  all $i\in \mc{N}$, $\mbox{\texttt{CODE2}}_{pub}$ continues mimicking
  $\mbox{\texttt{CODE1}}_{pub}$ until the end of the algorithm. In
  this case, the asymptotic behavior of $\dot{T}^\ast_{\mc{S}}$ is
  obviously still governed by Corollary~\ref{limits}.

\item otherwise, $\mbox{\texttt{CODE2}}_{pub}$ deviates from
  $\mbox{\texttt{CODE1}}_{pub}$ by further processing each level 2
  queue $Q_{\mc{S}}$ in subphase 2.1 mentioned in
  Section~\ref{user3}. An inspection of the \texttt{ACTFB2} procedure
  indicates that, during the combining of a level 2 queue
  $Q_{\mc{S}}$ with $Q_{\mc{N}}$, the actions regarding indices
  $K^i_{\mc{S}}$ are \textit{identical} to \texttt{ACTFB1} (in fact,
  the only difference between \texttt{ACTFB1} and \texttt{ACTFB2} lies
  in the handling of indices $K^i_{\mc{N}}$). Since each level 2 queue
  is still processed until all its $K$ indices become zero, we
  conclude that, if we denote with $T^\ast_{\mc{S}}$ the
  \textit{total} number of slots required for the processing of
  $Q_{\mc{S}}$ during phase 2 \textit{and} subphase 2.1,
  Corollary~\ref{limits} still holds. However, the value of
  $\dot{K}^i_{\mc{N}}$ at the beginning of phase 3 will be different
  than the corresponding value under $\mbox{\texttt{CODE1}}_{pub}$ due
  to the interjection of subphase 2.1.
\end{itemize}

Denote with $\tilde{t}_3$ the beginning of phase 3, equivalently the end of phase 2 or subphase 
2.1 (if the latter occurred). Since $\mbox{\texttt{CODE2}}_{pub}$ again mimics 
$\mbox{\texttt{CODE1}}_{pub}$ during phase 3, Corollary~\ref{limits} implies, under the
obvious substitutions, that
\beq \label{gather1}
\lim_{n\to \infty} \frac{\dot{T}^\ast_{\mc{N}}}{n}= \max_{i\in \mc{N}} \left[ \lim_{n\to \infty}
  \left( \frac{\dot{K}^i_{\mc{N}}(\tilde{t}_3)}{n} \right)\, \frac{1}{1-\epsi_i} \right] \quad a.e. ,
\eeq
provided that the rightmost limit exists w.p.~1 (this will be shown later). It then follows that
\beq \label{gather2}
\lim_{n\to \infty} \frac{\dot{T}^\ast}{n}= \sum_{\mc{S}: \abs{\mc{S}}\leq 2} \max_{l\in \mc{S}} 
  (\hat{f}^l_{\mc{S}} R_l) + \lim_{n\to \infty} \frac{\dot{T}^\ast_{\mc{N}}}{n} \quad a.e. ,
\eeq
so that we hereafter concentrate on the computation of the last limit, which clearly depends
on the specific decision at $t^\ast_2$.

Denote with $\dot{T}^\ast_{i,\mc{S}}$ the number of slots it takes $\mbox{\texttt{CODE1}}_{pub}$ 
(or $\mbox{\texttt{CODE2}}_{pub}$, if we consider both phase 2 and subphase 2.1)
to process a level 2 queue $Q_{\mc{S}}$ until $K^i_{\mc{S}}$ becomes 0. It clearly holds
$\dot{T}^\ast_{\mc{S}}= \max_{i\in \mc{S}} \dot{T}^\ast_{i,\mc{S}}$; if we also define
$\dot{T}^\dag_{\mc{S}}= \min_{i\in \mc{S}} \dot{T}^\ast_{i,\mc{S}}$, we can combine 
Lemma~\ref{just_recur} and Corollary~\ref{limits} to deduce
\beq \label{TT2}
\bar{T}^\dag_{\mc{S}}\stackrel{\vartriangle}{=} \lim_{n\to \infty} \frac{\dot{T}^\dag_{\mc{S}}}{n} 
  = \min_{i\in \mc{S}} \; (\hat{f}^i_{\mc{S}} R_i)\quad a.e. ,
\eeq
in addition to
\beq \begin{split} \label{lims2}
& \bar{T}^\ast_{i,\mc{S}} \stackrel{\vartriangle}{=} \lim_{n\to \infty} \frac{\dot{T}^\ast_{i,\mc{S}}}{n}
  = \hat{f}^i_{\mc{S}} R_i \\
& \bar{T}^\ast_{\mc{S}}\stackrel{\vartriangle}{=} \lim_{n\to \infty} \frac{\dot{T}^\ast_{\mc{S}}}{n} 
  = \max_{i\in \mc{S}} \; (\hat{f}^i_{\mc{S}} R_i) \quad a.e., 
\end{split} \eeq
which we already used in \eq{gather2}.

We next find an expression for $\dot{K}^i_{\mc{S}}(t^\ast_2)$, for all $\mc{S}$ with 
$\abs{\mc{S}}=2$, since this will affect the branching decision made by 
$\mbox{\texttt{CODE2}}_{pub}$ at $t^\ast_2$. The following relation is true for all 
$\mc{S}$ with $\abs{\mc{S}}=2$ and describes the total decrease of each $K$ index of 
a level 2 queue in the interval $[t^\ast_1 \;\; t^\ast_2]$.
\beq \label{dec12}
\dot{K}^i_{\mc{S}}(t^\ast_2)= \dot{K}^i_{\mc{S}}(t^\ast_1)- \sum_{l=1}^{\dot{T}^\dag_{\mc{S}}} 
  \mathbb{I}[K^i_{\mc{S}} \mbox{ reduced by 1 during } l\mbox{-th slot of processing } Q_{\mc{S}} 
  \mbox{ in phase 2}] .
\eeq
Dividing by $n$ and using \eq{TT2} we conclude that
\beq \begin{split} \label{llim}
k^i_{2,\mc{S}} & \stackrel{\vartriangle}{=} \lim_{n\to \infty} 
  \frac{\dot{K}^i_{\mc{S}}(t^\ast_2)}{n}= \left( \lim_{n\to \infty} 
  \frac{\dot{K}^i_{\mc{S}}(t^\ast_1)}{n} \right) - \left( \lim_{n\to \infty} 
  \frac{\dot{T}^\dag_{\mc{S}}}{n} \right) \Pr(K^i_{\mc{S}} \mbox{ reduced by 1 during proc. } Q_{\mc{S}}) \\
&= \hat{f}^i_{\mc{S}} R_i (1-\epsi_{\mc{N}-(\mc{S}-\{i\})}) - \min_{l\in \mc{S}} \; 
  (\hat{f}^l_{\mc{S}} R_l) (1-\epsi_{\mc{N}-(\mc{S}-\{i\})}) \quad a.e. \; \forall\, 
  \mc{S}: \abs{\mc{S}}=2 ,  
\end{split} \eeq
where we used Lemma~\ref{just_recur} (which is still applicable at $t^\ast_1$) for the 
asymptotic behavior of $\dot{K}^i_{\mc{S}}(t^\ast_1)/n$. The subscript $2$ emphasizes that 
the quantity refers to a limit of a random variable at $t^\ast_2$.

For $\mc{S}=\{i,j\}$, \eq{llim} can be written as
\beq \label{llim2}
\lim_{n\to \infty} \frac{\dot{K}^i_{\{i,j\}}(t^\ast_2)}{n}= k^i_{2,\{i,j\}}= \left[ 
  \hat{f}^i_{\{i,j\}} R_i - \hat{f}^j_{\{i,j\}} R_j \right]^+ 
  (1-\epsi_{\mc{N}-\{j\}}) \quad a.e. ,
\eeq
which motivates us to define
\beq \label{dom}
r^i_{\{i,j\}}(\vec{R})= \left[ \hat{f}^i_{\{i,j\}} R_i - \hat{f}^j_{\{i,j\}} R_j \right]^+ ,
\eeq
where $[x]^+\stackrel{\vartriangle}{=}\max(x,0)$ and we explicitly
state the $\vec{R}$ dependence of $r^i_{\{i,j\}}$. The binary relation $i\succ j$ is 
introduced to denote the inequality $\hat{f}^i_{\{i,j\}} R_i > \hat{f}^j_{\{i,j\}} R_j$ 
(equivalently, $r^i_{\{i,j\}}>0$) which, using the definition of $\hat{f}^i_{\mc{S}}$, 
can be expanded to
\beq \label{condi2}
\hat{f}^i_{\{i,j\}} R_i > \hat{f}^j_{\{i,j\}} R_j \Leftrightarrow R_i \left( \frac{1}{1-\epsi_{\mc{N}-\{j\}}} 
  -\frac{1}{1-\epsi_{\mc{N}}} \right) > R_j \left( \frac{1}{1-\epsi_{\mc{N}-\{i\}}}-
  \frac{1}{1-\epsi_{\mc{N}}} \right) .
\eeq
We also write $i\succeq j$ iff $\hat{f}^i_{\{i,j\}} R_i \geq
\hat{f}^j_{\{i,j\}} R_j$ and $i \asymp j$ if $\hat{f}^i_{\{i,j\}} R_i=
\hat{f}^j_{\{i,j\}} R_j$ (note that all relations $\succ$, $\succeq$,
$\asymp$ implicitly depend on $\vec{R}$), whence the following result
follows.
\begin{lemma} \label{order}
Consider any $\vec{R}> \vec{0}$ and distinct $i,j,k\in \mc{N}$. If
$i\succ j$ and $j\succeq k$, then $i\succ k$. Similarly, if $i\succeq
j$ and $j\succ k$, then $i\succ k$.
\end{lemma}

\begin{IEEEproof}
We prove by contradiction only the first part since the second one
follows similarly. We assume that $k\succeq i$, so that it holds
\beq \begin{split} \label{useful}
& \hat{f}^i_{\{i,j\}} R_i > \hat{f}^j_{\{i,j\}} R_j \Leftrightarrow R_i \left( 
  \frac{1}{1-\epsi_{\mc{N}-\{j\}}}- \frac{1}{1-\epsi_{\mc{N}}} \right) > R_j \left(
  \frac{1}{1-\epsi_{\mc{N}-\{i\}}}- \frac{1}{1-\epsi_{\mc{N}}} \right) , \\
& \hat{f}^j_{\{j,k\}} R_j \geq \hat{f}^k_{\{j,k\}} R_k \Leftrightarrow R_j \left(
  \frac{1}{1-\epsi_{\mc{N}-\{k\}}} -\frac{1}{1-\epsi_{\mc{N}}} \right) \geq R_k \left(
  \frac{1}{1-\epsi_{\mc{N}-\{j\}}} -\frac{1}{1-\epsi_{\mc{N}}} \right) , \\
& \hat{f}^k_{\{i,k\}} R_k \geq \hat{f}^i_{\{i,k\}} R_i \Leftrightarrow R_k \left(
  \frac{1}{1-\epsi_{\mc{N}-\{i\}}}- \frac{1}{1-\epsi_{\mc{N}}} \right) \geq R_i \left(
  \frac{1}{1-\epsi_{\mc{N}-\{k\}}}- \frac{1}{1-\epsi_{\mc{N}}} \right) .
\end{split} \eeq
The terms in parentheses above are non-negative by construction. In
fact, the term
$\frac{1}{1-\epsi_{\mc{N}-\{j\}}}-\frac{1}{1-\epsi_{\mc{N}}}$ is
positive, since otherwise we would conclude that $0$ is strictly
larger than a non-negative number. We can then use a similar reasoning
and the fact that $\vec{R}>\vec{0}$ to show that all terms in
parentheses are positive. Hence, we can multiply the 3 equations by
sides and arrive at a contradiction that a number is strictly larger
than itself.
\end{IEEEproof}

Using the notation of \eq{indic}, we can find the value of $K^i_{\mc{N}}$ at
$t^\ast_2$ as
\beq
\dot{K}^i_{\mc{N}}(t^\ast_2)= \sum_{l=1}^{\lceil n R_i \rceil} \mathbb{I}[
   \dot{D}^i_{\{i\},l}= \mc{N}] +\sum_{\substack{\mc{S}:i\in \mc{S}\\ \abs{\mc{S}}=2}}
   \sum_{l=1}^{\dot{T}^\dag_{\mc{S}}} \mathbb{I}[\dot{D}^i_{\mc{S},l}= \mc{N}] ,
\eeq
where the first, second term is the number of tokens moved during phase 1, 2, 
respectively. Using a procedure similar to Lemma~\ref{just_recur}, we can find
\beq \begin{split} \label{endp2}
k^i_{2,\mc{N}} & \stackrel{\vartriangle}{=} \lim_{n\to \infty} \frac{\dot{K}^i_{\mc{N}}(t^\ast_2)}{n}= 
  R_i \Pr(\dot{D}^i_{\{i\}}=\mc{N}) + \sum_{\substack{\mc{S}:i\in \mc{S} \\
  \abs{\mc{S}}=2}} \left( \lim_{n\to \infty} \frac{\dot{T}^\dag_{\mc{S}}}{n} \right) 
  \Pr(\dot{D}^i_{\mc{S}}=\mc{N}) \\
&= \hat{f}^i_{\{i\}} R_i\, p_{\{i\},\mc{N}-\{i\}} + \sum_{\substack{\mc{S}:i\in \mc{S} \\ \abs{\mc{S}}=2}} 
  \min_{l\in \mc{S}} \; (\hat{f}^l_{\mc{S}} R_l) \, p_{i,\mc{N}-\mc{S}} .
\end{split} \eeq
Any variation of $K^i_{\mc{N}}$ between $t^\ast_2$ (end of phase 2)
and $\tilde{t}_3$ (beginning of phase 3) under
$\mbox{\texttt{CODE2}}_{pub}$ can only be due to subphase 2.1 or the
continuation of processing level 2 queues if $\dot{S}u(l)=1$ for all
$l\in \mc{N}$. Hence we conclude:
\beq \label{cases3}
\dot{K}^i_{\mc{N}}(\tilde{t}_3)= \left\{ \begin{array}{l@{\quad}l} \dot{K}^i_{\mc{N}}
  (t^\ast_2) & \mbox{if } \dot{S}u(l)=0 \;\;\forall\, l\in \mc{N} , \\
  \dot{K}^i_{\mc{N}}(t^\ast_2)+ \sum_{\substack{\mc{S}:i\in \mc{S}\\ \abs{\mc{S}}=2}}
  \sum_{l=1}^{\dot{T}^\ast_{i,\mc{S}}- \dot{T}^\dag_{\mc{S}}} 
  \mathbb{I}[\dot{D}^i_{\mc{S},l}=\mc{N}] & \mbox{if } \dot{S}u(l)=1\;\; \forall\, 
  i\in \mc{N} , \\
  \left[ \dot{K}^i_{\mc{N}}(t^\ast_2) + \sum_{\substack{\mc{S}:i\in \mc{S}\\ \abs{\mc{S}}=2}}
  \sum_{l=1}^{\dot{T}^\ast_{i,\mc{S}}-\dot{T}^\dag_{\mc{S}}} \mathbb{I}[\dot{D}^+_{\mc{S},l}] 
  - \sum_{\substack{\mc{S}:i\in \mc{S}\\ \abs{\mc{S}}=2}}
  \sum_{l=1}^{\dot{T}^\ast_{\mc{S}}-\dot{T}^\ast_{i,\mc{S}}} \mathbb{I}[\dot{D}^-_{\mc{S},l}] 
  \right]^+ & \mbox{otherwise}, \end{array} \right.
\eeq
where $\mathbb{I}[\dot{D}^+_{\mc{S},l}]\stackrel{\vartriangle}{=}
\mathbb{I}[K^i_{\mc{N}} \mbox{ increased during } l\mbox{-th slot of
    processing } Q_{\mc{S}} \mbox{ in subphase 2.1}]$ with a similar
definition for $\mathbb{I}[\dot{D}^-_{\mc{S},l}]$ (replacing increased
with decreased).

At this point, it is convenient to consider the following two
complementary cases and individually examine each of them.
\begin{itemize}
\item it holds $r^l_{\mc{S}}=0$ for all $\mc{S}$ with $\abs{\mc{S}}=2$ and 
$l\in \mc{S}$. Equivalently, it holds $i\asymp j \asymp k$.

\item it holds $r^l_{\mc{S}}>0$ for at least one $l\in \mc{S}$ with $\abs{\mc{S}}=2$.
\end{itemize}

\subsubsection{The case $i\asymp j\asymp k$} Equations \eq{TT2}, \eq{lims2} 
imply that 
\beq
\lim_{n\to \infty} \frac{\dot{T}^\ast_{\mc{S}}-\dot{T}^\ast_{i,\mc{S}}}{n}= 
\lim_{n\to \infty} \frac{\dot{T}^\ast_{i,\mc{S}}-\dot{T}^\dag_{\mc{S}}}{n}=0 \quad a.e. ,
\eeq
so that, examining all 3 cases in \eq{cases3}, we conclude that
\beq
\lim_{n\to \infty} \frac{\dot{K}^i_{\mc{N}}(\tilde{t}_3)}{n}= \lim_{n\to \infty} 
  \frac{\dot{K}^i_{\mc{N}}(t^\ast_2)}{n} \quad a.e. ,
\eeq
which implies, through \eq{gather1}, \eq{gather2}, that
$\mbox{\texttt{CODE1}}_{pub}$ and $\mbox{\texttt{CODE2}}_{pub}$ have
the same asymptotic performance (meaning that
$\bar{T}^\ast(\vec{R})=\lim_{n\to \infty} \dot{T}^\ast/n$ is the same
function under both algorithms) for all $\vec{R}$ such that $i\asymp
j\asymp k$. Hence, defining the set
$\tilde{\mc{R}}\stackrel{\vartriangle}{=} \{ \vec{R}\geq \vec{0}:
i\asymp j \asymp k\}$, we conclude $\mc{R}_{\mbox{\scriptsize
    \texttt{CODE1}}_{pub}} \cap \tilde{\mc{R}}=
\mc{R}_{\mbox{\scriptsize \texttt{CODE2}}_{pub}} \cap
\tilde{\mc{R}}$. Furthermore, it holds $\tilde{\mc{R}} \subseteq
\mc{R}_{ord}$, where $\mc{R}_{ord}$ was defined in \eq{deford}, so
that
\beq
\mc{C}^{out} \cap \tilde{\mc{R}}= \mc{C}^{out} \cap \mc{R}_{ord} \cap 
  \tilde{\mc{R}}= \mc{R}_{\mbox{\scriptsize \texttt{CODE1}}_{pub}} \cap \mc{R}_{ord} 
  \cap \tilde{\mc{R}} \subseteq \mc{R}_{\mbox{\scriptsize \texttt{CODE2}}_{pub}}
  \cap \tilde{\mc{R}} ,
\eeq
where the last set equality is due to
Theorem~\ref{theo:genopt}. Hence, $\mbox{\texttt{CODE2}}_{pub}$
achieves all rates in $\mc{C}^{out} \cap \tilde{\mc{R}}$.

\subsubsection{The case $r^l_{\mc{S}}>0$ for at least one $l\in \mc{S}$ 
with $\abs{\mc{S}}=2$} Let $\mc{S}=\{i,j\}$ and assume
  $r^i_{\mc{S}}>0$, so that $i\succ j$. Then, two mutually exclusive
  cases exist according to Lemma~\ref{order} (in the following,
  $i,j,j$ are distinct):
\begin{itemize}
\item it holds $k\succeq i$, so that $k\succ j$.

\item it holds $i\succ k$.
\end{itemize}
In the first case, it follows from \eq{llim2} that it holds w.p.~1
\beq \label{intlim}
\lim_{n\to \infty} \frac{\dot{K}^i_{\{i,j\}}(t^\ast_2)}{n} >0,  \quad
  \lim_{n\to \infty} \frac{\dot{K}^j_{\{i,j\}}(t^\ast_2)}{n}=0 , 
\eeq

\beq
\lim_{n\to \infty} \frac{\dot{K}^k_{\{j,k\}}(t^\ast_2)}{n}>0, \quad
  \lim_{n\to \infty} \frac{\dot{K}^j_{\{j,k\}}(t^\ast_2)}{n}=0 ,
\eeq 
so that, but the definition of limit, there exists some $n_1$ such
that for all $n> n_1$ it holds $\dot{S}u(i) \geq 1, \; \dot{S}u(k)
\geq 1, \; \dot{S}u(j)=0$. In the second case, \eq{intlim} is still
true and it also holds
\beq
\lim_{n\to \infty} \frac{\dot{K}^i_{\{i,k\}}(t^\ast_2)}{n}>0, \quad
  \lim_{n\to \infty} \frac{\dot{K}^k_{\{i,k\}}(t^\ast_2)}{n}=0 ,
\eeq 
which implies via a similar argument that there exists some $n_2$ such
that $\dot{S}u(i)=2, \; \dot{S}u(j) \leq 1, \; \dot{S}u(k) \leq 1$,
for all $n> n_2$.

Hence, in both cases there exists a sufficiently large
$n_0$ such that for all $n>n_0$, the first two branches in \eq{cases3}
are excluded. Hence, it holds
\beq \begin{split}
\lim_{n\to \infty} \frac{\dot{K}^i_{\mc{N}}(\tilde{t}_3)}{n}= \Bigg[ 
  \lim_{n\to \infty} \frac{\dot{K}^i_{\mc{N}}(t^\ast_2)}{n} & + 
  \sum_{\substack{\mc{S}: i\in \mc{S} \\ \abs{\mc{S}}=2}} \left( \hat{f}^i_{\mc{S}} 
  R_i - \min_{l\in \mc{S}} \; (\hat{f}^l_{\mc{S}} R_l) \right) p_{\{i\},\mc{N}-\mc{S}} \\
& - \sum_{\substack{\mc{S}: i\in \mc{S} \\ \abs{\mc{S}}=2}} \left( \max_{l\in \mc{S}} \;
  (\hat{f}^l_{\mc{S}} R_l)- \hat{f}^i_{\mc{S}} \right) (1-\epsi_i) \Bigg]^+ \quad a.e. ,
\end{split} \eeq
which can also be written as
\beq
\lim_{n\to \infty} \frac{\dot{K}^i_{\mc{N}}(\tilde{t}_3)}{n}= \Bigg[ 
  \lim_{n\to \infty} \frac{\dot{K}^i_{\mc{N}}(t^\ast_2)}{n} + \sum_{\substack{\mc{S}: 
  i\in \mc{S} \\ \abs{\mc{S}}=2}} \mathbb{I}[r^i_{\mc{S}}>0] p_{\{i\},\mc{N}-\mc{S}}
  - \sum_{\substack{\mc{S}: i\in \mc{S} \\ \abs{\mc{S}}=2}} \mathbb{I}[r^i_{\mc{S}}=0]
  (1-\epsi_i) \Bigg]^+ .
\eeq
It is now a matter of case distinction, depending on the values of
$r^i_{\mc{S}}$, and simple algebra to verify that
$\mbox{\texttt{CODE2}}_{pub}$ also achieves all rates in $\mc{C}^{out}
\cap \tilde{\mc{R}}^c$, so that it achieves $\mc{C}^{out}$.

\bibliographystyle{IEEEtran}
\bibliography{IEEEabrv,erasure}

\end{document}

%% file: example.tex
We next provide a concrete example of execution for
$\mbox{\texttt{CODE1}}_{pub}$ that illustrates some of the points
mentioned in Sections~\ref{algdec}, \ref{prop_code1}. We consider the
case of 3 users with 10 packets destined to each of them and stored at
the transmitter. We denote the sets of packets destined for user 1, 2,
3 as $\mc{K}_1=\{u_1,\ldots,u_{10}\}$,
$\mc{K}_2=\{v_1,\ldots,v_{10}\}$, $\mc{K}_3=\{w_1,\ldots,w_{10}\}$,
respectively. We also introduce an upper index notation to denote the
set of users that have received a packet, e.g.~$u^{(23)}_1$ denotes
that packet $u_1$ was received by users 2, 3 only.

The initialization of $\mbox{\texttt{CODE1}}_{pub}$ is trivial: all
packets of set $\mc{K}_i$ are placed in queue $Q_{\{i\}}$, the indices
are initialized as $T^1_{\{1\}}(0)=T^2_{\{2\}}(0)=T^3_{\{3\}}(0)=10$
(all other indices are zero) and the basis sets are initialized as
$\mc{B}^{(1)}_{\{1\}}(0)=\mc{B}^{(2)}_{\{2\}}(0)
=\mc{B}^{(3)}_{\{3\}}(0)= standard\_basis(\mathbb{F}^{10})$ (all other
sets are empty). We denote with $\vec{e}_i$ the standard basis vector
which has its $i$-th component set to 1.

$\mbox{\texttt{CODE1}}_{pub}$ executes Phase 1, in which the queues
$Q_{\{1\}}$, $Q_{\{2\}}$, $Q_{\{3\}}$ are sequentially processed in
this order. The random erasure events that occur in each slot are
shown in Table~\ref{phase1:er}, where R/E stands for Received/Erased,
respectively, and X denotes an unimportant value (i.e. X can be either
R or E but, in any case, does not affect the algorithm's actions). For
example, the ERE for slot 2 of $Q_{\{1\}}$ denotes a transmission that
was received only by user 2. We also use the following conventions in
Table~\ref{phase1:er}:
\begin{itemize}
\item for simplicity, we omit any slots in which the packet must be
  retransmitted due to steps \ref{retrans}, \ref{nonewtok} of
  \texttt{ACTFB1}. Hence, the slot number (1,2, etc) should not be
  interpreted as physical time but rather as an ordinal indicating
  slots in which no retransmission was required. In other words, slots
  1, 2 need not be contiguous in time.

\item due to the imposed order of processing, queues $Q_{\{2\}}$,
  $Q_{\{3\}}$ are actually processed in slots 11--20 and 21--30,
  respectively. The reader should interpret the rows corresponding to
  $Q_{\{2\}}$, $Q_{\{3\}}$ accordingly.
\end{itemize}

\begin{table}[htbp]
\centering
\caption{$\mbox{\texttt{CODE1}}_{pub}$ execution. Erasures and queue contents at end of phase 1.}
\label{phase1:er}
\begin{tabular}{|c|c|c|c|c|c|c|c|c|c|c|} \hline
\multicolumn{11}{|c|}{Phase 1 execution} \\ \hline
Slot & 1 & 2 & 3 & 4 & 5 & 6 & 7 & 8 & 9 & 10 \\ \hline
$Q_{1}$ & RXX & ERE & ERR & ERE & EER & ERR & RXX & EER & RXX & ERR \\ \hline
$Q_{2}$ & REE & EER & XRX & REE & EER & XRX & RER & RER & XRX & EER \\ \hline
$Q_{3}$ & ERE & REE & REE & XXR & ERE & XXR & ERE & ERE & REE & RRE \\ \hline
\multicolumn{11}{|c|}{Queue status at end of phase 1} \\ \hline
Packets decoded by users & \multicolumn{10}{|c|}{user 1: $u^{(1)}_1$, $u^{(1)}_7$, $u^{(1)}_9$, \hspace{0.25cm}
user 2: $v^{(2)}_3$, $v^{(2)}_6$, $v^{(2)}_9$, \hspace{0.25cm} user 3: $w^{(3)}_4$, $w^{(3)}_6$} \\ \hline
$Q_{\{1,2\}}$ contents & \multicolumn{10}{|c|}{$u^{(2)}_2$, $u^{(2)}_4$, $v^{(1)}_1$, $v^{(1)}_4$}
 \\ \hline
$Q_{\{1,3\}}$ contents & \multicolumn{10}{|c|}{$u^{(3)}_5$, $u^{(3)}_8$, $w^{(1)}_2$, $w^{(1)}_3$, $w^{(1)}_9$}
 \\ \hline
$Q_{\{2,3\}}$ contents & \multicolumn{10}{|c|}{$v^{(3)}_2$, $v^{(3)}_5$, $v^{(3)}_{10}$, $w^{(2)}_1$, 
$w^{(2)}_5$, $w^{(2)}_7$, $w^{(2)}_8$} \\ \hline
$Q_{\{1,2,3\}}$ & \multicolumn{10}{|c|}{$u^{(23)}_3$, $u^{(23)}_6$, $u^{(23)}_{10}$, $v^{(13)}_7$, 
$v^{(13)}_8$, $w^{(12)}_{10}$} \\ \hline
Basis sets at end of phase 1 & \multicolumn{10}{|c|}{$\mc{B}^{(1)}_{\{1,2\}}=\{ \vec{e}_2,\vec{e}_4 \}$, 
$\mc{B}^{(2)}_{\{1,2\}}=\{ \vec{e}_1,\vec{e}_4 \}$, $\mc{B}^{(1)}_{\{1,3\}}= \{ \vec{e}_5,\vec{e}_8 \}$,
$\mc{B}^{(3)}_{\{1,3\}}= \{ \vec{e}_2,\vec{e}_3,\vec{e}_9 \}$} \\ 
 & \multicolumn{10}{|c|}{$\mc{B}^{(2)}_{\{2,3\}}= \{ \vec{e}_2, \vec{e}_5, \vec{e_{10}} \}$, 
$\mc{B}^{(3)}_{\{2,3\}}= \{ \vec{e}_1,\vec{e}_5, \vec{e}_7, \vec{e}_8 \}$, $\mc{B}^{(1)}_{\{1,2,3\}}=\{ \vec{e}_3,
 \vec{e}_6, \vec{e}_{10} \}$} \\ 
 & \multicolumn{10}{|c|}{$\mc{B}^{(2)}_{\{1,2,3\}}= \{ \vec{e}_7, \vec{e}_8 \}$, $\mc{B}^{(3)}_{\{1,2,3\}}=
\{ \vec{e}_{10} \}$} \\ 
 & \multicolumn{10}{|c|}{$\mc{B}_{D_1}= \{ \vec{e}_1,\vec{e}_7,\vec{e}_9 \}$, $\mc{B}_{D_2}= \{ \vec{e}_3, 
\vec{e}_6, \vec{e}_9 \}$, $\mc{B}_{D_3}= \{ \vec{e}_4, \vec{e}_6 \}$} \\ \hline
\end{tabular}
\end{table}

The transmitter starts processing $Q_{\{1\}}$ and sends the uncoded
packet $u_j$ in slot $j$. Similarly, when queues $Q_{\{2\}}$,
$Q_{\{3\}}$ are processed, packet $v_j$, $w_j$ is transmitted,
respectively, in slot $j$. This packet selection policy complies with
Criterion~\ref{minass2}. Specifically, if, at slot $j$, the packet
$u_j$ is received by user 1, then its corresponding $\vec{b}^{(1)}_j$
vector (i.e.~$\vec{e}_j$) is removed from set $\mc{B}^{(1)}_{\{1\}}$
and added to $\mc{B}_{D_1}$. If $u_j$ is erased by user 1 and received
by all users in set $\mc{S}$, then vector $\vec{e}_j$ is moved from
$\mc{B}^{(1)}_{\{1\}}$ to $\mc{B}^{(1)}_{\mc{S}}$. Similar actions are
taken for packets $v_j$, $w_j$.

The queue contents at the end of phase 1 are also shown in
Table~\ref{phase1:er}. Some packets have already been decoded by their
respective destinations, while the rest have been distributed among
the virtual queues. The indices at the end of phase 1 are as follows:
$T^1_{\{1,2\}}=T^2_{\{1,2\}}=2$, $T^1_{\{1,3\}}=2$, $T^3_{\{1,3\}}=3$,
$T^2_{\{2,3\}}=3$, $T^3_{\{2,3\}}=4$, $T^1_{\{1,2,3\}}=3$,
$T^2_{\{1,2,3\}}=2$ and $T^3_{\{1,2,3\}}=1$. For the sets
$\mc{B}^{(i)}_{\mc{S}}$, the policy of sending uncoded packets in
phase 1, combined with \texttt{ACTFB1}, implies that
$\mc{B}^{(i)}_{\mc{S}}(t_1)$ (where $t_1$ denotes the end of phase 1)
contains the unit basis vectors corresponding to the packets stored in
$Q_{\mc{S}}$ at the end of the phase.


The algorithm now executes phase 2, in which the queues $Q_{\{1,2\}}$,
$Q_{\{1,3\}}$, $Q_{\{2,3\}}$ are sequentially processed in this
order. Criterion~\ref{minass2} cannot be satisfied by sending uncoded
packets only, so the transmitter selects a proper linear combination
of all packets in the queue currently being processed. Hence, the
packet $s_n$ transmitted at slot $n$ of phase 2 has the form
$s_n=\sum_{p\in Q_{\mc{S}}} a_{s_n}(p) p$, where $Q_{\mc{S}}$ is the
queue being processed at slot $n$ and $a_s(p)$ satisfy
Criterion~\ref{minass2}. The erasures that occur in phase 2 and the
corresponding \texttt{ACTFB1} actions, as well as their results, are
shown in Table~\ref{phase2:er} (again, the slot number should be
interpreted as ordinal instead of actual time).

\begin{table}[htbp]
\centering
\caption{$\mbox{\texttt{CODE1}}_{pub}$ execution. Erasures and queue contents at end of phase 2.}
\label{phase2:er}
\begin{tabular}{|c|c|c|c|c|c|c|c|c|c|c|} \hline
\multicolumn{11}{|c|}{Phase 2} \\ \hline
Processing & \multicolumn{3}{|c|}{$Q_{\{1,2\}}$} & \multicolumn{3}{|c|}{$Q_{\{1,3\}}$} & 
  \multicolumn{4}{|c|}{$Q_{\{2,3\}}$} \\ \hline
Slot & 1 & 2 & 3 & 4 & 5 & 6 & 7 & 8 & 9 & 10 \\ \hline
Erasure event & REE & ERR & RER & ERE & ERE & ERE & REE & ERR & ERR & EER \\ \hline
Applicable \texttt{ACTFB1} actions & 3 & 3,4 & 3,4 & 4 & 4 & 4 & 4 & 3 & 3 & 3 \\ \hline
  & \multicolumn{3}{|c|}{$T^1_{\{1,2\}}$, $T^2_{\{1,2\}}$} & 
  \multicolumn{3}{|c|}{$T^1_{\{1,3\}}$, $T^3_{\{1,3\}}$} & 
  \multicolumn{4}{|c|}{$T^2_{\{2,3\}}$, $T^3_{\{2,3\}}$} \\ \hline
Index value at end of slot & 1,2 & 0,1 & 0,0 & 1,2 & 0,1 & 0,0 & 2,3 & 1,2 & 0,1 & 0,0 \\ \hline
\multicolumn{11}{|c|}{Queue contents at end of phase 2} \\ \hline
$Q_{\{1,2,3\}}$ & \multicolumn{10}{|c|}{$u^{(23)}_3$, $u^{(23)}_6$, $u^{(23)}_{10}$, $v^{(13)}_7$, $v^{(13)}_8$,
$w^{(12)}_{10}$, $s^{(23)}_2$, $s^{(13)}_3$, $s^{(2)}_4$, $s^{(2)}_5$, $s^{(2)}_6$, $s^{(1)}_7$} \\ \hline
Basis sets for $Q_{\{1,2,3\}}$ & \multicolumn{10}{|c|}{$\mc{B}^{(1)}_{\{1,2,3\}}= \{ \vec{e}_3,\vec{e}_6,\vec{e}_{10},
\vec{b}^{(1)}_{s_2}\in span(\vec{e}_2,\vec{e}_4), \vec{b}^{(1)}_{s_4}, \vec{b}^{(1)}_{s_5} \in span(\vec{e}_5,
\vec{e}_8)$} \\ 
at end of phase 2 & \multicolumn{10}{|c|}{$\mc{B}^{(2)}_{\{1,2,3\}}= \{ \vec{e}_7,\vec{e}_8, 
\vec{b}^{(2)}_{s_3}\in span(\vec{e}_1,\vec{e}_4), \vec{b}^{(2)}_{s_7}\in span(\vec{e}_2,\vec{e}_5,\vec{e}_{10}) \}$} \\
 & \multicolumn{10}{|c|}{$\mc{B}^{(3)}_{\{1,2,3\}}= \{ \vec{e}_{10}, \vec{b}^{(3)}_{s_4}, \vec{b}^{(3)}_{s_5},
\vec{b}^{(3)}_{s_6}\in span(\vec{e}_2,\vec{e}_3,\vec{e}_9), \vec{b}^{(3)}_{s_7}\in span(\vec{e}_1,\vec{e}_5,\vec{e}_7,
\vec{e}_8) \}$} \\ \hline
Packets received by 1 & \multicolumn{10}{|c|}{$u_1$, $u_7$, $u_9$, $s_1$, $s_3$, $s_7$} \\ \hline
$\vec{b}^{(1)}$ received by 1 & \multicolumn{10}{|c|}{$\vec{e}_1$, $\vec{e}_7$, $\vec{e}_9$, $\vec{b}^{(1)}_{s_1}
\in span(\vec{e}_2, \vec{e}_4)$, $\vec{b}^{(1)}_{s_3}\in span(\vec{e}_2,\vec{e}_4)$, $\vec{b}^{(1)}_{s_7}=
\vec{0}$} \\ \hline
Packets received by 2 & \multicolumn{10}{|c|}{$v_3$, $v_6$, $v_9$, $s_2$, $s_4$, $s_5$, $s_6$, $s_8$, $s_9$} \\ \hline 
$\vec{b}^{(2)}$ received by 2 & \multicolumn{10}{|c|}{$\vec{e}_3$, $\vec{e}_6$, $\vec{e}_9$, $\vec{b}^{(2)}_{s_2}\in
span(\vec{e}_1,\vec{e}_4)$, $\vec{b}^{(2)}_{s_4}=\vec{b}^{(2)}_{s_5}=\vec{b}^{(2)}_{s_6}=\vec{0}$} \\ 
  & \multicolumn{10}{|c|}{$\vec{b}^{(2)}_{s_8},\vec{b}^{(2)}_{s_9} \in span(\vec{e}_2,\vec{e}_5,\vec{e}_{10})$} 
  \\ \hline
Packets received by 3 & \multicolumn{10}{|c|}{$w_4$, $w_6$, $s_2$, $s_3$, $s_8$, $s_9$, $s_{10}$} \\ \hline
$\vec{b}^{(3)}$ received by 3 & \multicolumn{10}{|c|}{$\vec{e}_4$, $\vec{e}_6$, $\vec{b}^{(3)}_{s_2}=
  \vec{b}^{(3)}_{s_3}=\vec{0}$, $\vec{b}^{(3)}_{s_8},\vec{b}^{(3)}_{s_9},\vec{b}^{(3)}_{s_{10}} \in span
(\vec{e}_1,\vec{e}_5,\vec{e}_7,\vec{e}_8)$} \\ \hline
\end{tabular}
\end{table}

The first 6 slots of phase 2 illustrate some of the finer points of
the algorithm. Specifically, in slot 1 of phase 2, the transmitted
packet $s_1$ is only received by user 1. Since the vector
$\vec{b}^{(1)}_{s_1}$, corresponding to packet $s_1$, belongs to the
span of the vectors $\{ \vec{b}^{(1)}_p: p\in Q_{\{1,2\}} \}$, it
follows that $\vec{b}^{(1)}_{s_1} \in
span(\mc{B}^{(1)}_{\{1,2\}}(t_1))$, i.e.~$\vec{b}^{(1)}_{s_1} \in
span(\vec{e}_2,\vec{e}_4)$. The packets received by user 1 up to now
span the space $span(\mc{B}_{D_1}(t_1))=span(\{ \vec{e}_1, \vec{e}_7,
\vec{e}_9 \})$, so that $s_1$ brings innovative information for this
user. Hence, $T^1_{\{1,2\}}$, which counts the number of innovative
tokens that user 1 has yet to recover from $Q_{\{1,2\}}$, must be
decreased by one.

\begin{table}[t]
\centering
\caption{$\mbox{\texttt{CODE1}}_{pub}$ execution. Erasures and queue contents at end of phase 3.}
\label{phase3:er}
\begin{tabular}{|c|c|c|c|c|c|c|c|} \hline
\multicolumn{8}{|c|}{Phase 3} \\ \hline
Processing & \multicolumn{7}{|c|}{$Q_{\{1,2,3\}}$} \\ \hline
Slot & 11 & 12 & 13 & 14 & 15 & 16 & 17 \\ \hline
Erasure event & RRR & RER & RRR & ERR & RRE & RER & RRE \\ \hline
  & \multicolumn{7}{|c|}{$T^1_{\{1,2,3\}},T^2_{\{1,2,3\}},T^3_{\{1,2,3\}}$} \\ \hline
Index value at end of slot & 5,3,4 & 4,3,3 & 3,2,2 & 3,1,1 & 2,0,1 & 1,0,0 & 0,0,0 \\ \hline
\multicolumn{8}{|c|}{Queue contents at end of phase 3} \\ \hline
Packets received by 1 & \multicolumn{7}{|c|}{$u_1$, $u_7$, $u_9$, $s_1$, $s_3$, $s_7$, $s_{11}$, $s_{12}$, 
$s_{13}$, $s_{15}$, $s_{16}$, $s_{17}$} \\ \hline
$\vec{b}^{(1)}$ received by 1 & \multicolumn{7}{|c|}{$\vec{e}_1,\vec{e}_7,\vec{e}_9,\vec{b}^{(1)}_{s_1}, 
\vec{b}^{(1)}_{s_3} \in span(\vec{e}_2,\vec{e}_4), \vec{b}^{(1)}_{s_7}=\vec{0}$} \\ 
  & \multicolumn{7}{|c|}{$\vec{b}^{(1)}_{s_{11}},\vec{b}^{(1)}_{s_{12}},\vec{b}^{(1)}_{s_{13}},
\vec{b}^{(1)}_{s_{15}},\vec{b}^{(1)}_{s_{16}},\vec{b}^{(1)}_{s_{17}}\in span(\vec{e}_3,\vec{e}_1,\vec{e}_{10},
\vec{b}^{(1)}_{s_2},\vec{b}^{(1)}_{s_4},\vec{b}^{(1)}_{s_{10}})$} \\ \hline
Packets received by 2 & \multicolumn{7}{|c|}{$v_3$, $v_6$, $v_9$, $s_2$, $s_4$, $s_5$, $s_6$, $s_8$, $s_9$, $s_{11}$,
$s_{13}$, $s_{14}$, $s_{15}$, $s_{17}$} \\ \hline
$\vec{b}^{(2)}$ received by 2 & \multicolumn{7}{|c|}{$\vec{e}_3$, $\vec{e}_6$, $\vec{e}_9$, $\vec{b}^{(2)}_{s_2}
\in span(\vec{e}_1,\vec{e}_4)$, $\vec{b}^{(2)}_{s_8},\vec{b}^{(2)}_{s_9}\in span(\vec{e}_2,\vec{e}_5,\vec{e}_{10})$} \\
  & \multicolumn{7}{|c|}{$\vec{b}^{(2)}_{s_{11}},\vec{b}^{(2)}_{s_{13}},\vec{b}^{(2)}_{s_{14}},
\vec{b}^{(2)}_{s_{15}},\vec{b}^{(2)}_{s_{17}}\in span(\vec{e}_7,\vec{e}_8,\vec{b}^{(2)}_{s_3},\vec{b}^{(2)}_{s_7})$} 
\\ \hline
Packets received by 3 & \multicolumn{7}{|c|}{$w_4$, $w_6$, $s_2$, $s_3$, $s_8$, $s_9$, $s_{10}$, $s_{11}$, $s_{12}$,
$s_{13}$, $s_{14}$, $s_{16}$} \\ \hline 
$\vec{b}^{(3)}$ received by 3 & \multicolumn{7}{|c|}{$\vec{e}_4,\vec{e}_6,\vec{b}^{(3)}_{s_8},\vec{b}^{(3)}_{s_9},
\vec{b}^{(3)}_{s_{10}}$} \\
  & \multicolumn{7}{|c|}{$\vec{b}^{(3)}_{s_{11}},\vec{b}^{(3)}_{s_{12}},\vec{b}^{(3)}_{s_{13}},\vec{b}^{(3)}_{s_{14}},
\vec{b}^{(3)}_{s_{16}} \in span(\vec{e}_{10},\vec{b}^{(3)}_{s_4},\vec{b}^{(3)}_{s_5},\vec{b}^{(3)}_{s_6},
\vec{b}^{(3)}_{s_7})$} \\ \hline
\end{tabular}
\end{table}

In slot 2, the transmitted packet $s_2$ is received by users 2,
3. Using a similar argument as for user 1 in slot 1, we conclude that
user 2 gains an innovative token (since $\vec{b}^{(2)}_{s_2} \in
span(\{ \vec{b}^{(2)}_p: p\in Q_{\{1,2\}} \}=
span(\vec{e}_1,\vec{e}_4)$ and $\mc{B}_{D_2}= \{ \vec{e}_3,\vec{e}_6,
\vec{e}_9 \}$) and the $T^2_{\{1,2\}}$ index must be accordingly
reduced. It is important to note that, since at the time of
transmission of $s_2$ it holds $T^1_{\{1,2\}}>0$, $s_2$ is also an
innovative token for user 1. Additionally, $s_2$ is a token for users
1, 2 (due to Lemma~\ref{cruc}) and 3 (since user 3 received $s_2$), so
it is moved to queue $Q_{\{1,2,3\}}$. Hence, user 1 can now recover
the innovative token corresponding to packet $s_2$ from queue
$Q_{\{1,2,3\}}$ instead of $Q_{\{1,2\}}$, so that the $T^1_{\{1,2\}}$,
$T^1_{\{1,2,3\}}$ indices are modified accordingly. Notice that,
though $s_2$ becomes a token for user 3, it is \textit{not} innovative
for user 3 since it holds $\vec{b}^{(3)}_{s_2}=\vec{0}$. A similar
interpretation can be given for the actions in slot 3 by swapping the
roles of users 1, 2.

In slots 4, 5, 6, the transmitted packets are only received by user 2,
so that step 4 of \texttt{ACTFB1} is applicable and all 3 transmitted
packets are moved to $Q_{\{1,2,3\}}$. By construction of the
algorithm, it also holds
$\vec{b}^{(1)}_{s_4},\vec{b}^{(1)}_{s_5},\vec{b}^{(1)}_{s_6} \in span(
\mc{B}^{(1)}_{\{1,3\}}(t_1))$. Since, at the beginning of slot 4, the
vectors in $\mc{B}^{(1)}_{\{1,3\}}$ span a subspace of dimension
$T^1_{\{1,3\}}=2$ (due to Lemma~\ref{trueind}), it follows that
$\vec{b}^{(1)}_{s_4},\vec{b}^{(1)}_{s_5},\vec{b}^{(1)}_{s_6}$ are
linearly \textit{dependent even} though $s_4,s_5,s_6\in
Q_{\{1,2,3\}}$. The last statement clearly demonstrates the true
meaning of sets $\mc{B}^{(i)}_{\mc{S}}(t)$: these sets contain the
vectors $\vec{b}$ corresponding to tokens that remain to be received
by user $i$ from queue $Q_{\mc{S}}$ at slot $t$. It is exactly due to
the fact that the packets stored in $Q_{\mc{S}}$ are not
simultaneously innovative for all users $i\in \mc{S}$ that the sets
$\mc{B}^{(i)}_{\mc{S}}$ must be introduced in the first place.

At the end of phase 2 (denote this time as $t_2$), the indices for
$Q_{\{1,2,3\}}$ are as follows: $T^1_{\{1,2,3\}}=6$,
$T^2_{\{1,2,3\}}=4$, $T^3_{\{1,2,3\}}=5$. In phase 3, the transmitter
sends linear combinations of all packets stored in $Q_{\{1,2,3\}}$
until all $T$ indices become zero. Table~\ref{phase3:er} shows the
erasures that occurred and queue contents at the end of phase 3 (note
that only step 3 of \texttt{ACTFB1} is now applicable and the slot
numbering in phase 3 continues from where phase 2 stopped). At the end
of phase 3, each user has collected 10 innovative tokens
(i.e. linearly independent equations) and can decode its packets by
solving a linear system.